\documentclass[12pt,twoside]{reedthesis}
\usepackage{graphicx,latexsym} 
\usepackage{amssymb,amsthm,amsmath}
\usepackage{longtable,booktabs,setspace} 
\usepackage{listings}
\usepackage{chemarr} %% Useful for one reaction arrow, useless if you're not a chem major
\usepackage[hyphens]{url}
\usepackage{rotating}
\usepackage[numbers]{natbib}
\usepackage{xspace}
\usepackage[dvipsnames]{xcolor}
\usepackage{lucaudill}
\usepackage{aastexmacros}

\definecolor{codegreen}{rgb}{0,0.6,0}
\definecolor{codegray}{rgb}{0.5,0.5,0.5}
\definecolor{codepurple}{rgb}{0.58,0,0.82}
\definecolor{backcolour}{rgb}{0.95,0.95,0.92}

\lstdefinestyle{mystyle}{
    backgroundcolor=\color{backcolour},   
    commentstyle=\color{codegreen},
    keywordstyle=\color{magenta},
    numberstyle=\tiny\color{codegray},
    stringstyle=\color{codepurple},
    basicstyle=\ttfamily\footnotesize,
    breakatwhitespace=false,         
    breaklines=true,                 
    captionpos=b,                    
    keepspaces=true,                 
    numbers=left,                    
    numbersep=5pt,                  
    showspaces=false,                
    showstringspaces=false,
    showtabs=false,                  
    tabsize=2
}
\usepackage{ulem}
\title{The Missing Baryon Problem via Cosmological Zoom-in Simulations
}
\author{Lucas H. Caudill}
\date{May 2022}
\division{Mathematics and Natural Sciences}
\advisor{Johnny Powell}
\department{Physics}
\begin{document}

  \maketitle
  \frontmatter % this stuff will be roman-numbered
  \pagestyle{empty} % this removes page numbers from the frontmatter

% Acknowledgements (Acceptable American spelling) are optional
% So are Acknowledgments (proper English spelling)
    \chapter*{Acknowledgements}
	I would like to thank my thesis advisor, Johnny Powell, whose guidance and friendship over the %last three 
	years have been an integral part of my experience at Reed. I don't think I can fully express how meaningful Johnny's mentorship has been to me, %on a personal level
	not just in terms of my academic career, but also my personal development. I will always cherish our many ecstatic discussions of physics, and of life in general.
	
	I would like to thank my friends Dan, Amir, Emma, Arthur, Khalil, and Teo for their support and for making my last year here so memorable. After %an incredibly difficult junior year, I didn't think that my college experience was salvageable. 
	some incredibly difficult experiences as a junior, I didn't think there was much hope for my senior year. 
	I can't tell you how grateful I am that you proved me wrong, I love you guys.
	%Thank you so much for proving me wrong. I love you guys. 
	
	Finally, I would like to thank my family; those who brought me into this world and who have experienced it alongside me. You have provided me with so much, and I wouldn't be who am I today without you. I love you, always.
	
	%mentorship throughout my time at Reed has helped to shape the trajectory of my academic career. %and, in all likelihood, my life after graduation. 
	
	%, and more generally my life. 

% The preface is optional
% To remove it, comment it out or delete it.
    %\chapter*{Preface}
	%This is an example of a thesis setup to use the reed thesis document class.

    \chapter*{List of Abbreviations}
		%\printglossary

	\begin{table}[h]
	\centering % You could remove this to move table to the left
	\begin{tabular}{ll}
		%\textbf{ABC}  	&  American Broadcasting Company \\
		%\textbf{CBS}  	&  Columbia Broadcasting System\\
		%\textbf{CDC}  	&  Center for Disease Control \\
		%\textbf{CIA}  	&  Central Intelligence Agency\\
		%\textbf{CLBR} 	&  Center for Life Beyond Reed\\
		%\textbf{CUS}  	&  Computer User Services\\
		%\textbf{FBI}  	&  Federal Bureau of Investigation\\
		%\textbf{NBC}  	&  National Broadcasting Corporation\\
		\textbf{AGN} & Active Galactic Nucleus \\
		\textbf{AGORA} & Assembling Galaxies Of Resolved Anatomy \\
		\textbf{AMR} & Adaptive Mesh Refinement \\ 
		\textbf{CGM} & Circumgalactic Medium \\
		\textbf{COS} & Cosmic Origins Spectrograph \\
		\textbf{ChaNGa} & Charm N-body GrAvity solver \\
		\textbf{IDE} & Integrated Development Environment \\
		\textbf{IGM} & Intergalactic Medium \\
		\textbf{ISM} & Interstellar Medium \\
		\textbf{$\mathbf{\Lambda}$CDM} & Lambda Cold Dark Matter \\
		\textbf{QSO} & Quasi-Stellar Object (a.k.a. quasar) \\
		\textbf{SPH} & Smoothed Particle Hydrodynamics 
	\end{tabular}
	\end{table}

    \tableofcontents
% if you want a list of tables, optional
    %\listoftables
% if you want a list of figures, also optional
    \listoffigures

% The abstract is not required if you're writing a creative thesis (but aren't they all?)
% If your abstract is longer than a page, there may be a formatting issue.
    \chapter*{Abstract}
	This thesis explores the missing baryon problem in a computational context. An overview of the problem is given, along with a discussion regarding the relevance of the Circumgalactic Medium (CMG) and cosmological Zoom-in simulations. The mechanisms underlying the N-body code ChaNGa (H. Menon, et al., Computational Astrophysics and Cosmology \textbf{2}, 1 (2015), arXiv:1409.1929), as well as the data visualization and analysis tools yt (M. J. Turk, et al., \textbf{192}, 9 (2011), arXiv:1011.3514) and trident (Hummels, et al., \textbf{847}, 59 (2017), arXiv:1612.03935) are presented at a conceptual level. Finally, a series of synthetic quasar absorption spectra produced by using trident on a ChaNGa dataset from (S. Roca-F\`{a}brega, et al., \textbf{917}, 64 (2021), arXiv:2106.09738) at redshift of $z\sim4$ are shown. The low relative flux exhibited by these spectra render absorption features indistinguishable from background noise, and possible explanations for this phenomena such as high redshift are discussed. Though the resulting spectra exhibit serious obstacles for both qualitative and quantitative interpretation, they provide a ``proof-of-concept'' for future work, demonstrating trident's compatibility with ChaNGa's data format. Future prospects for using trident to analyze the CGM as simulated by ChaNGa are discussed, as well as possible extensions of this project. 
	
	\chapter*{Dedication}
	To my mother, who made all of this possible.
	%I dedicate this thesis to my mother, who made everything possible.

  \mainmatter % here the regular arabic numbering starts
  \pagestyle{fancyplain} % turns page numbering back on

  \mainmatter % here the regular arabic numbering starts
  \pagestyle{fancyplain} % turns page numbering back on
%\begin{figure}
    %\centering
    %\begin{tikzpicture}
        %\draw[lightgray, fill] (0,0) circle (0.5);
        %\draw (0,0) node[anchor=north] {$+$};
    %\end{tikzpicture}
    %\caption{Caption}
    %\label{fig:bohr}
%\end{figure}

%\begin{tikzpicture}
    %\feynmandiagram [layered layout, horizontal=a to d] {
  %a -- [scalar,red] b -- [fermion,blue] c -- [gluon,orange] d,
  %b -- [photon, half left] c, % this produces the curved photon line
%};
%\end{tikzpicture}
%\cite{barnes-hut-princeton}
%\cite{romeel}
%\cite{eagle}
%\cite{tumlinson}
%\cite{joel}
%\cite{einstein}
%\cite{butsky}
%\cite{danforth}
%\cite{whim}
%\cite{trident}
%Appendix~\ref{appendix:final}
    \chapter*{Introduction}
         \addcontentsline{toc}{chapter}{Introduction}
	\chaptermark{Introduction}
	\markboth{Introduction}{Introduction}
	
	%\cite{FUCK}
	
	%\cite{springel2005}
	%\cite{mitch}
	%\cite{beckett}
	%\cite{finnish}
	
	%As I understand it, the introduction should provide some basic background information for the reader who is not particularly acquainted with the field of study. \cite{yt}
	
	%\note{This thesis explores computer simulations as a means of probing a particular problem in astrophysics}
	
	%The subject of this thesis is %concerned with 
	%This thesis is concerned with \emph{computational astrophysics}, or the application of computer technology to problems in the field of astrophysics. Specifically, this thesis focuses in of N-body simulations, which are used to simulate the motion of physical bodies under the influence of their mutual gravitational force\footnote{Among other forces, see Sec.~\ref{sec:sph}.}, as well as the application of these simulations to a particular open question in astrophysics. It is my hope that I have provided a discussion of the topic at hand as well as the rather specific sub-field that it falls under
	%that can help to illuminate both the problem at hand
	
	This thesis %presents a project that %in computational astrophysics 
	explores the application of computer simulations (specifically those involving galaxies) to a particular open question in the broader field of astrophysics. Though the reader will not walk away from this text with a comprehensive understanding of astrophysics, computer simulations, or even the specific problem at hand, it is my hope that the discussion provided herein can at least help to illuminate some important concepts relating to all three, and maybe even inspire the reader to pursue the subject in a %way that extends beyond this document
	capacity that extends beyond this document.
	
	%Though I have tried my best to %present information on the subject 
	%structure this document so that information on the subject is presented in a linear fashion, the topic at hand --in my experience-- %is 
	%has proven to be a conceptual labyrinth. As such, there are a number of places where the document breaks continuity, referring the reader to previous or later sections 
	
	%I have not done this to confuse you, but rather as an attempt to 
	
	The purpose of this chapter is to introduce a number of concepts relevant to understanding the document as a whole. 
	%give some of the motivation for the use of computer simulations in the study of astrophysics, as well as to equip the reader with a few fundamental concepts that pertain to this work. While the former is intended go give the reader a sense of the driving force behind some of the studies and techniques described herein (and could therefore arguably be of some value to anyone interested in the field), the latter constitutes discussions of a number of concepts scattered across a range physics subfields, from rudimentary classical mechanics to spectroscopy. %The value that these discussions offer the reader will inevitably vary by background
	%Because of this, the value that readers will draw from these discussions will inevitably be determined by their background and level of familiarity with physics. Thus, readers are invited (if not actively encouraged) to skip through this section as they see fit. 
	%This section is 
	Sections \ref{sec:newton}-\ref{sec:cmb} are intended as a kind of ``crash course'' in a number of physics concepts that are relevant to understanding this thesis. Readers who are relatively unfamiliar with field will get the most value out of these sections, while those who are already well-versed may be inclined to skip them altogether. Sec.~\ref{sec:observation-simulation} aims to situate this project in its appropriate context relative to %the broader field of 
	astrophysics as a whole. Finally, Sec.~\ref{sec:simulation-types} describes common astrophysical simulation types\footnote{More specifically, N-body simulation types.}, including Zoom-in simulations, which are of primary importance to this project. %Finally, Sec.~\ref{sec:personal-interest} provides an explanation of my own personal interest in pursuing this project which is --hopefully-- not overly self-indulgent.
	
	%give a sense of the role of this project, and computational studies
	
	%aims to provide 
	
	%\footnote{I assure you this is not a bid to avoid getting fact-checked by more knowledgeable readers.%, by all means skip!}. 
	
	\section{Newton's Laws}\label{sec:newton}
	
	The first topic %that bears discussion 
	that warrants discussion is that of Newton's laws, which are foundational to modern classical mechanics \cite{classical}. They are among the first concepts taught in a physics curriculum, and %the reader has no doubt encountered them at some point
	readers of all backgrounds have doubtlessly already encountered them in some form. %Their ubiquity is certainly well-earned, and while they could well be considered assumed knowledge for this 
	While they could be taken to be assumed knowledge in the context of this thesis, %their relevance to the field is such that 
    they bear repeating by virtue of their relevance to the field, if only for those who are not deeply immersed in physics in their daily lives. %Expressed in somewhat simplified terms, 
	Drawing from \cite{classical} and \cite{bob}, Newton's laws are:
	
	\begin{enumerate}
	    \item A body at rest will remain at rest, and a body in motion will remain in motion (at constant speed), unless acted upon by an external force.
	    
	    \item The net force on a body is proportional to its mass and its acceleration, which can be expressed mathematically for a set of $n$ forces as
	    \begin{equation}\label{eq:n2}
	        \fvec_\text{net}=\sum_{i=1}^n \fvec_i=m\avec.
	    \end{equation}
	    
	    \item For every action there is an equal and opposite reaction; forces act in opposing pairs. For two bodies, this law reads
	    \begin{equation}\label{eq:n3}
	        \fvec_{12}=-\fvec_{21},
	    \end{equation}
	    where $\fvec_{12}$ is the force exerted on object 1 by object 2, and $\fvec_{21}$ is the force exerted on object 2 by object 1. Note that the minus sign tells us that these forces oppose one another. %$\fvec_{21}$ is just as \emph{strong} as $\fvec_{12}$
	\end{enumerate}
	
	In the context of this thesis, particularly with regards to %solving for gravitational motion 
	simulation techniques (see Ch.~\ref{ch:changa}), it is most important to understand laws 2 and 3. By providing a direct relation between the total force acting on a body and its acceleration, the second law allows us to derive the equations that are used in the calculation for gravitational motion (namely those for position and velocity, again, refer to Ch.~\ref{ch:changa}).
	
	The third law tells us that forces act in opposing pairs. The example that was used in my high school physics class was that if you go over to your bedroom wall and start pushing on it, the wall will not move because it is also ``pushing back'' on you with the same amount of force in the opposite direction. This is true of all forces, not just those of the bedroom-wall variety, and as was the case for the second law, it is most relevant to this document in the gravitational context.
	%, and, again, is 
	%and in the context of this document --as was the case with the second law-- it is particularly relevant when it comes to gravity. %Gravitational force
	
	%mutual gravitation
	
	%(see Sec.~\ref{sec:gravity}). Gravitational forces act in pairs
	
	%which are ultimately used in 
	%Eq.~\ref{eq:gravity vec sum}
	%\cite{classical}
	
	\section{Gravity}\label{sec:gravity}
	
	Gravity is one of the main forces that drives the structural evolution of the cosmos, playing a role at all levels %from the formation of
	including cosmic filaments, galaxies, and stars \cite{bob}. Naturally this means that it is also of primary importance for many astrophysical simulations. 
	%We all have an intuitive understanding of gravity from our daily lives, but a robust physical description of the phenomenon is far less intuitive than mere observation of its existence.
	%All humans have an intuitive understanding of gravity
	%We are all familiar with the concept of gravity from our daily lives --that objects fall when we drop them is a fact that is experimentally verified on a daily basis.
	A mathematical description of the gravitational force between two massive bodies was first derived by Isaac Newton through the application of his three laws (see Sec.~\ref{sec:newton}) to Kepler's laws of planetary motion\footnote{Which themselves were the result of careful study of the observations of the first state-funded astronomer Tycho Brahe (1546-1601) \cite{bob}.} \cite{bob}. For two bodies with masses $m_1$ and $m_2$, this description takes the form
	\begin{equation}\label{eq:gravity magnitude}
	    F=G\frac{m_1m_2}{r^2},
	\end{equation}
	where $r$ is the distance between the two bodies and $G$ is known as the \emph{gravitational constant}, with a value of roughly $6.673\times10^{-11}\si{\newton\,\meter^2\,\kilo\gram^{-2}}$ \cite{bob}. In Newtonian gravity, $G$ is understood to be a constant of nature which necessarily appears in all gravitational force calculations \cite{bob}.
	%, which is the gravitational model 
	%The physical meaning of this constant is beyond the scope of this document, and to the reader it can simply be understood as a number that necessarily appears in calculations of gravitational force. 
	
	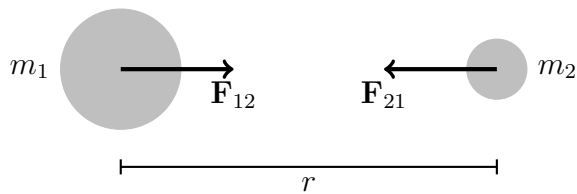
\begin{figure}[p!tb]
	    \centering
	    \begin{tikzpicture}
	        \filldraw[color = lightgray, fill = lightgray] (0,0) circle (0.8);
	        \filldraw[color = lightgray, fill = lightgray] (5,0) circle (0.4);
	        %\draw[thick, dashed] (0,0) -- (5,0);
	        \draw[ultra thick, ->] (0,0) -- (1.5,0) node[anchor = north]{$\fvec_{12}$};
	        %\draw (1.5,0) node[anchor = north]{$\fvec_{21}$};
	        \draw[ultra thick, ->] (5,0) -- (3.5,0) node[anchor=north]{$\fvec_{21}$};
	        
	        \draw (-0.8,0) node[anchor=east]{$m_1$};
	        \draw (5.4,0) node[anchor=west]{$m_2$};
	        
	        \draw[thick] (0,-1.3) -- (5,-1.3);
	        \draw[thick] (0,-1.2) -- (0,-1.4);
	        \draw[thick] (5,-1.2) -- (5, -1.4);
	        \draw (2.5,-1.3) node[anchor=north]{$r$};
	    \end{tikzpicture}
	    \caption{A depiction of the mutual gravitational interaction between two massive bodies $m_1$ and $m_2$ separated by a distance $r$.}
	    \label{fig:gravity}
	\end{figure}
	
	Eq.~\ref{eq:gravity magnitude} is referred to as Newton's law of universal gravitation, and can be applied to all massive bodies (not just large ones) \cite{bob}. There are a few things worth noting about this equation before moving on. First, the appearance of the mass terms $m_1$ and $m_2$ in the numerator means that gravitational force is stronger when larger masses are involved. This is why we are generally much more aware of the gravitational influence of the Earth than the other objects that happen to be around us. %as we go about our day. 
	Secondly, the $r^2$ term in the denominator means that gravitational force drops off drastically with distance, which is why we tend not to be too concerned about the gravitational force exerted on us by Pluto. Finally, according to Newton's third law, Eq.~\ref{eq:gravity magnitude} is really a statement about a \emph{pair} of forces acting with the same strength in opposite directions. We exert the same amount of force that the Earth exerts on us, but since the Earth is so big this doesn't affect it that much. This is why the term gravity is often interchanged with mutual gravitation, as it causes massive bodies mutually attract each other. This is shown in Fig.~\ref{fig:gravity}.
	
	%This is why gravity is sometimes referred to as a mutual force between objects
	
	%the force of mutual gravitation between objects
	%The gravitational force that you exert on the Earth is actually 
	%Eq.~\ref{eq:gravity magnitude} is really a statement about the \emph{magnitude} or strength of the gravitational force between two bodies, 
	
	\section{Light and Matter}\label{sec:light-matter}
	
	Interactions between light and matter %constitute 
	are a fundamental concept %for understanding this thesis
	in this thesis, so they warrant a brief review. I should note that in this section, and throughout the document, I use the term ``light'' as shorthand for electromagnetic radiation, a category that includes the light that is visible to the human eye as well as other forms of radiation such as radio waves, microwaves, and x-rays \cite{bob}. Precisely \emph{what} this radiation is is a concept from electromagnetism that is beyond the scope of this document\footnote{See Ch. 9 of \cite{griffiths} for a treatment of electromagnetic radiation at the undergraduate level.}, but readers with a scientific background have likely already encountered this topic before, and others probably %even 
	possess a somewhat intuitive understanding of it from life in the modern world\footnote{Consider, for instance, cell phones and wifi routers, both of which transmit information remotely using light in the radio wave range \cite{fda}. Alternatively, consider the ultraviolet light from the sun, responsible for giving humans sunburns. %These are phenomena that most of us are likely quite familiar with from our ordinary lives, and 
	While our experiential understandings of these phenomena cannot hope to usurp the robust descriptions offered by science, they do provide a baseline %for understanding how to think about light
	understanding of light, which I hope to build upon in the coming discussion.}.
	
	In modern physics, light is understood to simultaneously exhibit particle and wave-like properties \cite{tzd}. For instance, light is known to come in discrete ``packets'' or quanta (referred to as photons), which is characteristic of a particle. At the same time, light has both an associated frequency and wavelength, which are unequivocally wave properties. This apparent paradox regarding the nature of light is known as the \emph{wave-particle duality}, and as counter intuitive as it may be, it has been verified time and again by an overwhelming body of scientific work \cite{tzd}. \emph{How} it is that this can be the case is an incredibly deep question in physics, one which I will not get into here. Instead, I will %explain 
	outline how the dual nature of light helps to inform our understanding of the way it interacts with matter, which in turn will give background for understanding how observational techniques in astrophysics work.
	
	Before exploring what the wave-particle duality tells us about how light interacts with matter, I must first \emph{talk} about matter. As many readers are likely aware, matter is composed of minuscule units called atoms. Atoms consist of even smaller electrically charged particles called protons (positive charge), neutrons (neutral charge), and electrons (negative charge) \cite{tzd}. A robust and physically accurate description of atomic structure falls under the purview of quantum mechanics, which lies beyond the scope of this discussion\footnote{Readers who are interested in the quantum description of the atom may refer to an undergraduate quantum mechanics textbook such as \cite{griffiths-qm}.}. %Instead, 
	I will focus on a simplified classical interpretation, specifically the Bohr model of the atom \cite{tzd}, %\footnote{A frankly outdated interpretation, though it serves well for conceptually understanding light-matter interactions at a basic level. Specifically, this description follows closely with the Bohr model of the atom \cite{tzd}.}
	that draws from the theory of electrodynamics \cite{griffiths} as well as experimental and theoretical inquiry from the early twentieth century \cite{tzd}. %Though this interpretation is outdated from the perspective of modern science, it serves well as a conceptual introduction to atomic theory, which is foundational to the light-matter interactions that are most relevant to this thesis. 
	In the interest of simplicity, %I will at times make statements regarding atomic theory that are misleading
	I will make a number of statements throughout this discussion that are misleading with respect to the modern understanding of atomic physics. In these cases I will provide footnotes which attempt to clarify %the misleading 
	precisely \emph{which} aspects of these statements are misleading, though it should be explicitly noted that these footnotes will not act as substitutes for a robust description of the modern physical theory. %and also direct the reader to sources that can provide a more complete understanding 
	%I will also use these footnotes to direct readers to sources 
	Readers who are interested in understanding modern atomic models and their relation to quantum mechanics might start by looking at an undergraduate chemistry text such as \cite{chem}, which provides a treatment of the subject appropriate for someone with little background in physics. Readers looking for a more physics-oriented approach can refer to Ref.~\cite{tzd}, appropriate for a second year physics undergraduate, or Ref.~\cite{griffiths-qm} for a more rigorous treatment at the advanced undergraduate level. I should finally note that, although the interpretation I discuss is outdated from the perspective of modern science, it serves well as a conceptual introduction to the aspects of atomic theory %, which is 
	%foundational to the light-matter interactions 
	that are most relevant to this thesis, which is precisely the intention of this section.
	
	%I will be using footnotes to provide clarification for some of these misleading statements, which will also include sources that readers are encouraged to refer to for information on \emph{modern} atomic theory, 
	
	At a basic level, atoms can be thought of as having two structural components: a central body known as the nucleus, which is made of protons and neutrons, and a set of electrons orbiting said central body \cite{tzd}. According to the classical interpretation of the Bohr model, the orbits of these electrons are very similar to those of the planets around the Sun\footnote{I should point out that the word \emph{classical} is doing a lot of work in this statement. The orbits of electrons as described by quantum mechanics are quite different from gravitational models, both in terms of shape and their physical meaning \cite{chem} \cite{griffiths-qm}.} (i.e. gravitational orbits), 
	since classical electrodynamics prescribes an attractive force between bodies of opposite electric charge\footnote{Which is precisely what the nucleus and electrons are. Recall that the nucleus is composed of positive and neutral particles, and therefore has a net positive charge, while electrons have negative charge.} similar to that of Newtonian gravity (Eq.~\ref{eq:gravity magnitude}) \cite{griffiths}. There is one critical difference\footnote{Again, this is in reference to the classical description. The quantum mechanical description is \emph{fundamentally} different.} between electron orbits and those resulting from gravitational force: whereas gravitational systems exhibit a \emph{continuous} set of orbits (i.e. if one body can orbit another at a distance $r$ as well as at a distance $2r$, then it can also orbit at all distances in between), the electrons in an atom can only occupy a discrete set of orbits \cite{tzd}.
	
	Each orbit has an associated energy that an electron must possess in order to occupy it, with higher energies being associated with ``farther out'' orbits, and vice versa \cite{tzd} \cite{chem}. 
	In order to move from one orbit to another, an electron must 
	experience a change in energy (either negative or positive) \emph{exactly} equal to the energy difference between the two orbits in question \cite{chem}. \emph{How} it is that these changes in energy occur is a question that brings us back to our discussion of light. 
	
	The primary mechanism by which electrons in atoms gain or lose energy, at least in the context of this thesis, is through light. As previously mentioned, light comes in discrete quanta called photons, which also have associated wave properties of frequency and wavelength \cite{tzd}. What I have not yet mentioned is that photons also carry a specific associated energy, related to frequency $\nu$, or wavelength $\lambda$ by the equation \cite{tzd}
	\begin{equation}\label{eq:light}
	    E=h\nu=\frac{hc}{\lambda}.
	\end{equation}
	When an electron in an atom encounters a photon whose energy is equal to the difference in energy between the orbit it currently occupies and a higher (or ``farther out'') orbit, it absorbs that photon and ``jumps'' to the higher orbit \cite{chem}. This process is appropriately referred to as \emph{absorption}. Conversely, electrons ``want'' to occupy the lowest energy orbit available to them\footnote{The idea of orbits being ``available'' is a concept that also emerges from quantum mechanics, refer to discussions of electron orbitals (\emph{not} orbits) and the Pauli exclusion principle in \cite{tzd} and/or \cite{chem}.}. If a lower energy orbit is available, an electron will quickly ``jump down'' to that orbit, emitting a photon that carries away the exact amount of energy required to make the transition \cite{chem}. This process is referred to as \emph{emission}. Fig.~\ref{fig:emission} provides a visual interpretation of these processes for the classical model.
	
	%Emission and absorption are fundamental concepts in this thesis
	At the beginning of this section when I said that interactions between light and matter are fundamental concepts to this thesis, I was referring to precisely the phenomena of emission and absorption. These physical processes are at the heart of a particular observational technique that is of primary importance in the context of this project, which will be discussed in the following section (Sec.~\ref{sec:spectroscopy}).
	
	\begin{figure}
        \centering
        \begin{tikzpicture}
                \draw[lightgray, fill] (0,0) circle (0.3);
                \draw (0,0) node[font = \large]{$\mathbf{+}$};
                
                \draw (0,0) circle (0.8);
                \draw (0,0) circle (1.8);
                \draw (0,0) circle (3);
                
                \draw (0,0.8) node[anchor = south]{$n=1$};
                \draw (0,1.8) node[anchor = south]{$n=2$};
                \draw (0,3) node[anchor = south]{$n=3$};
                
                \draw[gray, fill] (0.7297,-0.3278) circle (0.1);
        \end{tikzpicture}
        \caption{The Bohr model of the atom, with a single orbiting electron. The central nucleus is depicted as a gray circle labelled with a ``plus'' sign, indicating its electric charge. The first three orbits are shown and numbered, and the orbiting electron is depicted as a dark gray dot in the first orbit. This image was adapted from a similar figure that was uploaded to Wikipedia by the user JabberWock, which can be found by visiting \href{https://en.wikipedia.org/wiki/Bohr_model}{https://en.wikipedia.org/wiki/Bohr\_model}.}
        \label{fig:bohr}
    \end{figure}
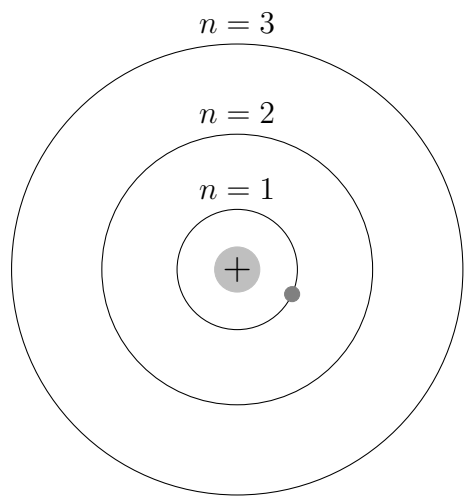
	
	\begin{figure}
        \centering
        \begin{subfigure}[b]{0.49\textwidth}
            \centering
            \begin{tikzpicture}
                \draw[lightgray, fill] (0,0) circle (0.3);
                \draw (0,0) node[font = \large]{$\mathbf{+}$};
                
                \draw (0,0) circle (0.8);
                \draw (0,0) circle (1.8);
                \draw (0,0) circle (3);
                
                \draw[thick, ->] (1.0653, 1.4509) -- (1.68667, 2.29732);
                
                \draw[gray, fill] (1.0653, 1.4509) circle (0.1);
                \draw[gray, fill] (1.7754, 2.4182) circle (0.1);
                
                %\feynmandiagram[horizontal = a to b] {a -- [photon] -- b};
                \draw[thick, ->, snake=coil,segment aspect=0] (3.6,1.6) node[anchor=west]{$\gamma_a$} -- (1.8,1.6);
            \end{tikzpicture}
            \caption{}
            \label{blah1}
        \end{subfigure}
        \hfill
        \begin{subfigure}[b]{0.49\textwidth}
            \centering
            \begin{tikzpicture}
                \draw[lightgray, fill] (0,0) circle (0.3);
                \draw (0,0) node[font = \large]{$\mathbf{+}$};
                
                \draw (0,0) circle (0.8);
                \draw (0,0) circle (1.8);
                \draw (0,0) circle (3);
                
                %\draw[fill] (0.7297,-0.3278) circle (0.1);
                
                \draw[thick, ->] (1.77544, 2.41823) -- (1.15403, 1.57185);
                
                \draw[gray, fill] (1.0653, 1.4509) circle (0.1);
                \draw[gray, fill] (1.7754, 2.4182) circle (0.1);
                
                \draw[thick, ->, snake=coil,segment aspect=0] (1.8,1.6) -- (3.6,1.6) node[anchor = west]{$\gamma_e$};
            \end{tikzpicture}
            \caption{}
            \label{blah}
        \end{subfigure}
        \caption{A depiction of emission and absorption facilitating electron transitions between the second and third orbits for an atom, according to the classical Bohr model. On the left a photon $\gamma_a$, depicted as a squiggly line, is absorbed by the electron in the second orbit, causing it to transition to the third orbit. On the right, a photon $\gamma_e$ is emitted from an electron in the third orbit as it transitions to the second orbit. In both cases, the initial and final positions of the electrons are depicted as dark gray dots, with the arrow indicating the direction of the transition. These images were adapted from a similar figure that was uploaded to Wikipedia by the user JabberWock, which can be found by visiting \href{https://en.wikipedia.org/wiki/Bohr_model}{https://en.wikipedia.org/wiki/Bohr\_model}.}
        \label{fig:emission}
    \end{figure}
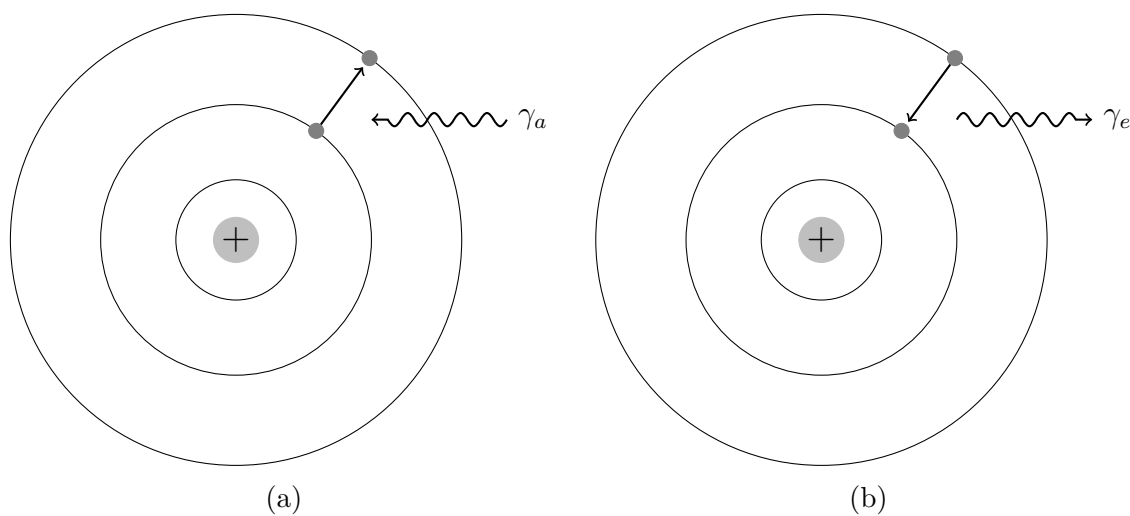
	
	%According to the wave-particle duality, all photons have associated wave properties (frequency and wavelength) that determine the \emph{type} of light they are (e.g. visible light, ultraviolet or
	
	\section{Spectroscopy}\label{sec:spectroscopy}
	
	%The previous section discussed the basics of atomic interaction with light. As its title suggests, this section is intended to introduce an observational technique known as \emph{spectroscopy} through which scientists are able to extract a great deal of information regarding matter in space based purely on the light that it absorbs or emits \cite{bob} \cite{tumlinson}.
	
\begin{figure}
	    \centering
	    \includegraphics[scale=0.5]{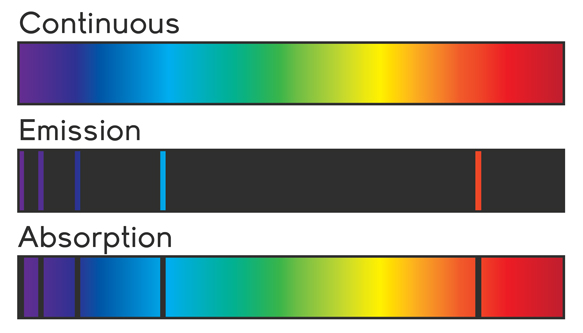}
	    \caption{A depiction of three visible light spectra. The top spectrum is continuous, with no distinct absorption or emission features. In the middle is an emission spectrum in which light is emitted at discrete wavelengths characteristic of the emitting element. At the bottom is an absorption spectrum, exhibiting dark absorption lines in an otherwise continuous spectrum at the same characteristic wavelengths as the emission spectrum. This image was borrowed from (\emph{Atomic Spectroscopy}, physicsopenlab.org, \href{https://physicsopenlab.org/2015/11/30/atomic-spectroscopy/}{https://physicsopenlab.org/2015/11/30/atomic-spectroscopy/}, accessed 27 Apr. 2022).}
	    \label{fig:spectral lines}
	\end{figure}

	The discussion of absorption in the previous section centered around the case of a single photon with exactly the right amount of energy hitting an electron and exciting it to a higher energy state. In reality, absorption generally occurs when a \emph{continuous spectrum} of light\footnote{Such as white light, which consists of a continuum of light from all the wavelengths on the visible spectrum \cite{chem} (400nm$<\lambda<$700nm \cite{tzd}).} passes through a material \cite{bob}. In this case, the light will exhibit an \emph{absorption spectrum} (Fig.~\ref{fig:spectral lines}) upon passing through the material, 
	%in which specific wavelengths of light corresponding to absorbed photons will %appear to be ``missing'' or ``blacked out'' from the original spectrum.
	which looks almost identical to the original continuous spectrum, but with a few dark lines corresponding to wavelengths of absorbed photons.
	In contrast, a material that is heated will lose energy through emission, resulting in an \emph{emission spectrum} (Fig.~\ref{fig:spectral lines}) that looks dark everywhere except for at a few specific wavelengths corresponding to %wavelengths of 
	emitted photons \cite{bob} \cite{chem}. In the mid 1800s scientists realized that %different chemical elements produce
	each chemical element %produces a distinct spectrum
	has a distinct set of wavelengths at which emission or absorption can occur\footnote{These wavelengths are the same for both processes \cite{bob} \cite{chem}.}, and that elements can therefore be identified by the spectra that they produce \cite{bob}. This had profound implications for the field of astronomy; for the first time ever scientists could make empirical determinations about the chemical composition of celestial bodies, a feat that seemed unimaginable just decades prior \cite{bob}.
	
	\emph{Spectroscopy} refers to the study of matter through its interactions with light \cite{chem}, and it still holds a crucial role in observational astrophysics in the modern day \cite{bob} \cite{tumlinson}. As an observational technique, spectroscopy relies on the \emph{spectroscope}, a piece of equipment that takes in light from a variety of ranges on the electromagnetic spectrum (depending on the design of the equipment) and separates it by wavelength\footnote{The light separation is done using components called diffraction gratings, see Ch. 5 of \cite{bob}.} so that its intensity can be measured \cite{bob}. The resulting spectroscopic data gives light intensity with respect to wavelength, which can provide insight into a number of properties of the source or intervening material such as chemical composition (as previously mentioned) and kinematics \cite{tumlinson}. %This thesis focuses heavily on \emph{absorption} spectroscopy
	Discussions of spectroscopy in this thesis focus primarily on absorption spectroscopy as a means of observing low-emission gas around galaxies (see Ch.~\ref{ch:cgm}), as it is most pertinent to the topic at hand. Spectroscopy as an observational technique is discussed further in Sec.~\ref{sec:cgm-observation}, and more details regarding its relevance %to the computational nature of this thesis is discussed 
	in the context of astrophysical simulations can be found in Sec.~\ref{sec:trident}.
	
	\
	
	\section{Dark Matter}\label{sec:dm}
	
	%At the time of writing, dark matter is among the oldest mysteries in physics that remains unsolved \cite{freese}.
	Dark matter (DM) is among the oldest mysteries in physics that still remains unsolved \cite{freese}. By virtue of intrigue it is a term that is likely well-recognized by modern readers. Dark matter is used to refer to matter which cannot be detected by conventional observational means, either due to it not emitting or absorbing electromagnetic radiation (i.e. light), as would be the case for \emph{baryonic} dark matter\footnote{It should be noted that for the remainder of this document, the term ``dark matter'' will be used as shorthand for \emph{nonbaryonic} dark matter, or matter which truly doesn't exhibit electromagnetic interaction. %This is a practice which is becomming increasingly conventional in the field as a whole \cite{carroll}.}
	This nomenclature is consistent with a growing convention in the field \cite{carroll}.}, or due to it interacting only weakly, if at all, with light \cite {bob} \cite{dmhistory}. In more straightforward terms, DM is matter which cannot be \emph{seen} using observational equipment, which relies almost exclusively on radiation originating from or passing through the material being observed.
	%since observational methods in astrophysics rely almost exclusively on 
	%incident light from incredibly distant celestial bodies, DM is matter
	%the detection of light, DM is matter that cannot be observed directly
	%DM is matter which astronomers are not able to \emph{see} 
	
\begin{figure}
	    \centering
	    \includegraphics[scale=1.8]{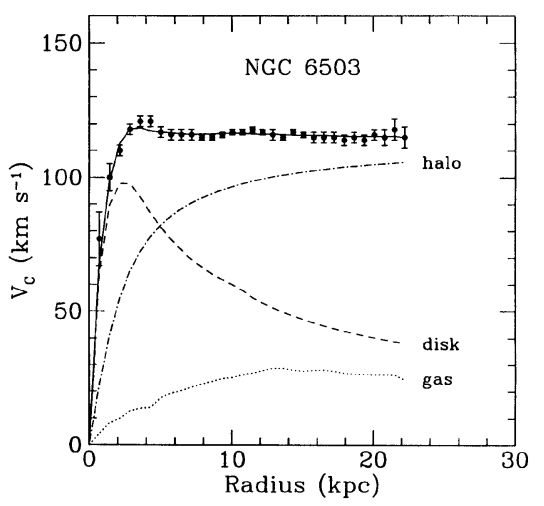}
	    \caption[A rotation curve graph for the galaxy NGC 6503. This figure was borrowed from (K. Freese, International Journal of Modern Physics D \textbf{26}, 1730012-223 (2017),
arXiv:1701.01840). Contributions from the galaxy's disk and gas, as well as the contribution from the dark matter halo are shown.]{A rotation curve graph for the galaxy NGC 6503. This figure was borrowed from (K. Freese, International Journal of Modern Physics D \textbf{26}, 1730012-223 (2017),
arXiv:1701.01840 [astro-ph.CO] ) and can be found along with more information at \href{https://ned.ipac.caltech.edu/level5/Sept17/Freese/Freese2.html}{https://ned.ipac.caltech.edu/level5/Sept17/Freese/Freese2.html}. Contributions from the galaxy's disk and gas, as well as the contribution from the dark matter halo are shown.}
	    \label{fig:rotation curve}
	\end{figure}

	A reasonable question arising from the fact that DM is not directly detectable is how scientists even know it exists to begin with. The answer is gravity. Specifically, 
	%beginning in the 1970s research into rotation curves revealed an apparent discrepancy in...
	advances in research into 
	%regarding 
	galactic rotation curves\footnote{Measurements of the rotation speed of orbiting matter in terms of distance from the galactic center, contributed largely by Vera Rubin \cite{beckett}.} in the 1970s revealed an apparent discrepancy between the observed motion of matter in galaxies and the Newtonian prediction given the mass distribution of said matter \cite{bob} \cite{dmhistory}. 
	%orbital motion of observed galactic matter and the 
	%led to the realization that the motion of 
	%matter within galaxies 
	%observed matter exhibited an unusual discrepancy with Newtonian mechanics \cite{bob} \cite{dmhistory}. 
	%Rotation curve measurements
	These observations showed that at greater distances from the galactic center, matter was rotating much faster than theoretical prediction indicated it should (Fig.~\ref{fig:rotation curve}) \cite{bob} \cite{dmhistory}.
	%matter orbits much faster at greater distances from the galactic center than would be expected
	This revelation led astrophysicists to postulate that a substantial fraction of the matter contained in these galaxies must exist in some unobserved form lying primarily beyond the galactic disk \cite{bob} \cite{dmhistory}. Scientists dubbed this matter ``dark matter'' owing to its apparent non-interaction with light. Dark matter has since been determined to constitute the majority of the matter not only in galaxies, but in the universe as a whole \cite{galaxies-in-universe} \cite{bob} \cite{tumlinson}.
	%about the presence of a significant amount of unobserved mass beyond the galactic disk \cite{bob} \cite{dmhistory}.
	%the theoretical prediction according to the observed distribution of luminous matter (Fig.~\ref{fig:rotation curve}, leading astrophysicists to postulate about the presence of a significant amount of unobserved mass beyond the galactic disk \cite{bob} \cite{dmhistory}. 
	%greater distances from the galactic center matter rotated much faster than the theoretical prediction according to the observed distribution of luminous matter (Fig.~\ref{fig:rotation curve}, leading astrophysicists to postulate about the presence of a significant amount of unobserved mass beyond the galactic disk \cite{bob} \cite{dmhistory}. 
	%, at large distances, galactic matter rotates much faster than would be expected according to Newtonian mechanics given the observed distribution of luminous matter \cite{bob} \cite{freese}. To account for this discrepancy, astrophysicists postulated that a significant amount of galactic 
	%An example of this phenomenon is depicted in Fig.~\ref{fig:rotation curve}, 
	%The modern understanding of Newtonian mechanics predicts a significant decline in rotation
	
	\
	
	Understanding the nature of dark matter is a goal that is widely pursued by astrophysicists as well as scientists in adjacent fields %like particle physics
	\cite{dmhistory}, and there are undoubtedly Nobel Prizes in store for the research team that is finally able to shed light %\footnote{Pun intended.} 
	on this enigma. 
	%unravels this mystery. 
	As exciting as the topic of dark matter is, however, unravelling a mystery which has perplexed generations of researchers would be a fairly ambitious project for an undergraduate thesis. Additionally, pragmatism dictates that research into other topics should also continue, 
	%must continue in spite of such tantalizing problems as that of dark matter
	and so it is the case that many astrophysicists pursue their work 
	%with an awareness of dark matter
	while dealing with dark matter primarily insofar as it pertains to these endeavors. 
	% primarily dealing with dark matter insofar as it aids their own endeavors
	% appropriate for their research pursuits
	This is precisely the approach adopted for this project: in light of another persistent astrophysical mystery, the missing baryon problem, we take dark matter for granted, plying our current understanding of the topic to try to understand a less widely known --but no less intriguing-- cosmological conundrum. Those interested in learning more about dark matter candidates and the current state of dark matter research should look at some of the citations used in this section, namely \cite{freese} and \cite{dmhistory}.
	%physical mystery.
	% cosmological conundrum
	%As exciting as the quest to understand dark matter is
	
	%unravelling the mystery of dark matter would be rather ambitious for an undergraduate thesis
	
	%perplexing mystery
	
	\section{The Cosmic Microwave Background (CMB)}\label{sec:cmb}
	
	\begin{figure}
	    \centering
	    \includegraphics[scale=0.7]{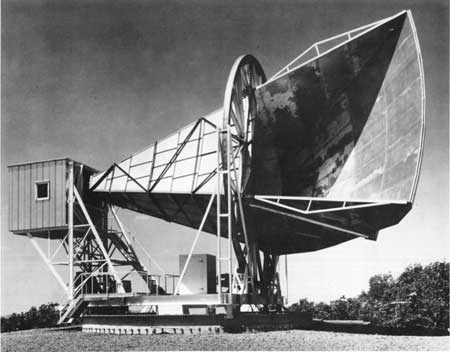}
	    \caption{A photograph of the Holmdel Horn, with which Penzias and Wilson were first able to detect the CMB. Photo credit: Bell Labs. Source: https://www.nps.gov/parkhistory/online\_books/butowsky5/astro4k.htm}
	    \label{fig:horn}
	\end{figure}\
	
	In the early 1960s, serendipity led to a monumental discovery in observational astrophysics: the cosmic microwave background (CMB). Bell Laboratories researchers Arno Penzias and Robert Wilson were studying the Milky Way using a radio telescope that had been repurposed from a satellite antenna (see Fig.~\ref{fig:horn}) \cite{bob} \cite{cmbhistory}, and as they were conducting their research they noticed that their equipment was picking up a persistent background noise in the microwave range ($\sim7.4\,\si{\centi\meter}$) which was causing interference in their measurements \cite{cmbhistory}. This background noise appeared to be coming from all directions, and after \emph{many} failed attempts at isolating its source\footnote{Which included relocating a pair of pigeons and scrubbing down the whole antenna in an effort to eliminate the possibility of interference from bird poop \cite{bob}; attempts at isolating the source were truly \emph{exhaustive}.} Penzias and Wilson were at a loss. This was the case until they became aware of theoretical work that had been done around the same time by Robert Dicke and his postdoctoral student P. J. E. Peebles regarding radiation left over from the formation of the universe during the Big Bang (which at the time was still a contentious subject in theoretical astrophysics) \cite{bob}. Penzias and Wilson invited Dicke to Bell Labs, who confirmed that the instrument was, in fact, picking up the theorized cosmic microwave background \cite{bob} \cite{cmbhistory}. This discovery earned Penzias and Wilson the Nobel prize in 1978 \cite{bob}
	
	%After \emph{many} failed attempts to isolate the source of the noise
	
	%an exhaustive
	
	%using a repurposed satellite antenna as a radio telescope
	
	\begin{figure}
	    \centering
	    \includegraphics[scale=0.2]{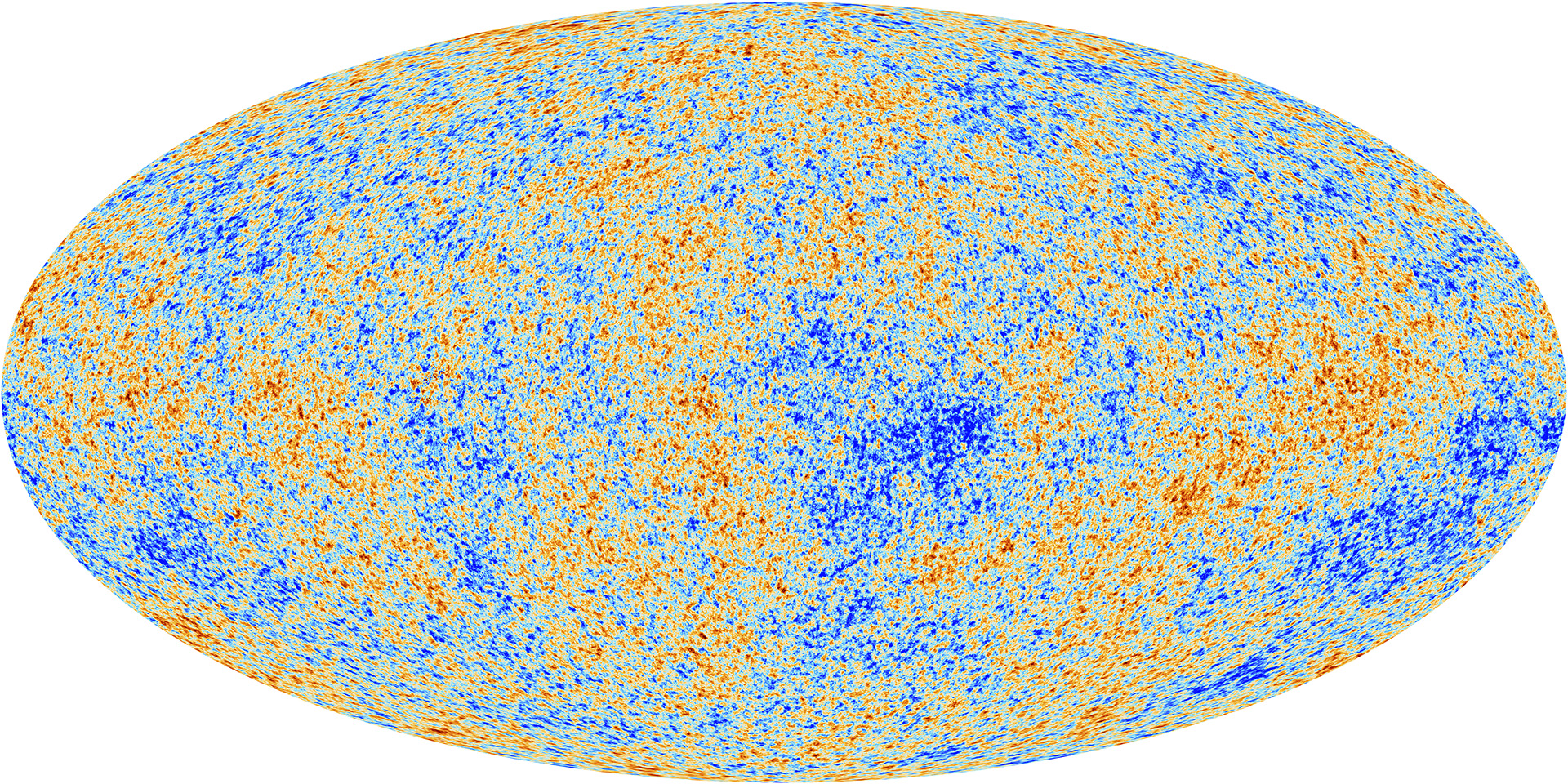}
	    \caption{CMB data from the Planck probe launched in 2009. Image courtesy of the European Space Agency https://www.esa.int/}%Science\_Exploration/Space\_Science/Planck/Planck\_and\_the\_cosmic\_microwave\_background\#TK1a}
	    \label{fig:my_label}
	\end{figure}\
	
	The discovery of the CMB is important for a number of reasons. Firstly, it helped to settle a crucial debate within cosmology as to whether or not the universe was steady-state (with no beginning or end). Armed with the evidence presented by the CMB, proponents of the Big Bang (the leading alternative at the time and the currently accepted theory) were able to put this debate to rest \cite{bob}. In addition to being a landmark discovery in terms of our understanding of the nature of the universe, the CMB also provides crucial insight into the state of the early universe. CMB observations pertain to this thesis in two fundamental ways. First, it is by studying the CMB\footnote{Or rather, by having others study the CMB so they can show us their results.} that we are able to produce initial conditions for astrophysical simulations. In essence, we base simulation starting conditions on CMB observations so as to help maximize the physical realism of our results\footnote{initial condition generation is really beyond the scope of this document, but the process is briefly mentioned in Sec.~\ref{sec:usersperspective}.} \cite{mitch}. The second reason that the CMB pertains to this thesis is that CMB observations have given astrophysicists an estimate of what is known as the cosmic baryon fraction. In brief, the term baryon in astrophysics refers to any matter that interacts with light, standing in contrast to dark matter \cite{bob} \cite{romeel}. Everything that you and I have ever seen in our lives has been made of baryonic matter, as is all matter that observationalists can directly detect through their telescopes. The cosmic baryon fraction, roughly speaking, tells us the fraction of baryonic matter present in the very early universe, a value which remains approximately constant over the lifetime of the cosmos \cite{dmhistory} \cite{tumlinson}. The importance of the cosmic baryon fraction to this thesis is discussed in greater depth in Ch.~\ref{ch:cgm}, but in essence it constitutes a fundamental part of the motivation behind this project.
	%(basically, where we tell our computers to start our simulations is informed by the CMB so as to maximize the physical realism of our results)
	%This specifically pertains to this 
	%and not fundamentally changing over time\footnote{At this point the expansion of the universe was a known phenomenon. To account for this expansion, a steady-state expanding universe })
	%finally settle a longstanding debate within cosmology 
	
	%Personal motivation
	
	%What would does the 4th reader need to hear in order to understand the information that is to come in the rest of the thesis. 
	
	%Why in GENERAL are people interested
	\section{Observation and Simulation}\label{sec:observation-simulation}
	
	An important and distinctive feature of physics as a discipline is the interplay between experiment and theory. Theoretical predictions can drive experimental inquiry, and experimental findings sometimes help to shape the development of new theoretical models. Oftentimes, work in the discipline is classified in terms of its alignment within this duality, being considered either experimental or theoretical. In astrophysics, however, the distinction is not always as clear as in other sub-fields (for instance, particle physics). Experimental work, %such as that which might be performed in an optics lab, 
	such as that which might be performed in an accelerator facility, relies heavily on laboratory environments where inquiries can be pursued in a controlled manner. Such controlled environments are essentially impossible to access for phenomena that occur at the cosmological or even galactic scale\footnote{This is not to say that astrophysics involves no experimentation. Spectroscopy experiments, for instance, have played a fundamental role in the development of the field \cite{bob}, but rather that we cannot perform \emph{controlled}  experiments relating to galaxies or the cosmos in the \emph{real world}.}. Because of this limitation, the empirical side of astrophysics generally takes the form of observation: the detection and measurement of light incident on the Earth from astrophysical objects. There is, of course, a theoretical side to astrophysics involving the development of mathematical models for describing observed phenomena as well, and --as is the case in all areas of the field-- the latter operates in concert with observation to advance our understanding of the nature of the universe.
	
	This thesis centers heavily around computer simulations of cosmological phenomena, putting it comfortably under the umbrella of what is referred to as \emph{computational astrophysics}. While there is a degree ambiguity at times in the precise %position 
	placement of computational astrophysics within the experiment-theory spectrum\footnote{In particular with respect to its simultaneous role as an extension of theoretical models and a means of testing them, which has been the cause of much personal deliberation over the course of my involvement in the field.}, for the purposes of understanding this thesis, computational astrophysics should be thought about as falling firmly under the category of theory\footnote{It can be thought of as such primarily due to the fact that its aim is \emph{not} empirical measurement of the real world.}. More important than categorizations of theory versus experiment and the placement of computation therein, however, is how these concepts can help inform an understanding of the \emph{role} of computation in the broader context of astrophysics. 
	
	I have already \emph{hinted} at the role of computational astrophysics, but in the interest of providing context and motivation for this project it bears stating more explicitly. Broadly speaking, computer simulations help to fill in the gap left by the impossibility of performing controlled experiments at astrophysical scales. To quote Kim, et al. 2019 \cite{agora2}:
	
	\begin{quote}
	    \emph{``Numerical experiments are often the only means to put our theory to a test, the result of which we can compare with observational data to validate the model's feasibility''}
	\end{quote}
	
	In essence, computer simulations offer a controlled environment through which researchers are able to probe the consequences of theoretical models for comparison against real-world observations, which can in turn help shed light on the meaning of observational results \cite{strawn} (as an example, see Fig. 6 in \cite{tumlinson} and its discussion in Sec. 4.2 of the same paper\footnote{See also: CLOUDY \cite{cloudy}, a code for simulating ionization, the results of which are used to inform observational analysis \cite{tumlinson}.}). It is thus the project of computational astrophysics, in a broad sense, to help reconcile theory with observation.

	\section{Types of simulation}\label{sec:simulation-types}
	
	%Computer simulations constitute a scientific tool that can be applied to a broad array of problems in a variety of disciplines (for 
	
	The term computer simulation is a broad one. In modern science, computational techniques are applied to a broad array of problems in a variety of disciplines, and even within astrophysics computer simulations can take on many different forms\footnote{See, for instance, \cite{madau} for an example of an astrophysical simulation of the universe's X-ray/UV background radiation.}. This document is specifically concerned with N-body simulations:
	those that model the evolution of matter under the influence of a gravitational field (as well as other physical forces, see Sec.~\ref{sec:sph}). 
	%in the universe over time.
	Such simulations start out with matter distributions that are based on early observations of the universe \cite{mitch} (see Sec.~\ref{sec:cmb}) and model the dynamics of that matter so as to (ideally) produce results which exhibit structures similar to those in the real universe\footnote{Ch.~\ref{ch:changa} goes into greater detail on techniques for accomplishing this}. %In a \emph{very} general sense
	For the purposes of this document, these types of simulation can be thought of as taking on one of three forms: 1.) isolated galaxy simulations, 2.) cosmological simulations, and 3.) zoom-in simulations. 
	
	Because they are arguably the simplest\footnote{Though admittedly calling something the simplest type of N-body simulation isn't really saying much.} both conceptually and computationally, and because 
	%For no other reason than because 
	they served as my own introduction to computational astrophysics, I will begin with isolated galaxy simulations. Isolated galaxy simulations are precisely what the name describes: N-body simulations modelling the evolution of a single galaxy in isolation\footnote{Incidentally, the physical approach to these simulations is very much the same as that described in Ch.~\ref{ch:changa}. The main difference responsible for producing an isolated galaxy is in the initial conditions \cite{agora2}.}. One of the main advantages to this type of simulation is that it provides a very high level of detail for a single galaxy for a relatively low computational cost. %The main drawback is physical realism
	The main drawback is that there are a great many physical processes that occur in the cosmos which are remain unaccounted for in this approach (such as accretion and ejection \cite{tumlinson}, or galactic mergers \cite{galaxy-formation}). As a result, isolated galaxy simulations are most applicable to special cases where galaxies emerge via \emph{secular evolution} (or evolution largely separated from the ``cosmological framework''), as is often the case for barred galaxies \cite{galaxy-formation}. 
	
	\begin{figure}
	    \centering
	    \includegraphics[scale=0.4]{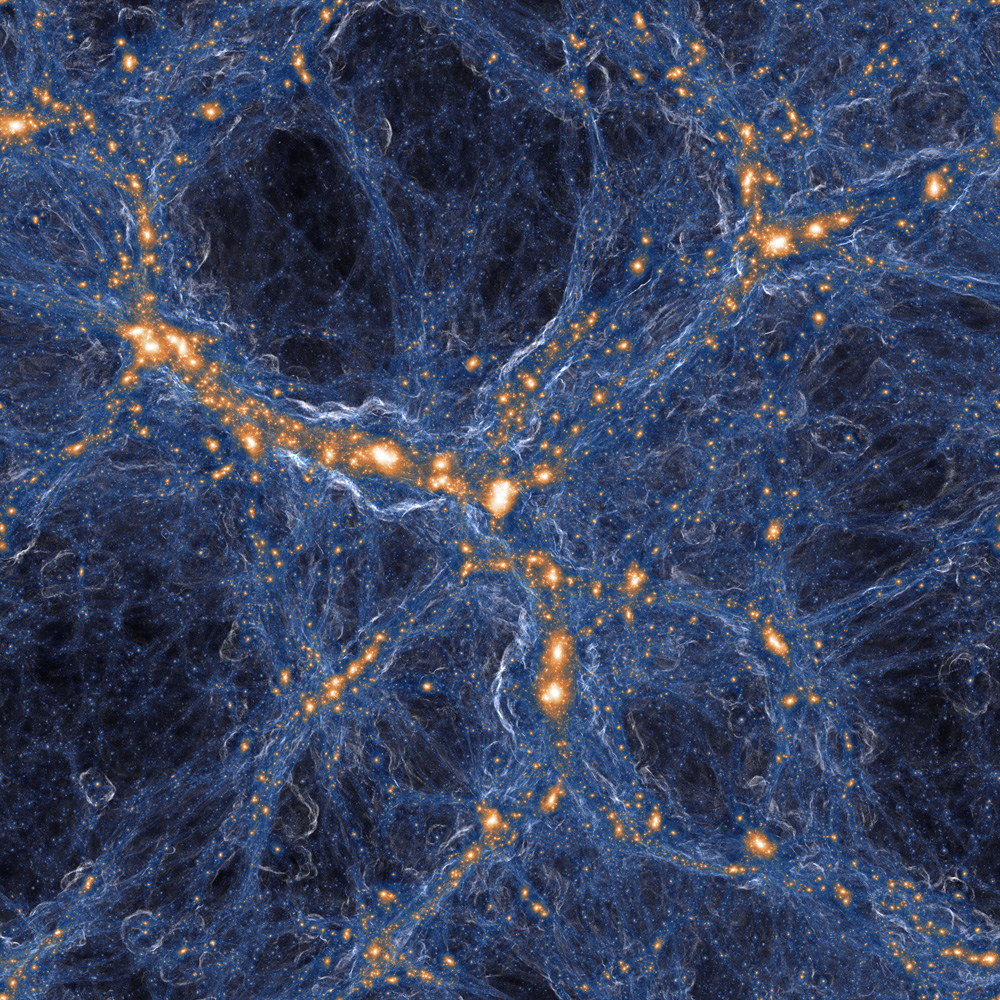}
	    \caption{A visualization of %simulated cosmic structure from the 
	    the results of a cosmological simulation performed as part of the IllustrisTNG project (J. P. Naiman, et al.
Monthly Notices of the Royal Astronomical Society 477, 1206 (2018).). The volume depicted is 100 kpc across, or approximately 300 million light years.}
	    \label{fig:tng}
	\end{figure}\
	
	Standing in contrast to the smaller-scale approach taken by isolated galaxy simulations, \emph{cosmological simulations} aim to model the large-scale structure of the universe. Fig.~\ref{fig:tng} depicts the results of one such simulation performed as part of the IllustrisTNG project \cite{illustris}. The structure shown is referred to colloquially as the ``cosmic web'' due to its shape, observations of which were first published by Margaret Geller and John Huchra in 1989 \cite{geller}. In addition to their aesthetic beauty and the insight they provide into the shape of the universe at massive scales, %they 
	such simulations also take into account crucial events %for both galaxy and cosmological evolution
	that drive both galactic and cosmic evolution, such as the aforementioned mergers. The trade-off is detail\footnote{IllustrisTNG being somewhat of an exception, having both large and incredible detail \cite{illustris}. TNG, however, was an incredibly ambitious project undertaken using state-of-the-art technology, so in general the statement holds.}. In general, cosmological simulations have lower spatial resolution\footnote{Spatial resolution, on a conceptual level, can be thought of as being analogous to the resolution of a digital image. Lower resolution means lower ``detail''.} than their isolated-galaxy counterparts, meaning much less data is available at the galactic scale. In essence, isolated galaxy simulations provide relatively detailed galaxies, and nothing else, while cosmological simulations provide detailed cosmological structure which is partially composed of low-detail galaxies\footnote{This is a very simplified explanation. Again, recall that the amount of data available in a given region of space is directly related to spatial resolution. Here ``detail'' means the amount of data available for a given structure.}.
	
	The third simulation type, the Zoom-in simulation, offers a compromise between the small-scale detail of isolated galaxy simulations and the more general physical realism of cosmological simulations. The fundamental principle behind the Zoom-in method involves the identification of interesting regions within a large-scale simulation so that the program can devote more computational resources to those regions \cite{finnish}. In the context of this project, this ``interesting region'' can simply be though of as a region which as been identified as having a galaxy of mass comparable to that the Milky Way, as this was the selection criterion for the AGORA data used in this thesis \cite{agora3}. 
	
	The technique used to perform Zoom-in simulations is a fairly complicated multi-step process, and I will not attempt to cover the full details here. I will, however, give a brief conceptual outline for the purpose of contextualizing the relevance of this type of simulation to the problem at hand. A Zoom-in simulation starts out as low-resolution cosmological simulation containing only dark matter\footnote{It should be noted that since dark matter constitutes the majority of matter in the universe \cite{bob} and because it doesn't interact hydrodynamically \cite{beckett}, a dark-matter-only approximation is justified.} (see Sec.~\ref{sec:dm}). A gravitational simulation is then performed on that dark matter down to a redshift of zero (which can be considered ``present day''), after which an algorithm known as a \emph{halo finder} is used to locate galaxies within the simulation volume (the specifics of halo finding algorithms are well beyond the scope of this document, but are discussed in \cite{finnish}). Once a sufficiently interesting galaxy is found, the code then restarts the simulation from the beginning, adding in baryonic matter\footnote{A.k.a. non-dark matter, see Sec.~\ref{sec:cmb} and Ch.~\ref{ch:cgm}.} and re-running the simulation with the identified volume-of-interest at a higher resolution \cite{finnish}. As previously mentioned, the result effectively combines the strengths of isolated galaxy and cosmological simulations, providing a higher resolution ``zoomed-in'' region within its appropriate cosmological context. This additional level of physical realism is particularly relevant for studies involving regions %beyond the galaxy itself\footnote{Or, more precisely, the galactic disk.} 
	such as the CGM (see Ch.~\ref{ch:cgm}) the evolution of %each of 
	which is driven by complex dynamical processes that extend well beyond the galaxy itself \cite{tumlinson}, or the IGM (e.g. \cite{whim}), which is entirely absent in isolated galaxy simulations \cite{ryden}. Since this project %is a theoretical inquiry into the CGM, 
	relates to the CGM, a Zoom-in simulation is what is used to produce the data being analyzed \cite{agora3}.

	\chapter{Missing Baryons and the CGM}\label{ch:cgm}
	
	\section{The Missing Baryon Problem}
	
	%This project can be best understood in the context of the question with which it is most directly concerned, that being the prevailing mystery that astrophysicists refer to as the ``missing baryon problem''. 
	As suggested by its title, one of the primary concerns of this thesis is a prevailing mystery in modern astrophysics that is often referred to as the \emph{missing baryon problem}.  
	This problem arises from the discrepancy between observations of the matter composition of the very early universe (specifically from the CMB, see Sec.~\ref{sec:cmb}) and the matter that is currently accounted for within galactic halos. In the vocabulary of this problem, matter can be divided into two categories: baryonic matter and dark matter. Readers who are well-versed in astrophysics lingo will likely already be familiar with the term ``baryonic matter,'' but for those who are not, baryonic matter constitutes all electromagnetically interacting matter, that is, any matter that interacts with light \cite{romeel}. In this sense, baryonic matter is what a non-astrophysicist might refer to simply as ``regular'' matter, as it describes everything that can be seen (or, more accurately, \emph{observed}). Readers may recall from Sec.~\ref{sec:dm} that the defining characteristic of the \emph{second} type of matter, dark matter, is that it doesn't interact with light (or that it only does so very weakly), putting it in contrast with baryonic matter.
	%In contrast, dark matter --as was discussed in Sec.~\ref{sec:dm}-- is understood to only interact very weakly with light, if at all.
	%the definining property of
	%In contrast, dark matter does not interact with light and has therefore only been detected through its gravitational influence on the baryonic matter that we \emph{can} see \cite{galaxies-in-universe}.
	%The distinction between these two types of matter is important because %the matter discrepancy between the early universe and 
	%they 
	
	Baryonic and dark matter constitute two of the fundamental ``building blocks'' of the universe from a cosmological perspective\footnote{The third being dark energy, see \cite{bob} and \cite{carroll}.} \cite{bob}, which gives them great appeal as a line of scientific inquiry. In the early 1970s, researchers turned their attention to the newly discovered CMB as a means of understanding the matter composition of the universe \cite{dmhistory}. Through these initial studies and decades of follow-up research, astrophysicists have been able to put increasingly precise constraints on the relative densities of baryonic and dark matter in the early universe as observed from the CMB \cite{dmhistory}. From these measurements, we derive what is known as the \emph{cosmic baryon fraction}, or the fraction of matter in the universe that is baryonic. By rough approximation (which is sufficient for understanding the problem in this context), modern measurements have the cosmic baryon fraction as being \cite{romeel}
	%Beginning in the early 1970s, observational studies began attempting to provide insight into the matter composition of the universe through measurements of the newly discovered CMB
	%Making them a prime line of inquiry for understanding the nature of the universe.
	%Naturally, this cosmological importance 
	%in modern cosmology the two are understood to 
	%it provides a framework for understanding why some of this matter is said to be \emph{missing}. As mentioned in Sec.~\ref{sec:cmb}, observations of the CMB have yielded measurements of the densities of baryonic and dark matter ($\Omega_\text{b}$ and $\Omega_\text{d}$, respectively) in our universe. By fairly rough approximation (which is sufficient for understanding the nature of the missing baryon problem) according to modern observation, the \emph{cosmic baryon fraction} is \cite{romeel}
	\begin{equation}
	    f_\text{b}\equiv\frac{\Omega_\text{b}}{\Omega_\text{b}+\Omega_\text{d}}\approx\frac{1}{6},
	\end{equation}
	Where $\Omega_\text{b}$ and $\Omega_\text{d}$ are the measured baryonic and drak matter densities, respectively. 
	
	Measurement of the cosmic baryon fraction is what brings us to the missing baryon problem. Based on the current cosmological model\footnote{Known as the $\Lambda$CDM model, the precise details of which are not crucial for understanding this document. See Chapters 29 and 30 of \cite{bob} for more information on modern cosmology.}, astrophysicists expect that baryonic matter is gravitationally drawn to dark matter halos, where it would eventually collapse in towards the center (i.e. the galactic disk). As a result of this process, observationalists should expect to find that the fraction of baryonic matter within the stars and Interstellar Medium (ISM) of galaxies should %roughly match that of the cosmos
	be close to the roughly one-sixth fraction exhibited by the cosmos \cite{romeel} \cite{tumlinson}. However, this is not observed to be the case. While quantities vary by galaxy mass, the stars and ISM %tend to only 
	account for no more than 20\% of the galaxy's expected baryon mass \emph{at best} \cite{tumlinson}. This begs the question: if so few of the baryons we expect to find are located in stars and interstellar gas, \emph{where are they?} This, in essence, is the missing baryon problem. We know what the fraction of baryons to total matter is in our universe, and our theoretical model of cosmology predicts that roughly the same fraction should be present in galaxies, but our observations fall significantly short of this expectation.
	%we know where we should expect to find the baryons, but we simply do not see them there. 
	
	%Faced with the shortage of observed baryonic matter within galaxies, scientists have been left to theorize about the location of the missing baryons and causes for their apparent absence.
	There exist three possible explanations for the shortage of observed baryonic matter within galaxies. First, it is possible that the missing baryons are present in the galaxies and have simply not yet been observed. In this case, it is likely that the baryons are in a phase that is difficult to detect due to low light emission, such as cold, hot, or low-density gas \cite{tumlinson}. Alternatively, the missing baryons could be absent from the galaxies entirely, in which case they were either accreted into the galactic halo, only to get ejected at some later point, or they failed to accrete altogether. It should be fairly clear that these are \emph{the} three possibilities for the missing baryons: the baryons are in the galaxies, or they aren't. In the latter case, they were either there at some point, or they never were to begin with. While it is important to be aware of the distinctions between these three processes, it is also important to note --as Tumlinson, et al. do in their review of modern CGM research \cite{tumlinson}-- that they are not necessarily mutually exclusive, and the true explanation for the missing baryons is most likely a combination of all three.
	%While these three processes are certainly distinct from one another, it is important to note, as Tumlinson, et al. did in their CGM review article \textbf{[CITE]}, that they are not necessarily mutually exclusive, and the true explanation for the missing baryons is most likely a combination of all three.%, though it is useful to present them as being distinct possibilities
	%While this may seem to complicate the search for the missing baryons, %, as we will see, there happens to be a certain region in galaxies that can help shed light on the problem, regardless of which process (or combination therein)
	%astrophysicists have recently turned their attention to a galactic component called the Circumgalactic Medium (CGM) for insight into the problem. As we will soon see, 
	
	\section{The Circumgalactic Medium}\label{sec:cgm}
	
	%We have seen the three possible explanations for the missing baryon problem, but how can this help us actually find them?
	We have seen that there are three possible explanations for the missing baryon problem, %but how does this theoretical knowledge translate into the actual search for the missing baryons?
	but it may not yet be clear how this knowledge alone brings us any closer to actually \emph{finding} the seemingly truant matter. What we now need is a way to observationally determine which of these explanations can account for the galactic baryon deficit. %To do this, 
	To accomplish this monumental task, astrophysicists have turned their attention to a galactic component known as the Circumgalactic Medium (CGM), which seems to hold much promise in the search for the missing baryons regardless of the cause of their absence \cite{tumlinson}. 
	
	\subsection{What is the CGM?}\label{sec:cgm def}
	
\begin{figure}
	    \centering
	    \includegraphics[scale=0.45]{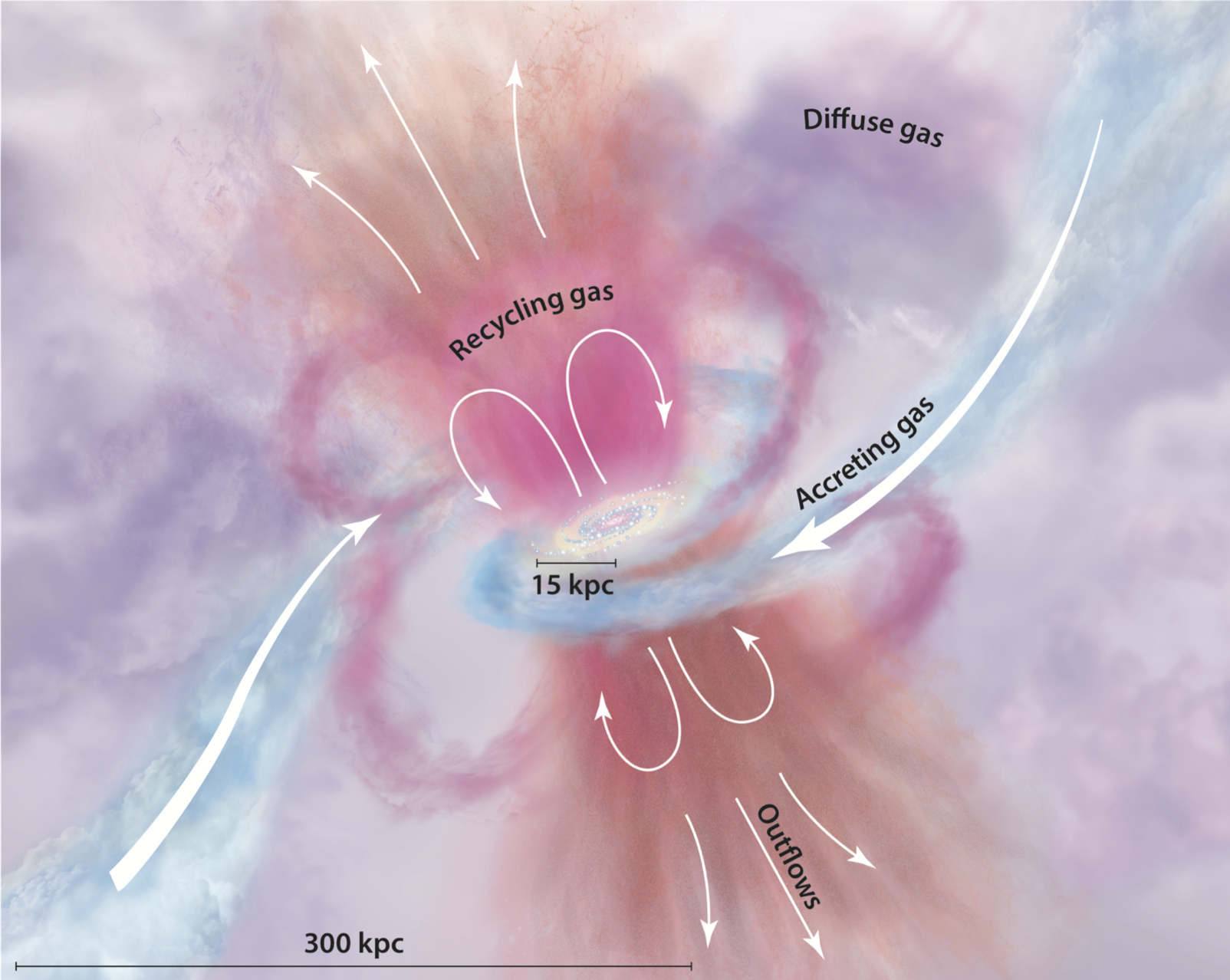}
	    \caption{An artistic depiction of the CGM and its dynamical processes taken from (J. Tumlinson, M. S. Peeples, and J. K. Werk, Ann. Rev. Astro. Astrophysics \textbf{55}, 389 (2017)). The galactic disk can be found at the center, labelled by its 15 kpc radius. Gas ejection is depicted in the outflows perpendicular to the galactic plane (top and bottom of image). Recycling from previous ejection is also shown along these outflow trajectories. Diffuse gas leftover from all three processes is depicted in purple for the surrounding region.}
	    \label{fig:cgm}
	\end{figure}

	As the name implies, the \emph{Circum}-galactic Medium is the medium of gas and dust directly surrounding a galaxy. More precisely, while there is some ambiguity with regards to the precise boundaries of the CGM, it is generally defined as residing outside of a galaxy's Interstellar Medium (or ISM, the gaseous medium between the stars of a galaxy) but still within its virial radius \cite{tumlinson} \cite{ryden}. The CGM is a highly dynamic region, being a site for accretion, ejection, and recycling of galactic matter \cite{tumlinson} (see Fig.~\ref{fig:cgm}).
	
	Already from its definition we can start to see the relevance of the CGM to the issue of the missing baryons. Being the medium that directly surrounds a galaxy's luminous disk, the CGM acts as an interface between the galaxy's ISM and the surrounding Intergalactic Medium (IGM) and a host for the processes that drive galaxy evolution \cite{tumlinson} (Fig.~\ref{fig:cgm}). In essence, any matter being accreted into or ejected out of the galaxy must necessarily pass through the CGM. If we recall that one of the potential explanations for the missing baryons is ejection from galaxies, it becomes clear that on their way out, these baryons must have traversed the CGM. Because of this, we can turn to the CGM for signs of baryon ejection over the lifetime of the galaxy. More specifically, ``signs of ejection'' refers to heavy elements, which are produced by stellar nucleosynthesis (the fusion of lighter elements within stars), and are propagated throughout a galaxy when their parent stars go supernova \cite{tumlinson}. The presence of these metals is significant because, along with stellar winds, %the supernovae that propagate them throughout galaxies are one of the main sources for the energy that causes matter to get ejected from galaxies. 
	supernovae are one of the main sources for the energy that causes matter to get ejected from galaxies. Therefore, observations of heavy elements (elements other than Hydrogen and Helium, also referred to as `metals') in the CGM can give estimates of the energetic output of supernovae over the galaxy's lifetime, which itself provides insight into the ejection of matter from the galaxy \cite{romeel} \cite{tumlinson}.
	
	% ROMEEL TALKS ABOUT METALS AS TRACERS AT 33:00
	
	% UV LIGHT EXTINCTION AT 39:00
	
	%\textit{Do the metals specifically reside in the CGM or can the tracers be found anywhere in the galaxy? Also does the presence of heavy elements in the CGM lend evidence to outflows? (Heavy elements as remnants of ejected matter?)}
	
	In the case of ejection (or, conversely, failure of accretion), the significance of the CGM is perhaps most apparent. As previously noted, it is the ``interface'' through which matter must pass as it enters and/or leaves the galaxy, and we have seen that metals originating from stars can serve as indicators of these processes \cite{tumlinson}. %In 
	This leaves the case of the matter being present but undetected in the galaxy, for which the significance of the CGM may seem less %glaringly significant. 
	immediately apparent. %However, %this possibility, too, gives us good cause to look towards the CGM. As 
	As previously stated, the most likely case for \emph{undetected} baryons is that they are present in a low-emission gas phase somewhere within the galaxy. It is this precisely fact which implicates the CGM as a possible location for missing baryons. %brings in the CGM as a likely candidate  %such as hot or diffuse (low density) gas.
	Because the CGM is composed primarily of diffuse gas, emission measurements are incredibly hard to obtain\footnote{This is because emission measurements for gas scale wit density squared \cite{tumlinson}. Qualitatively, low density leads to \emph{very} low emission measurements.} \cite{tumlinson}. Thus, in addition to being a site for the accretion and ejection of baryonic matter, the CGM could also hold its share of unobserved baryons.  
	%Measurements of gas emission are heavily dependent on density, scaling with density squared. As such, it can be very difficult to measure emissions from diffuse gas. 
	%and while the \emph{temperature} of CGM gas varies pretty significantly by phase, it remains diffuse throughout. The low density of CGM gas makes it very difficult to detect via emission, since measurements scale with density squared, which makes it an excellent place to look for undetected baryonic matter \cite{tumlinson} \cite{romeel}.
	
	To briefly summarize, regardless of the location of the baryons or the underlying cause of their observational shortage, the CGM appears to be the best place to start looking in order to close the observational ``baryon gap.'' If baryons have entered the galaxy only to be ejected at a later date, we can search for signs of this ejection in the form of CGM metals. Additionally, %we can directly search for baryons in the CGM, as its associated 
	the observational challenges associated with the CGM make it a prime location to look for undetected baryons that have not been ejected. 
	%Since the only other possibility is that baryon accretion has been prevented altogether, being able to estimate the fraction of the galactic baryon budget that is accounted for by ejected matter and diffuse CGM gas can also help us make educated guesses about the degree to which accretion is being ``blocked,'' though it should be emphasized that any such guesses are not substitute for actual observations that can be associated with accretion blocking.
	
	\subsection{Observing the CGM}\label{sec:cgm-observation}
	
	Having outlined the motivation for studying the CGM in the context of the missing baryon problem (Sec.~\ref{sec:cgm}), %it is relevant to discuss how 
	a discussion of \emph{how} these studies are performed is now warranted. First, the challenges associated with observing the CGM bear repeating, since they are crucial to understanding %the application of 
	the observational techniques in this context and their motivation. To reiterate from the previous section, %a characteristic of CGM gas is its low density, which 
	CGM gas is partially characterized by its low density, a fact which makes observation of emission incredibly difficult \cite{tumlinson}. %(one of the main tools of observational astrophysics \cite{bob}). 
	Since emission studies are one of the main avenues by which observational research is performed \cite{bob}, this challenge poses no small obstacle.
	%Ironically, this observational challenge constitutes part of the grounds on which we seek to study the CGM in the first place.
	At the same time, the observational difficulties associated with the CGM constitute a major part of the grounds on which we seek to study it in the first place, so finding a means to overcome them is important.
	
	\begin{figure}
	    \centering
	    \includegraphics[scale=0.4]{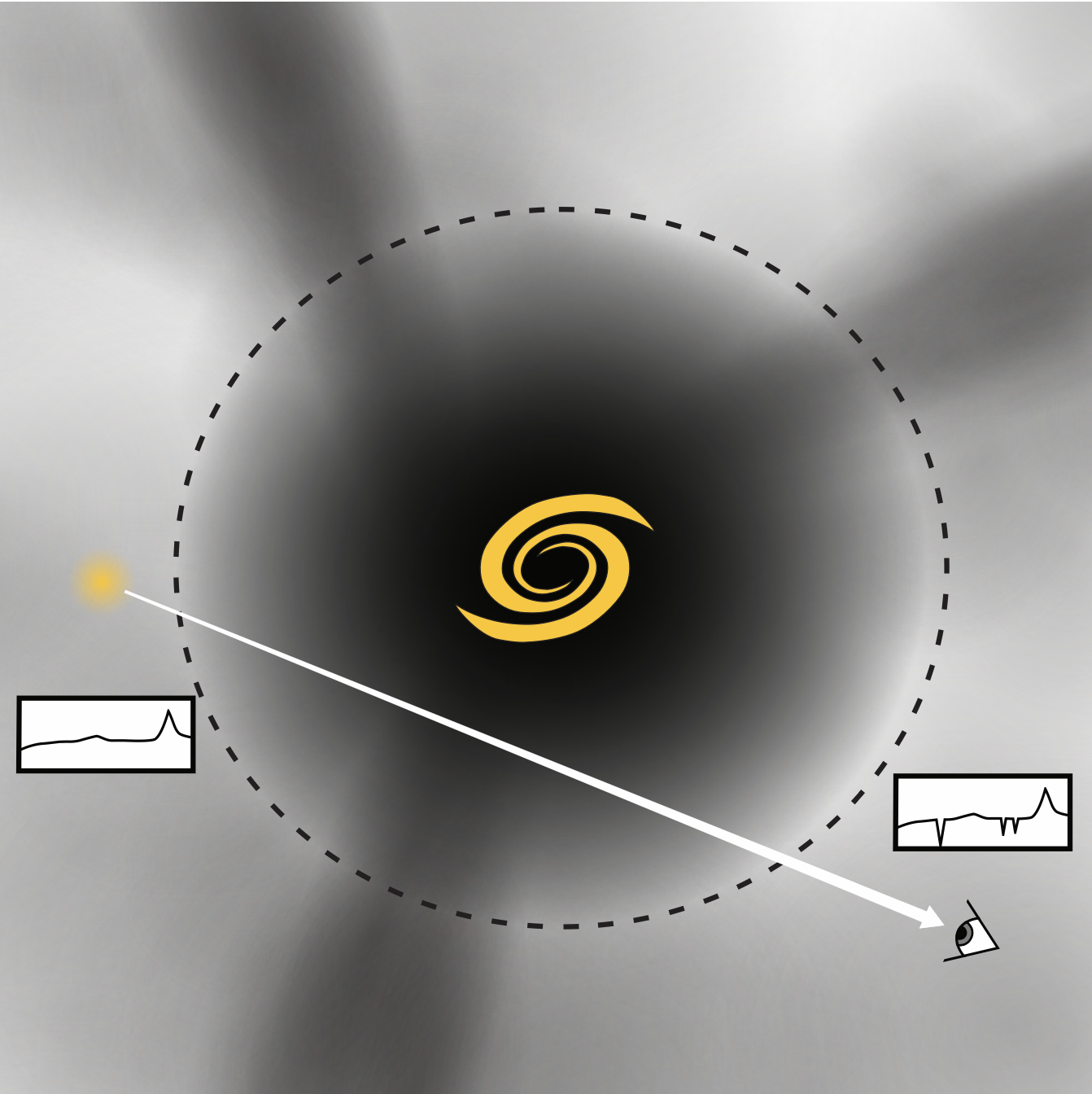}
	    \caption{A depiction of quasar absorption spectroscopy through the CGM of a galaxy borrowed from (C. B. Hummels, B. D. Smith, and D. W. Silvia, \textbf{847}, 59 (2017), arXiv:1612.03935
[astro-ph.IM]). The ``swirl'' in the center represents the galactic disk, the dotted circle depicts the outer limit of the CGM. The yellow diffuse dot on the left represents the quasar source, with its highly schematic emission spectrum shown below. The eye on the right represents the observer, and the observed spectrum is shown above.}
	    \label{fig:sightline}
	\end{figure}\
	
	There are a number of methods for collecting observational data on the CGM (see Sec. 3 of \cite{tumlinson}). This thesis focuses in on one in particular: quasar (QSO\footnote{The use of QSO as shorthand for quasar comes from the abbreviation of their alternate name: quasi-stellar object \cite{bob}.}) absorption studies (Fig.~\ref{fig:sightline}). A quasar is a type of active galactic nucleus (AGN), a dense galactic center exhibiting a significant amount of luminosity from sources other than starlight \cite{nutshell}. Specifically, quasars are the brightest type of AGN\footnote{They are also the brightest phenomena observed in the universe, oftentimes completely drowning out the luminosity of their host galaxy \cite{nutshell} \cite{galaxies-in-universe}} that are thought to be powered by supermassive black holes \cite{quasar} \cite{nutshell}. %They are the brightest phenomena observed in the universe, being so bright as to obscure their host galaxies \cite{nutshell} \cite{galaxies-in-universe}
	Quasar absorption studies take advantage of the brightness of these objects by using them as background sources to perform absorption spectroscopy on intervening matter, which is particularly useful for working around the low density of CGM gas \cite{tumlinson}. A distinct disadvantage to using quasars as background sources for CGM studies of galaxies \emph{other} than our own\footnote{Which I will focus on because it is the more complicated case. As noted by Tumlinson, et al. \cite{tumlinson}, Milky Way CGM studies can use ``essentially any quasar'' as a source.} is that we must rely on %naturally occurring quasar sightlines through galaxies, which is 
	the %unreliable 
	natural occurrence of quasar sightlines passing through galaxies (we obviously cannot choose where the quasars are or what their light passes through). While this is not exceptionally uncommon, it is rare enough that in most cases only one quasar sightline is available per observed galaxy \cite{tumlinson}. Despite having to rely on the essentially random occurrence of a quasar passing through a galaxy, the advantages offered by this method such as sensitivity to the low density of the CGM and consistency in detection irrespective of the brightness of the host galaxy \cite{tumlinson}.

	\section{Insight from the Simulated CGM}\label{sec:cgm-simulation}
	
\begin{figure}
	    \centering
	    \includegraphics[scale=0.6]{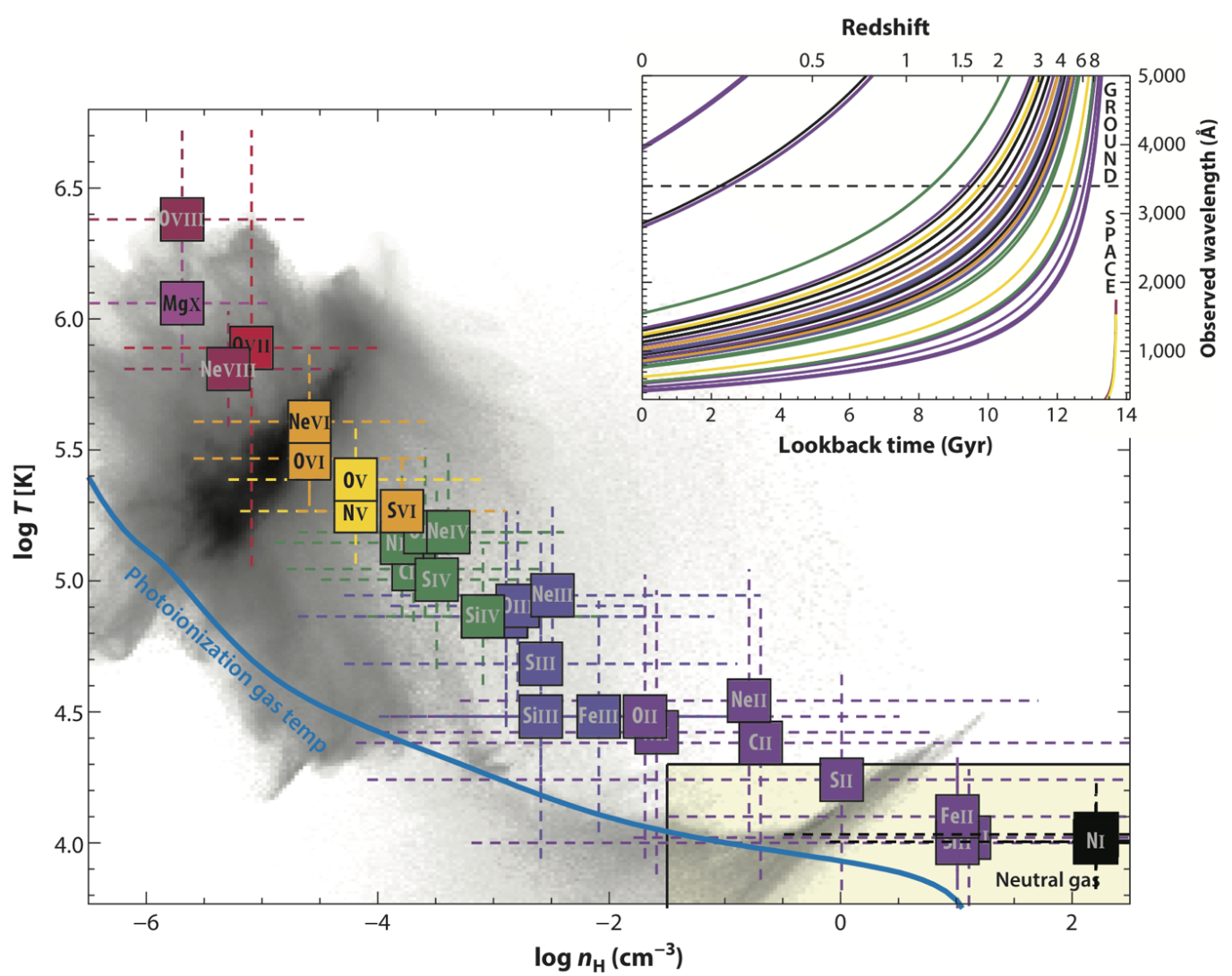}
	    \caption{A phase plot of temperature vs hydrogen density for gas within $R_\text{vir}$ for the EAGLE simulation at redshift zero (J. Schaye, in \textit{IAU General Assembly}, Vol. 29 (2015) p. 2250900). Ions are positioned on the graph according to their respective peaks in ionization fraction, and are colored according to ionization state. The range bars associated with each plotted ion give the range over which its ionization fraction is greater than half the determined maximum. The subgraph in the upper right shows the observed wavelength of absorption features with respect to lookback time and redshift. This image was taken from (J. Tumlinson, M. S. Peeples, and J. K. Werk, Ann. Rev. Astro. Astrophysics \textbf{55}, 389 (2017), arXiv:1709.09180 [astro-ph.GA]).}
	    \label{fig:eagle}
	\end{figure}

	Sec.~\ref{sec:observation-simulation} discussed the utility of astrophysical simulations for testing theory and providing insight into observational results. As a concrete example of the latter in the context of CGM research, this subsection discusses an important figure from \cite{tumlinson} (Fig.~\ref{fig:eagle}). This figure presents an analysis of data from the EAGLE simulation \cite{eagle} which gives a log-log plot of temperature ($T$) against hydrogen density ($n_\text{H}$) for the gas within the simulated galaxy's virial radius\footnote{Recall from Sec.~\ref{sec:cgm def} that the CGM is defined as being within the virial radius but outside of the luminous disk \cite{tumlinson}.} at a redshift of $z=0$ \cite{tumlinson}. %Points representing specific ions are graphed on this plot according to 
	The positions of these ions on the graph represent peaks in their ionization fraction peaks with respect to $T$ and $n_\text{H}$ according to theoretical ionization models\footnote{For more information on ionization models see \cite{tumlinson}.}. To paraphrase Tumlinson, et al. \cite{tumlinson}, this graph provides a guide as to which ions are most likely to be associated with which phases of the CGM, which can help inform analysis of observational data. 
	
	This is, of course, only one example of simulations being used to give additional context to observations. Those interested in learning more should refer to Section 4 of \cite{tumlinson}, Sections 4.2 and 4.5 in particular. I should also note that this figure has little direct bearing on this project, %obtaining results similar to those depicted in Fig.~\ref{fig:eagle}, 
	and that my primary motivation for showing this graph is because I personally find it to be an incredibly interesting application of simulation results to observation.
	
	%Sections 4 and 
	
	%\section{Baryon Content of the CGM}\label{sec:baryon content}
	
	%\ref{sec:baryon content}
	
	%\chapter{LOREM IPSUM}\label{ch:changa}
	
	%\section{ChaNGa}\label{sec:changa}
	
	\chapter{Simulation Code: ChaNGa}\label{ch:changa}
	
	%\note{Should write a snappy intro to the chapter}
	
	This thesis uses data from a zoom-in simulation (see Sec.~\ref{sec:simulation-types}) performed using an N-body code called ChaNGa (Charm N-body GrAvity solver) \cite{menon}. In this chapter I discuss a number of the techniques used by this program, as well as the motivations behind them. My hope in doing so is to provide the reader with a more robust understanding of what goes on under the hood of an N-body code, as well as a deeper appreciation for simulation codes from a technological perspective. At the same time, it is worth acknowledging that the subject matter of this chapter is very technical, and though I have not attempted to provide an exhaustive description of the code\footnote{One could easily write an entire thesis on that alone.}, the details that \emph{are} covered can be conceptually difficult. For the reader in a hurry, an outline of the chapter is as follows: Sections \ref{sec:code gravity}-\ref{sec:barnes-hut changa} discuss gravitational motion and the code's approach to simulating it. Sec. \ref{sec:boundary conditions} describes considerations for a finite simulation of an infinite (or at least incredibly large) universe. Sec. \ref{sec:force softening} describes a technique for reducing gravitational force at small distances so as to avoid problems that would otherwise arise in the calculation. Sec. \ref{sec:sph} describes the method by which ChaNGa accounts for the behavior of gas in the universe. Sec. \ref{sec:usersperspective} describes the process of actually using the code to get results.
	
	%\begin{itemize}
	    %\item Sections \ref{sec:code gravity}-\ref{sec:barnes-hut changa} discuss gravitational motion and the code's approach to simulating it.
	    
	    %\item Section \ref{sec:boundary conditions} describes considerations for a finite simulation of an infinite (or at least incredibly large) universe.
	    
	    %\item Section \ref{sec:force softening} describes a technique for reducing gravitational force at small distances so as to avoid problems that would otherwise arise in the calculation.
	    
	    %\item Section \ref{sec:sph} describes the method by which ChaNGa accounts for the presence of gas in the universe.
	    
	    %\item Section \ref{sec:usersperspective} describes the process of actually using the code to get results.
	%\end{itemize}
	
	%provide the audience with 
	
	\section{Gravity}\label{sec:code gravity}
	
	If we were to imagine ourselves sitting down at a computer, opening up our favorite IDE, and writing our own N-body code from scratch, one of the first considerations we'd need to make would be that of gravity\footnote{A bunch of free particles bouncing around a computational volume, while fun, would not make for a particularly interesting simulation, and would certainly not come anywhere close to what might be considered an accurate representation of the physical universe.
	%One could make a compelling argument for velocity: free particle motion is certainly simpler than gravitational motion. However, as we shall soon see, 
	%perhaps we would like to start with the absolute simplest case of a bunch of particles in a box, each moving in some direction with some initial velocity
	%However, as we shall soon see, 
	}. Gravity, after all, is precisely what makes an N-body simulation an N-body simulation! So where do we start in our treatment of gravity?
	
	Perhaps we would like to go with a maximally accurate treatment, in which case we would be tempted to follow the gravitational model provided to us by General Relativity (GR). GR is the most widely accepted gravitational theory in modern physics, and has been studied extensively since it was first published by Albert Einstein in 1916 \cite{einstein}. GR is a geometric theory which treats space and time as being fundamentally interlinked in a manifold\footnote{A manifold is a geometric mathematical construct, the rigorous definition of which is beyond the scope of this document. For the purposes of this discussion, it is sufficient to think of a manifold as being any geometry which ``looks'' euclidean on a small enough scale. For more information on manifolds, see Ch. 2.2 of \cite{carroll}.} %(something like a generalized surface) 
	known as spacetime. %, with light and matter travelling along geodesics\footnote{A geodesic being the shortest path between two points.} in the geometry defined by this spacetime manifold. 
	According to this theory, all gravitational motion that we observe in the three dimensional space of our universe occurs along geodesics\footnote{A geodesic being the shortest path between two points. The actual path of a geodesic, as one might expect, is heavily dependent on the shape of a geometry.} in the spacetime manifold. The fundamental equation governing the shape of this manifold, known as Einstein's equation, takes the form \cite{joel}\cite{carroll}
	\begin{equation}\label{eq:einstein}
	    R_{\mu\nu}-\frac{1}{2}g_{\mu\nu}R=\frac{8\pi G}{c^4}T_{\mu\nu},
	\end{equation}
	where $g_{\mu\nu}$ is a tensorial object known as the metric, which represents the curvature of the spacetime manifold (and therefore the shapes of the geodesics governing motion within said manifold), and $T_{\mu\nu}$ is known as the stress tensor or the energy-momentum tensor, which represents the mass/energy\footnote{In general relativity, mass and energy are understood to be essentially equivalent \cite{carroll}.} of the source generating the spacetime curvature (the phenomenon we call \emph{gravity}) \cite{joel}. %\note{say more about this tensor?}. 
	In essence, this equation expresses the relation between the energy/mass present in a spacetime manifold and the manifold's shape, which itself defines the shape of gravitational trajectories. In this sense, the equation represents a means of understanding gravitational motion in the context of general relativity, hence its central role in the theory.
	%In the sense that it can be used to relate the presence of gravitational sources to gravitational motion, it is analogous to Maxwell's equations in E\&M \note{Run this statement by Johnny}.
	
	With all this talk of manifolds and tensors --concepts that even an advanced undergraduate is likely to be unfamiliar with-- you might be getting a little concerned: how exactly are we supposed to write code that deals with such a mathematically rigorous gravitational model? The short answer is that we do not. While relativity and its implications constitute a fundamental part of the modern understanding of physics, 
	and are certainly interesting to consider in %a philosophy-of-science kind of way
	in terms of its more philosophical implications, in all but the most extreme cases (e.g. extremely high-mass objects such as black holes), it is generally sufficient to ignore Eq.~\ref{eq:einstein} altogether and stick with the classical treatment of gravity. Indeed, in cases of low mass and velocity\footnote{Here meaning objects significantly less massive than black holes travelling significantly slower than the speed of light, criteria that are not particularly hard to meet, even in a cosmological context.}, general relativity approximately reduces to Newtonian gravity\footnote{This is known as the weak field approximation \cite{joel}.}, so this simplification is far from unjustified \cite{finnish}. This is fortunate for us, as codes capable of treating the N-body problem relativistically do not currently exist %\footnote{So be not ashamed if you were unable to write your own.} 
	\cite{finnish}. Thus, we find that we have no choice but to engage in the time-honored physics tradition of approximation!
	
	\section{Newtonian Gravity and Numerical Integration}\label{sec:code integration}
	
	According to Newton's law of universal gravitation (Sec.~\ref{sec:gravity}), the gravitational force between two massive particles $m_1$ and $m_2$ in 3D space is given in vector notation by the equation
	
	\begin{equation}\label{eq:gravity vec}
	    \fvec_1=-\fvec_2=\frac{Gm_1m_2}{|\rvec_2-\rvec_1|^3}(\rvec_2-\rvec_1),
	\end{equation}
	where $\rvec_1$ and $\rvec_2$ denote the position vectors of the two particles. Note that Eq.~\ref{eq:gravity vec} is simply the vector form of the same equation expressing the \emph{magnitude} of gravitational force between two bodies presented in Sec.~\ref{sec:gravity} (Eq.~\ref{eq:gravity magnitude}). Of course, Eq.~\ref{eq:gravity vec} is only the $N=2$ case. Noting that forces are additive (that is, $\fvec_\text{net}=\sum_i\fvec_i$), we can write the equation for for the gravitational force acting on the $i$th particle in a configuration of three or more massive particles as
	\begin{equation}\label{eq:gravity vec sum}
	    \fvec_i=\sum_{i\neq j}\frac{Gm_im_j}{|\rvec_j-\rvec_i|^3}(\rvec_j-\rvec_i).
	\end{equation}
	Eq.~\ref{eq:gravity vec sum} gives us our Newtonian approximation of gravity, so we need only solve this second order differential equation for all the particles in our simulation and it's off to the races! Unfortunately%\note{"as mentioned in..."}
	, this equation has no analytical solution for $N>2$ \cite{beckett}, so rather than spending our careers trying to solve the unsolvable by determining the exact paths traced by our $N\gg2$ particles with respect to time, we need to use a numerical integration technique to approximate these paths. A particularly common numerical method used for force integration among N-body codes \cite{quinn}, which also happens to be the method adopted by ChaNGa \cite{beckett}, is known as ``leapfrog integration.'' %The first step\footnote{No pun intended} employed by ChaNGa in 
	%The first step\footnote{Pun intended.} in the leapfrog method is to divide time up into discrete time \emph{steps}, so that rather than dealing with the continuous motion of particles under the influence of a time-dependent force, we only consider their 
	As the reader with experience in numerical methods likely already knows, the first step to making such a problem as that presented by Eq.~\ref{eq:gravity vec sum} suitable for evaluation by a computer is \emph{discretization}, the division of a continuous phenomena into appropriately small (but still finite) discrete ``steps.'' While the programmer has some freedom in choosing how to handle the process \cite{beckett}, the approach adopted in the leapfrog method is time discretization, breaking up the continuous ``flow'' of time into individual ``chunks'' (or \emph{time steps}). Mathematically, the routine used for leapfrog integration (derived via Taylor expansion) is most commonly expressed using the equations \cite{beckett}
	%The basic principle underlying the leapfrog method is time discretization --the division of time into individual ``steps''-- 
	%such that you only consider the motion of particles from one time step to another rather than their continuous movement through space
	%such that the motion of particles is only considered between time steps, rather than continuously. The routine for leapfrog integration is most commonly expressed using the equations
	\begin{equation}\label{eq:leapfrog0velocity}
	    \vvec_{n+\frac{1}{2}}=\vvec_{n-\frac{1}{2}}+\avec_n\Delta t
	\end{equation}
	\begin{equation}\label{eq:leapfrog0position}
	    \rvec_{n+1}=\rvec_n+\vvec_{n+\frac{1}{2}}\Delta t.
	\end{equation}
	The form of Equations \ref{eq:leapfrog0position} and \ref{eq:leapfrog0velocity} gives a good impression of why the method is called ``leapfrog'' \cite{quinn}: given initial values for position and velocity, first you use Eq.~\ref{eq:leapfrog0velocity} to update velocity, then you use the new velocity in Eq.~\ref{eq:leapfrog0position} to update position, and then you repeat the process alternating between velocity and position in a manner similar to the game ``leapfrog''. 
	
	While Equations \ref{eq:leapfrog0velocity} and \ref{eq:leapfrog0position} give us the routine in a nice compact form \emph{and} provide insight into its rather fun nickname, they are not the equations that ChaNGa actually uses in it's implementation. Instead, ChaNGa uses what is known as the ``kick-drift-kick'' form of the routine, which can be written as \cite{beckett} \cite{mitch}
	\begin{equation}\label{eq:kick drift kick 1}
	    \vvec_{n+\frac{1}{2}}=\vvec_n+\avec_n \frac{\Delta t}{2},
	\end{equation}
	\begin{equation}\label{eq:kick drift kick 2}
	    \rvec_{n+1}=\rvec_n+\vvec_{n+\frac{1}{2}}\Delta t,
	\end{equation}
	\begin{equation}\label{eq:kick drift kick 3}
	    \vvec_{n+1}=\vvec_{n+\frac{1}{2}}+\avec_{n+1}\frac{\Delta t}{2}.
	\end{equation}
	Taking a closer look at these equations, we can begin to see where the name ``kick-drift-kick'' comes from. At each step, we first update the intermediate velocity term ($\vvec_{n+\frac{1}{2}}$) using the ``current'' gravitational acceleration (the acceleration acting as the ``kick''), then update the position ($\rvec_{n+1}$) according to this new velocity (the equation treats the particle as though it is moving in the absence of external forces, i.e. drifting), then update the velocity ($\vvec_{n+1}$) using the gravitational acceleration at the new spatial position. 
	
	There are a number of reasons why the kick-drift-kick format of the leapfrog method is favored by ChaNGa's approach\footnote{The second-to-last paragraph in Sec. 1.1.2 of \cite{beckett} has a nice discussion on why these equations are also preferable in terms of \emph{form}.}, paramount among these, however, is a concept known as multi-stepping \cite{menon}. %In essence, the number of timesteps occurring within a fixed time interval is largely what determines the physical accuracy of this algorithm
	In essence, the number of computations (Equations \ref{eq:kick drift kick 1}-\ref{eq:kick drift kick 3}) made within a fixed time interval is largely what determines the physical accuracy of the algorithm. Since the method effectively amounts to an approximation in which particles move with fixed velocity between time steps, the smaller the time step, the more ``true-to-life'' the paths become. However, as always there is a computational tradeoff in terms of runtime, and so rather than pursuing increasingly small timesteps, the goal is generally to find a step number relative to the size of the time interval for which the path accuracy is ``good enough.'' Since gravity is a spatially dependent force, the notion of what constitutes a ``good enough'' temporal resolution is closely linked with particle density. Because of this, 
	%modelling particle trajectories in high density and low density regions with the same level of physical accuracy will
	regions with higher particle density will naturally require smaller timesteps in order to achieve the same level of physical accuracy that can be attained in a lower density region with a larger timestep %\note{holy fuck revise this sentence pls} 
	\cite{beckett} \cite{mitch} \cite{springel2005}. 
	%found in lower density regions with a much smaller timestep
	%Regions of higher particle density will naturally require smaller time steps to accurat
	While one approach to this problem might be to increase the number of timesteps \emph{globally} to be suitable to the desired level of accuracy in high density regions, doing so would introduce a large number of extraneous calculations for low density regions and would thus be wasteful from a computational perspective \cite{springel2005}. The solution to this efficiency problem is to assign time steps on a per-particle basis, that is, each particle gets its own time step size in accordance with the density of surrounding particles (multi-stepping). The kick-drift-kick method ensures that such an individualized time-step assignment is possible, so long as the time step sizes of all particles are related by powers of two \cite{beckett}.
	
	%\note{Wanna say more here? Calculating acceleration twice per time step?}
	
	\section{The Barnes-Hut Algorithm}\label{sec:barnes-hut}
	
	In the previous two sections, we discussed the use of Newtonian gravity as a weak-field approximation to the far more complicated general relativistic formulation, and the use of the leapfrog integration method to numerically evaluate motion in a Newtonian gravitational field. One might reasonably expect that this brings us close to a complete understanding of ChaNGa's most rudimentary functionality: solving the N-body problem for purely gravitational interactions (readers who are concerned by my use of ``rudimentary'' here should be comforted to know that I explain myself in Sec.~\ref{sec:sph}). Unfortunately, owing to the ever-complicated nature of the N-body problem, we find ourselves in yet another snare, that being the issue of runtime efficiency. To understand the problem, first note that the \emph{net} acceleration terms in the leapfrog routine (Eq. \ref{eq:kick drift kick 1}-\ref{eq:kick drift kick 3}) are actually the result of the sum of gravitational interactions expressed in Eq.~\ref{eq:gravity vec sum}. From the form of the latter, it shouldn't be too hard to convince yourself that, in a simulation of $N$ particles, each particle experiences $N-1$ gravitational interactions. Thus, if we were to write a simulation code with just the information from Sections \ref{sec:code gravity} and \ref{sec:code integration}, at each time step our program would need to perform $N-1$ force calculations $N$ times which, %as readers familiar with asymptotic analysis in computer science 
	for the reader familiar with asymptotic notation in computer science, is $O(N^2)$ \cite{finnish}. Such asymptotic behavior might be acceptable for small-scale computations, the current state of computational astrophysics is such that simulations can have particle counts well in the millions and often evolve over thousands of time steps \cite{beckett} \cite{mitch}. Modern codes need to accommodate for this level of scale, which necessitates methods of speeding up the gravitational force calculation process. 
	
	The solution employed by ChaNGa is to use a modified version of the Barnes-Hut algorithm \cite{barnes-hut}, which offers an approximation of the gravitational force acting on a particle in logarithmic time ($O(\log N)$), making the total force calculation per timestep $O(N\log N)$, a substantial improvement from the polynomial time ``direct summation'' method. In the next few paragraphs, I will introduce the standard Barnes-Hut algorithm to familiarize the reader with the general approach to the problem, and will then proceed to discuss some of the modifications employed by ChaNGa and the motivations behind them. The reader searching for a more in-depth or specific treatment of Barnes-Hut might be interested in reading the original 1986 paper \cite{barnes-hut}, or some of my more undergraduate-friendly sources such as \cite{barnes-hut-princeton} and \cite{beckett}.
	%to simplify the gravitational force calculation for distant particles
	
	%\note{Probably include a figure here}
	
	The fundamental principle behind the Barnes-Hut algorithm is to reduce the number of gravitational acceleration calculations performed for each particle by applying a center-of-mass approximation for the force exerted by distant particles. 
	%approximating distant particle clusters as single point-sources.
	More explicitly, for distant clusters --particles that are relatively far from the particle whose acceleration we want to calculate but are relatively close to each other-- the algorithm determines the center of mass and total mass of the cluster, then performs a single gravitational force calculation using these values. %The justification for treating distant clusters as single ``pseudoparticles'' 
	As noted in \cite{beckett}, the nature of gravitation is such that for a sufficiently distant cluster of identical particles the force contribution associated with each particle will be roughly identical (both in terms of magnitude and direction), so the approximation is justified. Of course, distance constitutes the main underlying assumption for the approximation, meaning it does not hold for nearby particles, for whom direct summation is still performed. %, so for nearby particles the Barnes-Hut algorithm performs direct summation
	
	%The cursory description of the algorithm in the previous paragraph
	Of course, since this approach relies so heavily on distance, we need to establish a way to determine what constitutes ``far enough'' for a center-of-mass approximation and ``close enough'' for particles to be considered part of the same cluster. On cursory inspection, one may be tempted to simply compare the distances of all the particles in the simulation, but such a process would be $O(N^2)$ itself \cite{beckett}, %which would defeat the purpose of optimization. %and would therefore render the optimization
	no better than direct summation. 
	Barnes-Hut offers a rather clever solution to this problem, which starts with a process known as ``domain decomposition,'' or the repeated decomposition of the initial volume into smaller subvolumes. The traditional Barnes-Hut algorithm uses an Octree decomposition structure \cite{beckett} \cite{barnes-hut}, in involves recursively dividing cubical volumes into eight subvolumes\footnote{This is not the decomposition scheme ChaNGa opts for, but understanding the basic Octree decomposition method can be helpful as an introduction to what ChaNGa is doing.}, hence the name. A summary of Octree decomposition is as follows:
	
    \begin{enumerate}
        \item Begin with the volume containing all particles, this is the ``root node'' of the Octree.
        
        \item If there are no particles in the node, discard it.
        
        \item If two or more particles are present in the cell, divide it evenly into eight ``daughter nodes,'' then inspect each daughter node individually.
        
        \item repeat steps 2-3 until each node contains at most one particle.
    \end{enumerate}
    
\begin{figure}
        \centering
        \includegraphics[scale=0.3]{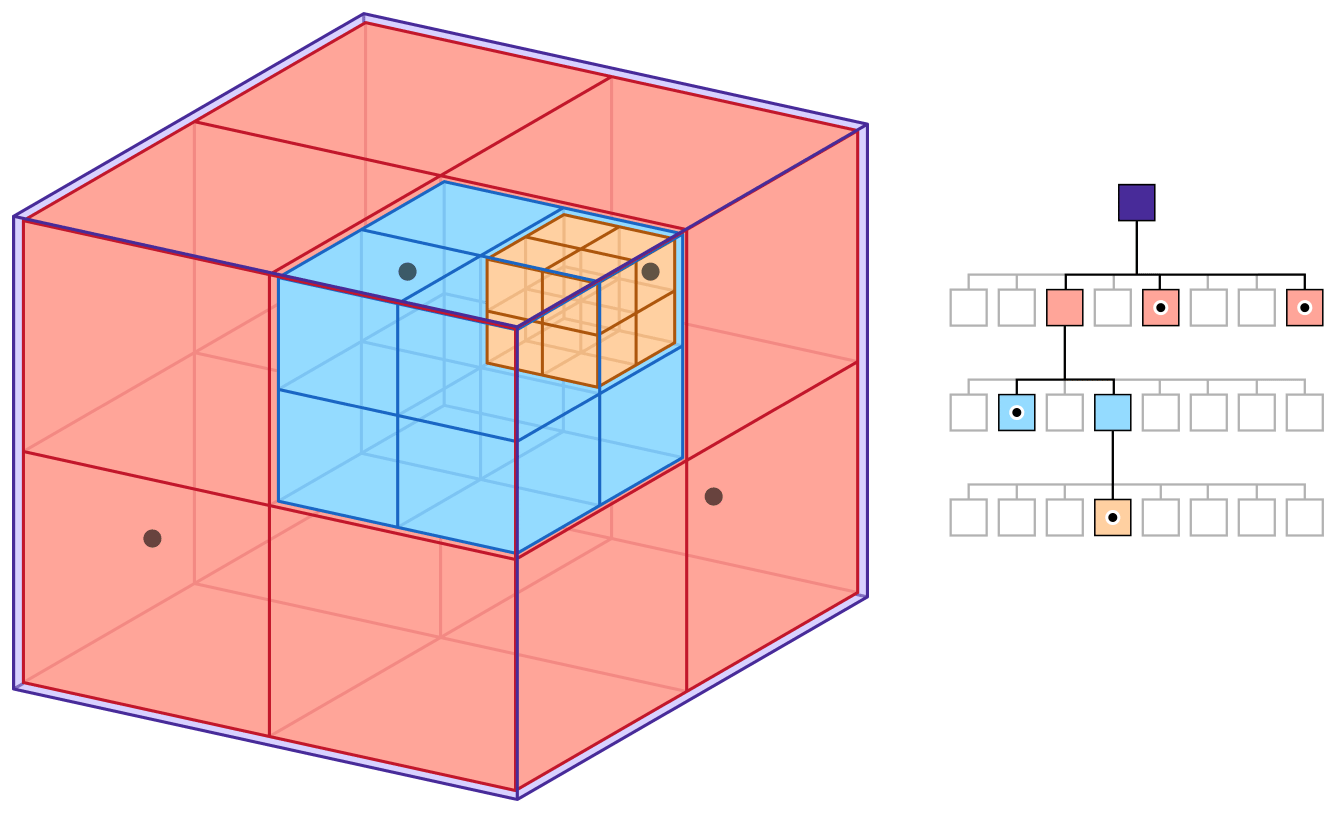}
        \caption[A visual representation of an Octree decomposition for four particles taken from Apple's GameplayKit documentation website \linebreak \href{https://developer.apple.com/documentation/gameplaykit/gkOctree}{https://developer.apple.com/documentation/gameplaykit/gkOctree}]{A visual representation of an Octree decomposition for four particles taken from Apple's GameplayKit documentation website \href{https://developer.apple.com/documentation/gameplaykit/gkOctree}{https://developer.apple.com/documentation/gameplaykit/gkOctree}.}
        \label{fig:octree}
    \end{figure}

    Fig.~\ref{fig:octree} gives a visual example of this process for a volume containing four particles. Note the nodes vary in size according to their corresponding ``level'' in the tree structure. Only boxes containing more than one particle are subdivided into daughter nodes, and division is performed recursively until each node contains one or zero particles.
    
    In total, the time complexity of constructing the Octree is $O(N\log N)$ \cite{barnes-hut}, so far we are still looking at an improvement on direct summation. We still need to discuss how exactly this structure expedites the force calculation, however, an insight which may yet elude 
    %is likely to elude 
    all but the most computer-savvy readers. 
    %To the reader who has never seen this algorithm before, it may still be unclear exactly how Octree domain decomposition helps with the force calculation. 
    The next step involves center of mass calculations. The implementation of the algorithm as described in the original paper \cite{barnes-hut} is such that the data structure representing a node in the tree has attributes corresponding to the total mass and center of mass (COM) for the particles contained therein. For nodes containing only one particle (dubbed ``leaf nodes'' in the language of the tree metaphor), these values are trivial, so it is most efficient to propagate this information backwards through the data structure, i.e. leaves-to-root, a process which is also $O(N\log N)$ \cite{barnes-hut}. 
    
    \begin{figure}
        \centering
        \includegraphics[scale=0.8]{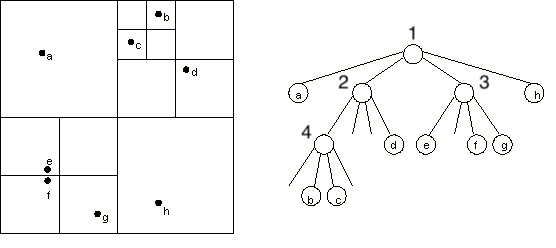}
        \caption{A 2D analogue to the Barnes-Hut Octree, known as a Quadtree. This image was borrowed from  (\href{https://www.cs.princeton.edu/courses/archive/fall03/cs126/assignments/barnes-hut.html}{\emph{Cos 126 programming assignment: Barnes-hut galaxy simulator}, Princeton University, (2003)}), which provides a very approachable description of Barnes-Hut in two dimensions.}
        \label{fig:quadtree}
    \end{figure}
    
    If we examine the analogous 2D Quadtree data structure depicted in Fig.~\ref{fig:quadtree}, we can see that the COM and total mass calculation would begin with node 4, whose corresponding attributes will be calculated using the positions and masses of particles a and b. We can then use these attributes of 4, as well as the mass and position of particle d, to compute the total mass and COM of node 2 and so on. Nodes containing more than one particle can then be treated as ``pseudoparticles'' with their own associated mass and position\footnote{Given by their total mass and center of mass attributes, respectively.}. 
    
    Once all the psuedoparticle nodes in the simulation have been assigned their mass and COM attributes, all the machinery necessary to begin force calculations is in place. The algorithm for calculating the force on a given particle $p$ involves a traversal of the Octree, where $\ell$ is the side length of the cubical region represented by the current node, $d$ is the distance between $p$ and the node's COM, and $\theta$ is an accuracy parameter set at the start of the simulation, usually ~1. With these values in mind, the algorithm for each particle goes as follows \cite{barnes-hut}
    \begin{enumerate}
        \item Start at the root node.
        
        \item If $\ell/d<\theta$, compute the gravitational force between the current node and $p$, and add it to the total force acting on $p$.
        
        \item Otherwise, traverse one layer down the tree and perform Step 2 for each daughter node.
        
        \item End.
    \end{enumerate}
    
    %\note{Too computer-science-y?}
    
    Once this algorithm has run to completion, the gravitational influence of each particle on $p$ will have been accounted for (either by direct force calculation or COM approximation), so it presents a reliable method of updating particle accelerations. For large $N$, this process involves performing order $\log N$ force calculations for each particle, so the asymptotic runtime of the algorithm is still $O(N\log N)$ overall!
    
    %in practice the algorithm is likely to be implemented such that cells are represented by data structures that are able to account for this information.
    
    %in practice the implementation of the algorithm is likely to be such that cells are represented by data structures that are able to account for this information. 

	\section{Barnes-Hut Modifications}\label{sec:barnes-hut changa}
	
	%The algorithm for computing gravitational interactions as I've described it so far is roughly in line with that outlined in the original paper \cite{barnes-hut}
	
	The description I've provided of the Barnes-Hut gravitational force computation has thus far followed closely with the original paper \cite{barnes-hut}, but as previously mentioned there are a number of changes that the ChaNGa code makes to adapt the algorithm to its approach. The first notable change is that ChaNGa doesn't actually use the center of mass of far away particle groups to approximate their gravitational influence. Instead, ChaNGa performs an operation known as a \emph{multipole expansion} for improved force accuracy \cite{menon}. A sufficiently advanced physics undergraduate is likely to have encountered the multipole expansion in an E\&M course (see Ch. 3.4 in \cite{griffiths}), where it is commonly used to model electric fields which cannot easily be calculated analytically \cite{beckett}. Due to the similarities between Newtonian gravity and the electric field, this method can also be used to approximate the gravitational potential produced by analytically difficult \emph{mass} distributions (e.g. clusters of massive particles).
	
	The multipole expansion effectively amounts to an infinite sum of terms (similar to a Taylor series) with increasing angular dependence \cite{multipole}. The first (or zeroth order) term is a monopole, or a point source\footnote{i.e. a point charge or point mass. The center of mass approximation used in the original Barnes-Hut algorithm is effectively just a zeroth order multipole expansion.}, which has no angular dependence. The first few higher order terms are the dipole, quadrupole, and hexadecapole terms. While a derivation or a precise mathematical description of the multipole expansion are beyond the scope of this document, readers can turn to \cite{griffiths} or \cite{multipole} for more detailed discussions in the E\&M context, or to \cite{stadel} for the N-body simulation context. For the reader in a hurry though, it is most important to know that ChaNGa performs a multipole expansion to third order, i.e. includes all terms up to and including hexadecapole \cite{menon}, and that this is both faster and more accurate than a quadrupole order expansion \cite{stadel}. 
	
	A second notable difference between the 1986 Barnes-Hut algorithm and that employed by ChaNGa is the precise structure of the spatial decomposition tree. As we saw in Sec.~\ref{sec:barnes-hut}, the algorithm as presented in the original paper recursively divides the spatial domain into even sections of eight, resulting in a data structure called an Octree (again, see Fig.~\ref{fig:octree}). The spatial decomposition performed by ChaNGa actually employs a binary tree rather than an Octree, meaning the domain is recursively divided into sections of two rather than eight. %ChaNGa performs a binary spatial decomposition, that is, recursive division of space into sections of two rather than eight. 
	This change is in line with the approach used by the N-body code PKDGRAV \cite{stadel}, from which ChaNGa inherits much of its gravitational force calculation \cite{menon}. It should be noted that %PKDGRAV uses 
	the justification for using a binary tree rather than an Octree %because 
	is that %doing so
	a binary tree structure offers advantages for \emph{code parallelization}, particularly in terms of the distributing the computation over an arbitrary number of processors \cite{stadel}, a feature which is clearly desirable in the context of high performance computing.
	
	\begin{figure}
	    \centering
	    \includegraphics[scale=0.35]{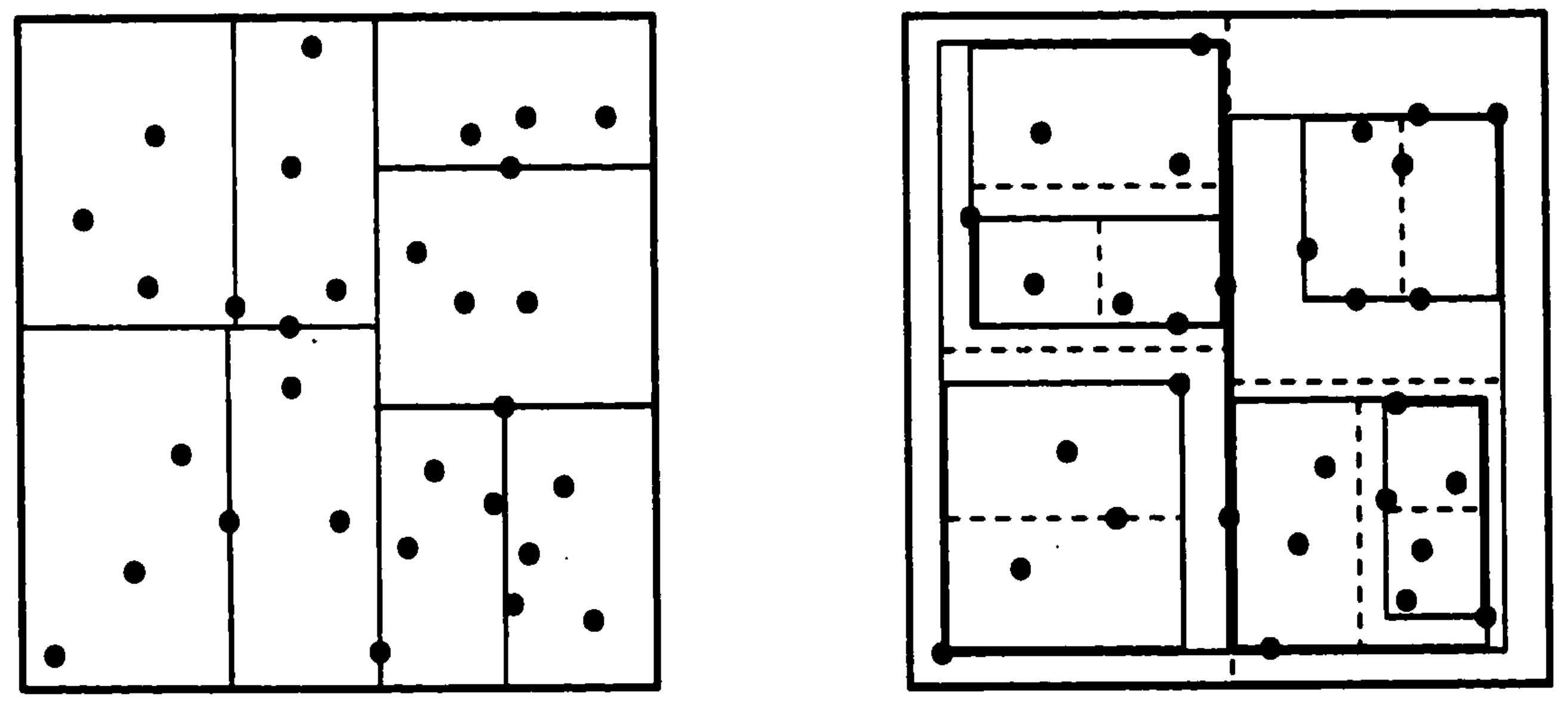}
	    \caption{Images representing a standard k-D tree domain decomposition (left) and the spatial binary tree decomposition used by PKDGRAV (right). This image is borrowed from (J. G. Stadel, \emph{Cosmological N-body simulations and their analysis}, Ph.D. thesis, University of Washington, Seattle (2001)).}
	    \label{fig:binary tree}
	\end{figure}
	
	While the precise details of ChaNGa's spatial binary tree implementation prove to be rather elusive in the literature, ChaNGa explicitly inherits much of its gravitational force calculation from PKDGRAV \cite{menon}, the details of which can be found in \cite{stadel}. Because of this inheritance, a brief discussion on PKDGRAV's %implementation of domain decomposition is warranted 
	domain decomposition method is warranted to provide a better understanding of what is happening in ChaNGa, if only by proxy. Aside from the overall difference in organizational structure, a crucial difference between the Barnes-Hut Octree and the binary trees used by modern simulation codes is that in the latter space is often not divided up evenly. Instead, daughter nodes are sized dynamically in accordance with some bisection scheme, often relating to the number of particles per daughter node. An example of such a binary tree algorithm (in two dimensions) is the k-D tree, depicted in the left hand side of Fig.~\ref{fig:binary tree}. This algorithm involves recursively bisecting nodes through their longest axis such that both daughter nodes contain approximately the same number of particles. In the figure, you can see that the first bisection is done such that each daughter node contains ~15 particles, then ~8, and finally ~4. %As previously stated, recursive bisection algorithms have advantages in terms of parallelization, 
	
	The k-D tree offers a good first look into spatial binary trees with dynamically sized nodes, but it also presents problems in terms of force error and runtime efficiency \cite{stadel}. The decomposition method used by PKDGRAV (again, one that is likely inherited in some capacity by ChaNGa \cite{menon}) is depicted (again, in 2D) on the right hand side of Fig.~\ref{fig:binary tree}. Comparing this diagram to its neighbor representing the k-D tree, it is clear that there's a bit more going on, and readers who are anything like myself will find themselves initially confused at what exactly is being represented. %However, it should be noted that the primary difference between these two algorithms 
	
	To better understand what's happening in the diagram, one should note that this algorithm really has only one fundamental difference from the k-D tree: the ``squeezing'' of daughter nodes to minimize the volume they represent. At every step, PKDGRAV's binary tree compresses the volume of node being examined so that %it only considers the bounding box for the particles
	represents the bounding box\footnote{i.e. the smallest rectangle --or rectangular prism in 3D-- that contains all the particles in the node.} for the particles. %The algorithm then considers the bounding box to be the new node, and 
	Since the bounding box is now considered to be the node's volume, it gets bisected through its longest axis into two daughter nodes containing roughly equal numbers of particles, %, before ``squeezing'' the daughter nodes and bisecting them if the number of particles exceeds the maximum for leaf nodes. 
	which are then squeezed themselves. This process is repeated until each node is under the maximum particle count for a leaf node, which for ChaNGa is usually around 8-12 \cite{menon}.
	
	I have so far discussed two of the fundamental modifications to the Barnes-Hut algorithm employed by ChaNGa, namely the use of a hexadecapole order multipole expansion for the gravitational force approximation at large distances, and the use of a binary tree structure rather than an Octree. These details, along with the description of Barnes-Hut provided in Sec.~\ref{sec:barnes-hut}, should give the reader a pretty good idea of how ChaNGa handles its gravitational force calculation, at least at the conceptual level, and I could --in fairly good conscious-- leave the discussion at that to move onto other aspects of ChaNGa, referring the reader whose curiosity has yet to be satiated to a number of more detailed resources. Before moving on, however, there is \emph{one} final aspect of the gravitational force calculation that I believe bears discussion in this section, if only briefly, namely ChaNGa's approach to parallelization, since large-scale parallelization is a primary raison d'\^{e}tre for the code \cite{massively-parallel}.
	
	%Since parallelization raison d'\^{e}tre
	A detailed description of the parallelization process is beyond the scope of this document, interested readers are encouraged to look at Section 4 of \cite{menon}. What I will provide in the next paragraph is a treatment at the conceptual level, with the hope %providing the reader with at least some understanding of
	that the reader will come away an impression of the role parallelization plays in the code and how it affects the force calculation process. To talk about parallelization in ChaNGa, I must first confess to a small lie\footnote{A revelation that may not come as a surprise for readers who have been following along with this chapter in a linear fashion.}. In previous paragraphs I described how the code uses a spatial binary tree to handle its gravitational force calculation, and while it is true that the binary tree is the fundamental data structure for this computation, ChaNGa's force calculation actually involves not one but a number of spatial binary trees, each representing a sub-region of simulation volume containing a subset of the overall particle count, divided up among the processors allotted to the computation \cite{menon}. To facilitate this approach, ChaNGa performs an initial domain decomposition step \emph{not} described by the spatial binary tree or Octree, in which the simulation volume is divided into a number of sub-regions containing an equal number of particles according to a space-filling curve (SFC) algorithm \cite{menon}. The precise details of this initial decomposition, again, are beyond the scope of this discussion, but it is important to emphasize that the resulting spatial decomposition looks different from a binary tree or Octree, and that this process occurs as an initial step to balancing the computational load across the processors being used. The code iterates through a number of potential decompositions until all bins (sub-regions) are deemed ``sufficiently optimal'' \cite{menon}. All particles in each bin are then assigned to a data object called a tree piece\footnote{Used to facilitate parallelization} so that each tree piece represents a subset of the overall volume. Each tree piece then performs a spatial binary decomposition of its assigned subvolume (using a bounding box method similar or identical to the one previously described), and the gravitational force calculation begins \cite{menon}. Note that, since particles are assigned to tree pieces in accordance with the initial SFC decomposition, no single processor has direct access to all the particles in the simulation, but all particles interact with each other gravitationally. Because of this, ChaNGa also facilitates communication across processors to allow remote access to nodes that are part of different tree pieces \cite{menon}.

	\section{Boundary Conditions and Ewald Summation}\label{sec:boundary conditions}
	
	%It is an open question in modern cosmology whether the universe is infinite
	Whether or not the universe is truly infinite is an open question in modern cosmology, but we know that the observable universe spans a distance of 46 billion light years \cite{infinite}. Even very large simulations such as those performed as part of the IllustrisTNG project \cite{illustris} have computational volumes with widths only\footnote{The concerned reader should be informed that I use ``only'' here with more than a little sarcasm.} in hundreds of millions (ly). %Indeed, with modern technology it would be impractical to 
	Thus, even exceptionally large simulations model regions that only constitute small fractions of the (very possibly infinite\footnote{In which case my prior use of the word ``fraction'' loses all meaning.}) universe. Seeing as how gravity is a force that acts at large distances, this limitation\footnote{If it can even be called such. Would that we could perform an infinite simulation. Alas.} presents a distinct problem for simulation realism. %This problem is overcome by a process known as Ewald Summation \cite{menon}, but before I formally introduce this concept, 
	Before discussing approach to this issue, however, it would be prudent to first draw the reader's attention to a concept that has become a staple of the currently accepted cosmological model: the cosmological principle. The cosmological principle states that the universe is both 1.) isotropic, and 2.) homogeneous \cite{bob}. What does this mean? Well, being isotropic means that the universe looks the same in all directions. Similarly, being homogeneous means that the universe looks the same at all locations. Readers who have looked at the night sky may object, clearly the universe doesn't look the same in all directions if we can make out distinct features like stars and constellations. Those who have peered through a telescope might attest that doing so only heightens this sentiment. Others might argue in a similar vein that the universe \emph{must} look at least a little different for someone sitting in the Andromeda galaxy. Such readers might be relieved to hear that these %issue is simply a matter of scale.
	grievances effectively boil down to the issue of scale. The cosmological principle doesn't state that on a  small scale (such as that of our observational perspective) the universe cannot contain distinct features, but that at sufficiently large scales (relative to our perspective) the two axioms hold \cite{uoregon-cosmology}. In effect, if we zoomed out and considered a scale on the order of millions of light years, our observations would be consistent with the cosmological principle \cite{uoregon-cosmology}.
	
	%As a consequence of the cosmological principle, to model the universe we can simply treat our simulation volume 
	The cosmological principle provides justification for modelling a simulation volume as a single cell in an infinite (or at least very large) 3D grid of perfectly identical 
	cubes,  
	%simulation volumes. %This is where Ewald Summation comes into play. 
	%treat such a 
	%To implement this model, codes like ChaNGa\footnote{Note that grid-based codes do not need to account for boundary conditions, as they are implicit in the grid model \cite{stadel} \cite{wadsley}.} implement periodic boundary conditions by splitting up the gravitational calculation over the infinite grid into short-range and long-range components \cite{mitch} \cite{stadel}.
	%treat the boundary of the simulation as being periodic,
	%treat the simulation as having periodic boundary conditions
	 %via Ewald summation \cite{stadel} which effectively splits
	 %periodic lattice (the repeating grid previously mentioned)
	%, which are then calculated in real space and Fourier space respectively \cite{mitch}.
	which is precisely the approach taken by ChaNGa and similar codes\footnote{Note that grid-based codes do not need to account for boundary conditions, as they are implicit in the grid model \cite{stadel} \cite{wadsley2003}, see Sec.~\ref{sec:sph} for a brief explanation of the term ``grid modell''.}. Such codes are referred to as having \emph{periodic boundary conditions}, since the structure of the simulation volume is repeated over a periodic lattice \cite{stadel}. This approach addresses the issues with realism that arise from finite simulation volumes, but an obvious consequence %of this approach 
	is that it is now necessary for these simulations to somehow treat the gravitational force calculation over this infinite lattice, which may seem no less daunting. As it turns out, however, there are ways to handle such a calculation, such as the solution adopted by ChaNGa: breaking up the calculation into long-range and short-range components \cite{mitch} \cite{stadel}. 
	%which as it turns out, is done by breaking up the calculation into long-range and short-range components \cite{mitch} \cite{stadel}
	%, for which simulations are referred to as having \emph{periodic boundary conditions}
	The short-range calculation is performed as an extension of the Barnes-Hut algorithm described in Sec.~\ref{sec:barnes-hut} by including a number of neighboring lattice cells (usually 26 of them \cite{stadel}) in the Barnes-Hut force calculation. The long-range gravitational contribution is then accounted for via Ewald summation, the details of which are beyond the scope of this document but can be read about in \cite{stadel} or the original paper \cite{ewaldsum} (though it should be noted that the version used by the former, and ChaNGa for that matter, differs slightly from the original). To briefly touch on the concept, the Ewald summation employed by ChaNGa models gravity over a periodic lattice as a Green's function, which is then truncated past a certain (reasonable) distance from the particle in question \cite{stadel}.
	
	%a process which is usually performed over the 26 neighboring volume cells \cite{mitch} \cite{stadel}. The calculation in reciprocal space  
	%, meaning the volume is treated as being a part of a ``periodic lattice'' \cite{stadel} (the grid described in the previous statement
	%meaning that gravitational forces that would generally extend beyond the simulation volume effectively ``loop'' back around and act on the particles being simulated. 
	
	%As a consequence of this feature of modern cosmology, we can treat our simulation volume as being a single cell in an infinite 3D grid of perfectly identical simulation volumes.
	
	\section{Force Softening}\label{sec:force softening}
	
	One final issue related to the gravitational force calculation, which is a consideration for all N-body codes \cite{stadel} is how to handle gravity between particles at small distances. Looking back at Eq.~\ref{eq:gravity vec}, we can see that because of the denominator term, the interparticle force blows up very quickly at small particle separations. This presents a problem because 1.) It can be difficult for codes to handle incredibly large forces computationally \cite{beckett}, and 2.) such forces would violate the assumptions underlying our model, leading to unphysical results \cite{stadel}. To counteract this effect, it is important for an N-body code to implement some kind of \emph{softening} to impose a limit on force magnitude. In ChaNGa this ``softening'' is closely related to the handling of Smoothed Particle Hydrodynamics (see the following section, Sec.~\ref{sec:sph}) and involves something called a \emph{spline softening kernel} \cite{menon} \cite{stadel}, which effectively ``cuts off'' gravitational force at an interparticle separation of zero, while maintaining usual Newtonian gravity (Eq.~\ref{eq:gravity vec}) at distances above some prescribed softening length \cite{wadsley2003}. 
	%the denominator term, $|\rvec_2-\rvec_1|^3$ becomes very small very quickly as the interparticle distance decreases. 
	
	\section{Smooth Particle Hydrodynamics (SPH)}\label{sec:sph}
	
	Evidently a robust computational treatment of gravity is no simple task. The entirety of this Chapter has so far been dedicated to this exact subject in the context of ChaNGa, and there remain a number of specific nuances that have only been discussed at the superficial level, or that have been omitted entirely\footnote{I once again urge the curious reader to look at resources such as the following for more detailed information: \cite{stadel} \cite{menon} \cite{wadsley2003} \cite{wadsley2017}.}. The reader should know that I do not intend to take my discussion of ChaNGa much further than %the gravitational force calculation
	what has already been provided in previous sections, as one could write an entire thesis just on the workings of a single sufficiently advanced N-body code\footnote{indeed this has been done at a doctoral level \cite{stadel}.}, and that is not my intention for this document. %However, there is one particular additional element of ChaNGa for which it would be absolutely negligent to not discuss at least briefly
	Nonetheless, there remains one final element of ChaNGa that --from my perspective-- I would be absolutely remiss to not discuss at least briefly: gas dynamics. 
	
	To reiterate: all that I've discussed about ChaNGa so far has been related to predicting the motion of infinitesimal point masses under the influence of a mutual gravitational force. Clearly this is crucial for compuational astrophysics: gravity is a dominant force in the cosmos. However, purely basing the motion of simulated particles on gravity would only be sufficient if we lived in a universe that consisted purely of discrete, massive bodies (such as stars and dark matter). In particular, a purely gravitational simulation would have to ignore the significant amounts of gas and dust present in our universe, which also play a significant role in galaxy formation and cosmological structure \cite{beckett}. Gravity clearly influences the motion of gas, but additional considerations need to be made for the dynamics of the gas itself. In order to handle this, N-body codes turn to fluid mechanics, treating the gas in a simulation as a continuous fluid medium, a reasonable approximation at a macroscopic scale \cite{beckett} \cite{fluid}. 
	
	In general, there exist two approaches to fluid dynamics: the Eulerian description and the Lagrangian description \cite{fluid}. The Eulerian description considers what is happening at fixed locations (or points) in space as a fluid medium flows through said points, whereas the Lagrangian description considers the dynamic evolution of individual pieces of matter representing ``fluid elements'' as they travel through space \cite{fluid}. This dichotomy within fluid mechanics is reflected in N-body simulation codes that treat gas dynamics. Broadly speaking, there are two types of codes\footnote{While this is an oversimplification, it is true that most major codes fall into one of two camps.}: grid-based (also referred to as mesh-based) codes, which follow the Eulerian description by considering a simulation grid and tracking the flow of gas through that grid, and particle-based codes %like ChaNGa
	that follow the Lagrangian description by tracking the spatial motion individualized of gas ``parcels'' \cite{wadsley2003}. A rigorous treatment of either of these techniques is beyond the scope of this document, and my only motivation for bringing the reader's attention to this ``dicotomy'' of codes is because it is a subject that comes up often in the literature, and because I have personally found it helpful in my own research to be at least superficially aware of this difference. 
	
	%Though I do not intend to provide much depth in my discussion of simulated fluid dynamics, there is one piece of information that 
	%Rather than taking a rigorous approach
	Though I will not be going into much depth, I would like to provide at least a conceptual description of ChaNGa's approach to implementing the fluid dynamics of gas. The first thing to note is that ChaNGa opts for the Lagrangian description: tracking individualized parcels of gas rather than focusing on the flow of matter through fixed points in space. One of the advantages to this approach is that it is a fairly natural extension of the Barnes-Hut based gravitational force calculation described in Sec.~\ref{sec:barnes-hut} \cite{menon}. In fact, the particles involved in the gravitational force calculation are \emph{also} used to discretize gas, effectively representing the ``parcels'' for the Lagrangian approach \cite{beckett}. This is accomplished through a process known as Smoothed Particle Hydrodynamics (SPH), which uses the particles present in the simulation to derive continuous fluid quantities (such as pressure, temperature, etc.) spanning the surrounding region of space \cite{beckett} \cite{mitch}. In this sense, the particles can be thought of as getting ``smoothed out'' (hence the name) over space, smudged into a continuous particle-goo that represents the fluid medium of the gas. 
	
	\begin{figure}
	    \centering
	    \includegraphics[scale=0.15]{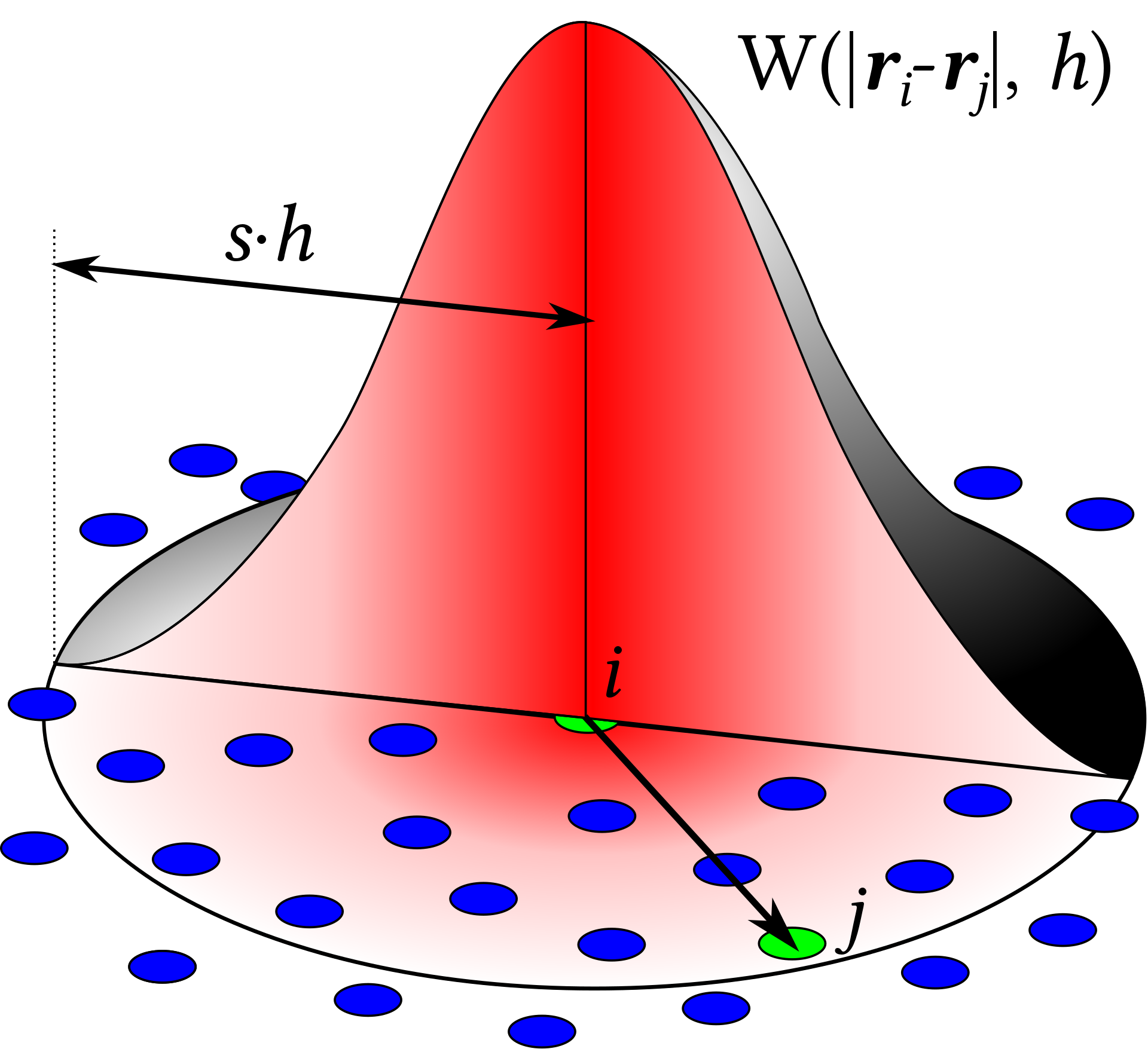}
	    \caption{A depiction of a smoothing kernel being applied to a discrete set of particles. This image was uploaded to Wikipedia by the user Jlcercos and is licensed under Creative Commons. Here the circles represent the particles, $h$ the smoothing length, and the 3D curve represents the value of the smoothing kernel for all points in space. }
	    \label{fig:sph}
	\end{figure}\
	
	For those who are more mathematically inclined, I will briefly provide a more technical description of the smoothing method used for SPH. Readers who are interested in learning more are encouraged to look at Section 5 of \cite{wadsley2003}, as ChaNGa's implementation of SPH follows closely with the description provided therein \cite{menon}. The method involves using a function called a \emph{smoothing kernel} to continuously map fluid quantities (provided by the representative particles) to the region of space around the particles \cite{beckett}. According to this method, the smoothing calculation is given by the equation \cite{mitch}
	\begin{equation}\label{eq:smooth}
	    \langle f(\rvec)\rangle = \int f(\rvec_j)W(|\rvec_i-\rvec_j|,h)dV(\rvec_j),
	\end{equation}
	where $f(\rvec_j)$ is the fluid quantity being mapped onto space (such as density) for a representative particle, $\langle f(\rvec)\rangle$ is its average value computed at location $\rvec$, $W$ is the smoothing kernel function, and $h$ is the smoothing length \cite{beckett}. Eq.~\ref{eq:smooth} can also be approximated as the following sum over particles \cite{mitch}
	\begin{equation}\label{eq:smooth sum}
	    \langle f(\rvec_i)\rangle=\sum_j V_jf_jW(|\rvec_i-\rvec_j|,h).
	\end{equation}
	A depiction of the smoothing process is shown in Fig.~\ref{fig:sph}. I should again note that I have only aimed to provide a conceptual overview of SPH as a method of modelling a continuous fluid medium using discretized particles (or ``parcels'')%converting from discretized particles (or ``parcels'') to a continuous fluid medium
	, and have only provided a brief look into the mathematical machinery underlying this process. I would like to state explicitly that understanding the math is not crucial %in the context of 
	for understanding this document, but the reader who is %personally 
	interested in learning more can refer to resources such as \cite{wadsley2003}, which provides a much more explicit description of the mathematical process.
	
	%my only intention in drawing the reader's attention to this ``dichotomy'' of simulation codes is because the difference in approaches is a subject that comes up often in the literature
	%This is an oversimplification, but most major codes fall into one of two categories.}
	
	%For the most part I intend my discussion to end 
	%either remain unaddressed or only
	
	%ChaNGa's implementation of SPH is comparable to that of Gasoline2, described in \cite{wadsley2017}

	% https://arxiv.org/abs/astro-ph/9710043
	
	%\begin{itemize}
	    %\item First, Newtonian gravity
	    
	    %\item Leapfrog integration
	    
	    %\item talk about scalability: $O(n^2)$ vs $O(n\log n)$ and the Octree solution
	    
	    %\item then new section: multipole expansion
	    
	    %\item new section: SPH
	    
	    %\item Should also talk about collisionlessness and boundary conditions.
	%\end{itemize}
	
	\section{The User's Perspective}\label{sec:usersperspective}
	
	So far this chapter has focused heavily on introducing N-body simulation methods at the conceptual level, and while I have personally found my research into the subject incredibly illuminating\footnote{I hope that I have succeeded in passing on some level of this illumination to the reader, if only a small fraction.}, 
	%Illumination which I hope I I have succeeded in passing onto the reader, if only by some small fraction.}
	%and while my sincere hope for this discussion is that it is able to provide the reader with a fraction of the illumination I have gained in researching it, 
	it bears acknowledging that the vast majority of us will actually never write such codes of our own (at least, not from scratch). While I don't think that this fact necessarily detracts from the value of learning about simulation techniques, it does invite discussion of N-body codes through a more practical lens. In this subsection, I will attempt to provide such a discussion by giving a brief overview of the \emph{user's} perspective when running a simulation in ChaNGa.
	
	On the most basic level, assuming that you have a computer with a Linux operating system, running a simulation in ChaNGa requires only three things: the code itself\footnote{Which, as of 2022 can be obtained in its latest release from https://github.com/N-BodyShop/changa/wiki}, a parameter file, and an initial conditions file. As previous sections in this chapter have sought to highlight, ChaNGa is an \emph{incredibly} complicated piece of software. %, and as such is supported by a number of dependencies.
	This complexity, as well as that of its dependencies, can --in my experience-- lead to considerable difficulties in the installation process. If you have gotten ChaNGa to successfully compile on your machine, be proud! It took me nearly half of my first research summer to accomplish exactly that. For those having trouble getting the software set up, I can only refer you to the github documentation \cite{changa-github} and offer my sympathy. 
	
	This chapter has hopefully provided a decent idea of what ChaNGa does, but to briefly summarize: it simulates the gravitational motion and gas dynamics of matter in space, producing results that aim to model galactic and/or cosmological structure. Assuming that you have the code downloaded and installed on your machine, you are well on your way to running your own astrophysical simulations! The next two things you need, as previously noted, are parameter and initial conditions files. The latter of the two is perhaps the easiest to intuitively understand: it is a binary file (i.e. a file that is not human readable) that provides the initial positions and velocities of all particles in the simulation \cite{mitch}, which as you may recall from Equations \ref{eq:kick drift kick 1}-\ref{eq:kick drift kick 3} are crucial for the numerical integration process. These initial conditions are not something that can be arbitrarily assigned, at least not if one is hoping to get accurate results, and must be computed in a manner that approximates observations of the early universe via the CMB (see Sec.~\ref{sec:cmb}) \cite{mitch}. In my own research, I have never been personally responsible for generating an initial conditions file, so I will not cover that process here. For discussion on one of the tools used for doing so, as well as the process itself, I would refer the reader to Chapter 1 of Mitchell Burdorf's 2021 thesis \cite{mitch}. On a practical level, what is most important to understand about a simulation's initial conditions file are that 1.) it provides the starting point for the computation by giving position and velocity values for all particles at $t=0$, and 2.) that these values are assigned so as to be approximately consistent with observation.
	
	\begin{figure}
	    \centering
	    \includegraphics[scale=0.4]{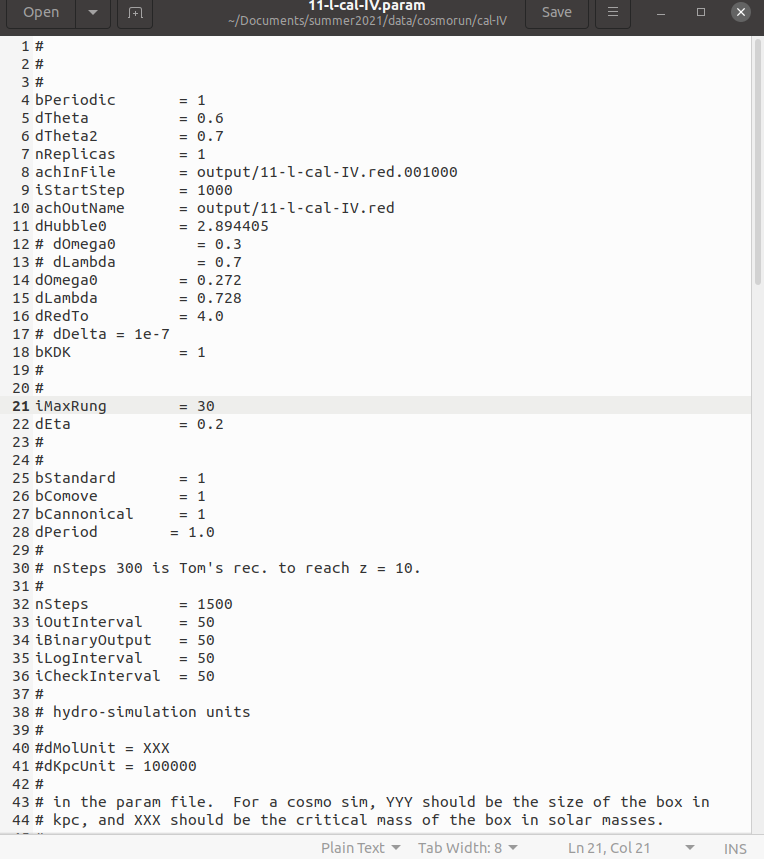}
	    \caption{The parameter file of a ChaNGa from AGORA's Cosmorun Paper (S. Roca-F\`{a}brega, et al. \textbf{917}, \href{https://arxiv.org/abs/2106.09738}{64 (2021),
arXiv:2106.09738 [astro-ph.GA] }), opened in a text editing program.}
	    \label{fig:parameter}
	\end{figure}
	
	The final piece of the puzzle is the parameter file, a human readable file with a \texttt{.param} extension that specifies a number of values (parameters) that are important for the computation. Fig.~\ref{fig:parameter} depicts a ChaNGa parameter file opened in a text editing program on a Linux machine. As you can see in the figure, all parameters (minus a few notable exceptions) are named such that their data type is specified by their first letter: \texttt{b} for boolean, \texttt{i} for int, and \texttt{d} for double. Some parameters worth noting are \texttt{dLambda} which represents the cosmological constant $\Lambda$ (for more information see Ch. 29 of \cite{bob}), \texttt{dHubble} which represents the Hubble Constant $H_0$ (again, refer to \cite{bob}), \texttt{nSteps} which gives the number of time steps to be performed before the simulation ends, and \texttt{iOutInterval} which gives the number of time steps between simulation outputs. Unlike the initial conditions file, since parameter files are human readable, values can be input by hand. However, it should be explicitly noted that, much like the initial conditions, parameters are chosen with physical realism in mind. 
	
	Once the user has a working version of ChaNGa, an initial conditions file, and a parameter file, performing the simulation is simply a matter of running the program, which itself involves a terminal command specifying the number of processor cores to use, as well as the location and name of the parameter file (note that the name of the initial conditions file is specified as one of the parameters).
	%a more practical 
	%the implementation of an N-body code at a conceptual level, and while %such discussions can be incredibly illuminating for those of us who already have experience working with such codes,
	%I have found my research into the topic incredibly illuminating, it bears noting that the vast majority of us will actually never write such codes of our own (at least, not from scratch). %N-body codes from scratch
    %This is especially true in the case of highly optimized codes such as ChaNGa
    %at the level of ChaNGa. 
	
	%\section{The Data}
	
	\chapter{Visualization and Analysis Tools}\label{ch:trident}
	
	This chapter primarily concerns tools and methods for data visualization and analysis. I should note that, while I will be introducing a number of the tools and techniques used to analyze data in this project, I will not be going into much detail on \emph{how} they are used (for instance, which commands perform which operations). My goal here is to present an introduction to the important aspects of data analysis rather than providing a guide for how to replicate my results. I would direct the reader who is interested in learning how to use these tools to their respective online documentation libraries \cite{yt-doc} \cite{trident-doc}, where you can find guides for performing different kinds of analysis at varying levels of complexity. These are the resources I used when I was getting started, and it is my opinion that the quality and comprehensiveness of their pedagogy eliminate the need for me to try to guide the reader through their use. 
	
	\newcommand{\yt}{\texttt{yt}\xspace}
	\section{Data Visualization - yt}
	
	The first analysis tool I will discuss, which serves as a foundation for the analytical work for this project (and in fact many other in computational astrophysics and similar fields), is a piece of software called \texttt{yt}. %As discussed in Sec.~\ref{sec:changa} \note{Don't forget to discuss this in Sec 2.2!!}, w
	When an N-body code such as ChaNGa performs a simulation, it outputs %simulation 
	data in the form of binary ``snapshot'' files containing information on all the quantities being tracked by the simulation at a given time (or redshift). Unsurprisingly, these files can contain a great deal of data (consider, for instance, the TNG simulations \cite{illustris}, some of the snapshots from which contain terabytes of information), but as they are encoded in binary (and are therefore not human readable), we need a way to extract the information in order to interpret it. This is where \yt comes in. \yt \cite{yt} is an open-source python package which allows the user to perform visualization and analysis on volumetric data sets (that is, data sets that represent a volume of three dimensional space), which is precisely the kind of data set output by simulation codes!
	
	The first thing that \yt does when loading a volumetric data set is create a specialized data object containing all the information it reads off of the binary file (I should note that from this point forward, when I use the term ``data object'' I am likely referring to precisely this python class) \cite{yt-doc}. This information is stored in what is known in the language of \yt as a ``field.'' A \yt field represents some spatially varying quantity related to the data being loaded, such as gas temperature or density \cite{yt-doc}. The package distinguishes between two types of field, native fields and derived fields. Native fields can be thought of as those inherent to the data set, they are quantities which the simulation specifically tracked and consecutively stored in the data output. These are fields that are described by the documentation as ``existing externally'' to \yt, often being stored on disk rather than in memory, and are accessible but immutable with \yt \cite{yt-doc}. Derived fields are those that \yt computes using other fields that are already present in the data object (most often, native fields). This is convenient for accessing quantities which a simulation doesn't directly track, but are nonetheless 
	relevant, or simply more reasonable to work with, %for the research being conducted. 
	in the context of the analysis being performed. There are a number of derived fields that \yt calculates automatically when loading a data set, but the package also enables the user to define their own derived fields to add to the \yt data object \cite{yt-doc}. While the creation of derived fields is obviously an integral part of the \yt package, for the purposes of this thesis there is no need to go into any greater detail. For the reader who is interested in creating their own derived fields and/or in learning more about how \yt loads and handles data, I would suggest having a look at the extensive online documentation that exists for the package \cite{yt-doc}. 
	
	% FIELDS: https://yt-project.org/doc/analyzing/fields.html#fields
	
	% DERIVED FIELDS: https://yt-project.org/doc/developing/creating_derived_fields.html?
	
	To briefly reiterate, \yt reads binary simulation data files and creates a specialized data object through which the user is provided access to a number of data fields corresponding to physical quantities tracked within the simulation \cite{yt-doc}. From here, it would reasonable to wonder what exactly one does with this data object and its constituent fields. After all, the process by which \yt reads and consecutively organizes data -- impressive though it may be -- would be for naught it it did not enable the user to the go on and perform meaningful analysis with said data. And herein lies the true power of \yt: it not only provides access to the contents of non-human-readable simulation files, it also provides a robust framework for extracting meaning from the raw information therein \cite{yt}. 
	
\begin{figure}
	    \centering
	    \includegraphics[scale=0.4]{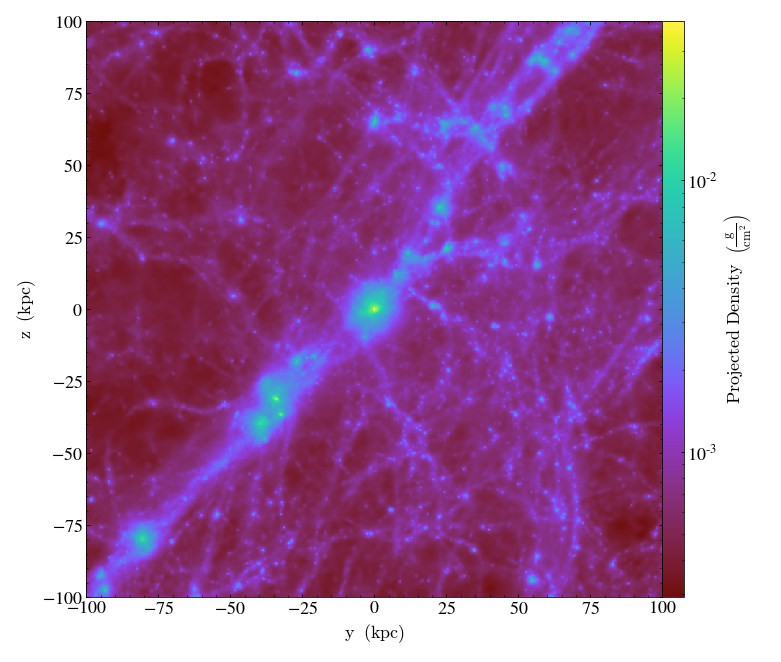}
	    \caption{A plot of gas density projected on the $y$-$z$ plane for an AGORA Cal IV data set.}
	    \label{fig:projection_plot_density}
	\end{figure}

	Among \yt's more notable data analysis features is the ability to filter data, either by field value (e.g. selecting all gas above a certain temperature threshold) or spatially using geometric objects such as boxes, spheres, and lines (e.g. selecting only the data within a spherical section of the total volume) \cite{yt-doc}. In the context of this project, \yt's feature of filtering data spatially was particularly useful, as the simulation snapshots I had covered distances many times larger than the radius of the galaxy that I was examining. Using a \yt sphere centered at the galaxy, I was effectively able to ``focus in'' on a smaller, more relevant volume, the scale of which also made the galaxy easier to make out visually (see Fig.~\ref{fig:projection_plot_density}). %Another \emph{crucial} feature of \yt is data visualization. 

	Mention of being able to make out features within simulation data \emph{visually} foreshadows another crucial ability granted by \yt, that being data visualization. There are a variety of means by which \yt allows the user to visualize data, %For instance, the user can perform a ``line query'' by specifying a start and end point for a \yt ray object, which then enables the user to perform analysis using only data from points along the ray (though this feature is not explicitly used in this project
	but those most relevant to this particular project are slice plots, projection plots, and phase plots\footnote{Readers who have spent some time within the field of astrophysics might also be interested to know that \yt can also produce profile plots, said readers can refer to the online documentation for more information \cite{yt-doc}.} \cite{yt-doc}. Creating a \yt slice plot involves taking either the full data set volume or a subvolume selected using one of the geometric spatial features discussed in the previous paragraph (e.g. a spherical subvolume centered at a galaxy), slicing that volume with some intersecting plane, and plotting some relation between field values for points within that plane \cite{yt-doc}. %\note{this is a complicated explanation, maybe I should include a picture of the process to give a visual aid. Also I should include a figure of an actual slice plot}. 
	Projection plots function similarly in that they begin with defining some plane that intersects the volume (or subvolume), but rather than plotting data only for points within that plane \yt projects all data within the volume onto the specified plane, either by integrating or summing over all field values along the lines perpendicular to said plane \cite{yt-doc}. %This projection is performed either by integrating or summing over values along lines perpendicular to the plane. %The result is a 2D representation of data within the full volume. 
	One thing I feel I should highlight, especially for readers overwhelmed with some of the more technical language in the previous few sentences, is that primary functional difference between projection and slice plots is what they represent spatially. Slice plots represent data within a two dimensional \emph{cross-section} of the volume, as though you took an incredibly thin slice out of space and had a look at what was going on just in that slice. Projection plots, on the other hand, give information about some field quantity throughout the \emph{whole} 3D volume. The process of projecting data quantities can be thought of as similar to ``flattening,'' as though you smushed down the data volume to an incredibly thin data-patty.
	
	\yt projection plots are particularly in that they provide a glimpse into what your data actually ``looks'' like. For instance, the gas density projection plot depicted in Fig.~\ref{fig:projection_plot_density} gives a good at-a-glance impression of how the matter is distributed throughout the data set volume. The resulting image is not dissimilar from what you would expect to see if the simulated volume was real, and you were looking at it through a telescope. 
	%, at the (generally acceptable) cost of ``losing'' the knowledge of how the matter
	Because of this, one of the first things I usually do when looking at a new data set in \yt is create a gas density projection plot to try to get a better idea of the kinds of astrophysical features I'm dealing with, and where they are. By extension, these plots are also incredibly useful when writing about or presenting on computational work, as it requires relatively little effort on the part of the audience to understand what a density projection plot is conveying to them\footnote{Plus, for the most part, they're pretty to look at!} \cite{agora3}. I should also note that because of their ease of interpretation, it is common practice to use density projection plots to demonstrate results, as exhibited by \cite{agora2} and \cite{agora3}. %\note{AGORA often uses density projection plots to show the galaxies they're simulating.}
	
	\begin{figure}
	    \centering
	    \includegraphics[scale=0.4]{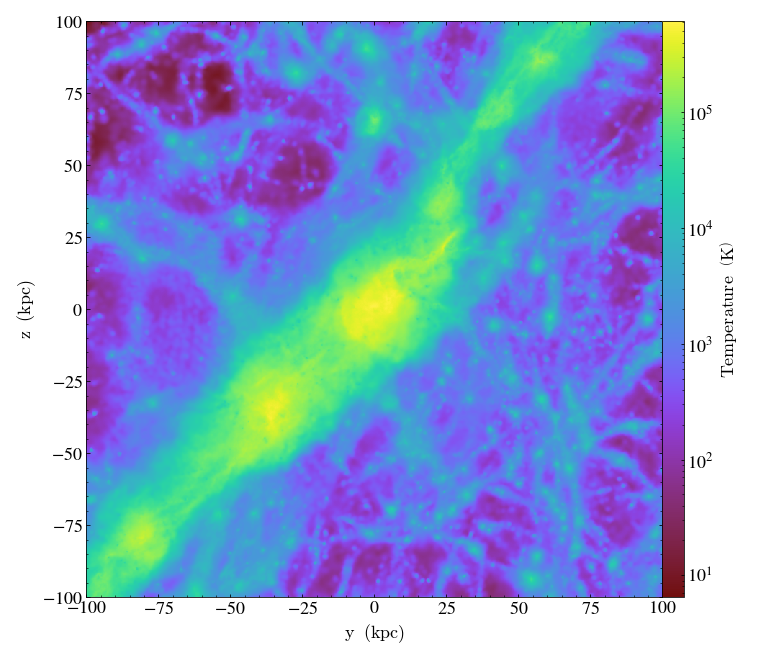}
	    \caption{A plot of gas temperature projected on the $y$-$z$ plane for the same data set and region depicted in Fig.~\ref{fig:projection_plot_density}.}
	    \label{fig:projection_plot_temp}
	\end{figure}
	
	Though gas density projection plots are an incredibly useful way of visualizing simulation data, and are among the most common one is likely to encounter in the literature, they are far from the only kind of projection plot that \yt is equipped to produce. On the contrary, \yt's projection plot function doesn't limit the user on which data fields they specify to be projected, so -- at least in theory -- one could create a \yt projection plot of just about any data field. As an example, Fig.~\ref{fig:projection_plot_temp} depicts a projection plot of the same region as the gas density plot in Fig.~\ref{fig:projection_plot_density}, but instead of density, the quantity being projected is gas temperature.
	
	%\note{Should I talk about phase plots too? They're an important feature of \yt, but I may not use any in the project.}
	
	%\note{Another thing to consider: \yt treats SPH datasets differently from AMR. Rather than directly loading an SPH dataset, \yt will convert the particle data into an Octree ``grid-like'' structure (trident paper talks about this in Sec. 2.1). This may be relevant to include in the \yt section but it would take some reading to fully understand.}
	
	It should be noted that \yt treats datasets from particle-based codes differently than it does for grid-based codes (see Sec.~\ref{sec:sph} for a discussion on these two approaches) \cite{yt}. This dichotomy is discussed further in Sec.~\ref{sec:trident}, but readers are also encouraged to look at Sec. 2.1 of \cite{trident} for more details on this difference.
	
	%some spatially varying quantity that has been tracked by the simulation and is consecutively stored in the data snapshot
	
	%Within the context of computational astrophysics, these data sets are simply the files produced from running an astrophysical simulation, containing such pertinent information as the positions and velocities of the particles in the simulation at a given time. %(though, depending on the complexity of the simulation, 
    %Originally designed for data sets from the AMR code Enzo, yt has since been adapted to handle data from a wide array of different codes, both AMR and SPH (a list of the different data sets yt is equipped to analyze can be found in the ``Loading and Examining Data'' section of yt's \href{https://yt-project.org/doc/}{online documentation}). My work with yt thus far has been confined to data sets from the SPH code ChaNGa \textbf{[Cite CHANGA]}, so the yt section of this document will focus on using yt to analyze Tipsy data sets (the data format used by ChaNGa).
	
	\newcommand{\trident}{\texttt{trident}\xspace}
	\section{Comparison with Observation - trident}\label{sec:trident}
	
	In the previous section, we discussed the use of \yt to perform visualization and analysis of the volumetric data in simulation output files. To these ends, \yt is invaluable due to the level of control it grants the user and its portability across multiple different simulation codes \cite{yt}. However, data from observational studies naturally do not take the same form as simulation output files. %That is to say, when we point our telescope at a particularly interesting region in space and begin to take measurements of some relevant quantity%\footnote{For instance, the relative intensity of incident radiation with respect to wavelength. \wink}
	%, we do not do so with the intention of determining the spatial position of each unit of gas within that region --we couldn't hope for such a level of three dimensional detail from our (cosmologically) fixed viewing position here on Earth. 
	Spectroscopic data, for instance, takes the form of light intensity and wavelength measurements, whereas simulation data takes the form of numerical quantities tracked over the simulated volume. 
	Since %the ultimate goal of computational work
	one of the primary goals of N-body simulations in astrophysics is to perform comparisons against observational results, researchers are left with the often daunting task of finding ways to make such comparisons in a meaningful way given their volumetric outputs.
	
	Luckily for us, in the case of comparing simulation results with those of spectroscopic studies, there is a tool called \trident that was created precisely %for facilitating such undertakings
	for the purpose reconciling these two types of data. \trident \cite{trident} is a python library that takes advantage of \yt's portability to produce synthetic spectroscopic absorption data (specifically, quasar absorption, see Sec.~\ref{sec:cgm-observation}) using output files from a rather impressive list of modern simulation codes \cite{trident}. Essentially, \trident allows the user to define two points within the volume of their 3D dataset (for convenience I will refer to these as the ``source'' point and ``viewing'' point), and produces an absorption spectrum that represents what an observer at the viewing point might expect to see if they pointed a spectroscope at a %bright object 
	quasar situated at the source point. %This is why we refer to the data that trident produces as ``synthetic'' or ``mock'' spectra: they are spectra produced using simulation data that 
	Note that we add the qualifier ``synthetic'' (or sometimes ``mock'') to the word ``spectrum'' to distinguish the data that \trident produces from real-world observational data. It is precisely \trident's capacity to produce such spectra that makes it such a powerful analytic tool, as it allows %for more direct comparison against observational without necessitating
	users to perform more direct comparisons against observation without making them perform any of the (rather painful) calculation and data manipulation that goes into producing synthetic spectra \cite{trident}.
	
	From the perspective of the user, creating a synthetic spectrum with \trident is simply a matter of specifying which ion species to consider (e.g. Mg, in which case absorptions from Magnesium and all of its ions will be included in the spectrum), then defining the start (source) and end (viewing) points of the sightline, with additional (but uncomplicated\footnote{At least, from the user's perspective.}) steps such as adding simulated noise to get the spectrum to more closely resemble observational data \cite{trident-doc}. \trident also allows the user to specify real-world observational equipment for outputs to ``mimic'' in terms of features such as wavelength range \cite{trident-doc}. For instance, users can specify that they want the synthetic spectrum to look like one that might be obtained from the G130M grating of Hubble's Cosmic Origins Spectrograph (COS) \cite{trident-doc}, in which case the output spectrum would exhibit features characteristic of that grating such as its 1150-1149 $\si{\angstrom}$ wavelength range \cite{cos}. Additionally, users can specify their own \emph{custom} wavelength ranges that don't necessarily correspond to a particular piece of instrumentation \cite{trident-doc}. This was particularly useful in my case, as the redshift of the simulation dataset that I used was such that absorption features would not be detectable in the wavelength ranges of the COS's far ultraviolet modes \cite{cos}. Instead, I was able to determine appropriate wavelength ranges for the dataset I was using through the equation \cite{trident-doc}
	\begin{equation}\label{eq:wavelength shift}
	    \lambda_\text{obs}=(1+z)\lambda_\text{rest},
	\end{equation}
	which gives the absorption wavelength observed at redshift zero ($\lambda_\text{obs}$) for an absorption feature with rest wavelength $\lambda_\text{rest}$ produced by a material at redshift $z$.
	%a material at redshift $z$ 
	%under the hood of this process is 
	
	While \trident is a remarkably simple tool from the user's perspective, it should come as no surprise that %under the hood \trident 
	it is doing a significant amount of non-trivial computation under the hood in order to produce its synthetic spectra. Though an in-depth discussion of \trident's code method is beyond the scope of this project, there are certain aspects of the process that feel important to highlight for the sake of providing a sufficiently robust qualitative understanding of how \trident operates. For readers interested in some of the finer details about how \trident operates, I would suggest having a look at the method paper \cite{trident}.
	
	%\note{you should mention somewhere in here that trident lets the user customize which instrument (if any) to mimic, which wavelengths to consider, etc.}
	
	\subsection{Accounting for missing ions}
	
	I have thus far introduced \trident's purpose: to generate synthetic absorption spectra with volumetric simulation data, but in order to properly understand how \trident does this, and whether we can trust its results, I must address a particularly significant challenge -- one born primarily out of the pragmatic realities of modern computation -- that \trident must overcome to achieve this end. As discussed in Sec.~\ref{sec:spectroscopy}, absorption spectra are the result of measuring the intensity of light after it passes through intervening material between source and observer. %In particular,
	Recall that certain sets of absorption features (intensity ``dips'') are associated with specific atomic species and ion stages, with each species and stage having its own characteristic absorption wavelengths\footnote{Of course, this wavelength is subject to change depending on the redshift between the observer and the intervening material \cite{tumlinson} \cite{trident-doc}.}. Thus, one crucial aspect of spectroscopy is identifying the elements associated with the absorption features present in a measurement \cite{bob}. The problem that emerges from generating synthetic spectra using simulation data is that there isn't perfect correspondence between what is present in a simulation and what exists in the physical universe. I am specifically talking about the presence of chemical elements in the simulation.
	%In the context of \trident, limitations on which chemical elements are \emph{present} within the simulation 
	%Specifically what I mean here 
	%Naturally, the matter in our physical universe encompasses all elements on the periodic table, as well as all of their ions \note{FUCKING fact check this bro, probably revise regardless}. So, for instance, if I wanted to look for iron (the 26th element on the periodic table) out in the universe, all I would need would be to know the absorption wavelength of the atom (which can be experimentally determined), and then point my spectroscope at various bright objects in space until I found a strong absorption feature at that wavelength. All that is to say, in a long-winded way, that there \emph{is} iron in our universe. 
	%This is not necessarily true of simulation data. 
	While most simulations will explicitly track lighter elements such as hydrogen and helium, it is often infeasible from a practical standpoint to keep track of each element we would like to consider for the purposes of analysis\footnote{This infeasibility arises from the limitations of computer technology \cite{trident}. In the context of scientific simulations, researchers must constantly navigate a delicate balance between producing results that are maximally realistic and writing simulations that will run to completion on a reasonable timescale.} (let alone all elements in the universe) \cite{trident}. So while knowing the abundance of %each element and its
	metals and their ions throughout the volume of a simulation dataset is \emph{crucial} for producing high-fidelity synthetic spectra, it is simply not %possible on the simulation side of things to keep track of all that data. 
	reasonable for a simulation to track that level of information and the ionization physics necessary to produce it while also performing the gravitational and hydrodynamic calculations characteristic of modern N-body codes (see Ch.~\ref{ch:changa}) \cite{trident}.
	%The solution to this dilemma is that \trident must extrapolate, from whatever data \emph{is} available regarding ion fields, \note{the presence of the ions that will ultimately contribute the the spectra being produced}
	
	\trident's solution to this dilemma is to extrapolate the presence of %the rest of 
	any missing ions desired for generating spectra from whatever ion data fields that \emph{have} been tracked by the simulation. % that you want to consider when generating your synthetic spectra.
	To do so, \trident makes the assumption that the number density of the $i$th ion of an element $X$ is given by \cite{trident}
	\begin{equation}\label{eq:trident1}
	    n_{X_i}=n_Xf_{X_i},
	\end{equation}
	where $f_{X_i}$ is the ionization fraction associated with that particular ion. It should be noted that the ionization fraction is \emph{not} a constant, and operating under the assumption of ionization equilibrium (as \trident does), its value is dependent on temperature, density, and the nature of the radiation field responsible for ionization \cite{trident}. %As such, the ionization fraction is a value which itself must be computed. Derivation of ionization fractions is a rather complicated task itself, often requiring its own 
	This means that it, too, is a value that must be computed, which unfortunately is not a trivial task. Often, simulations modelling ionization are needed in order to get a sense of appropriate values for the ionization fraction\footnote{Note that such simulations are entirely \emph{separate} from the N-body simulations that give us our data files. See, for instance, \cite{cloudy}.} \cite{trident}. Naturally, %users don't these taxing computer simulations every time they try to generate a synthetic spectrum, so the developers of \trident 
	%a computation that neither users or developers would want \trident to perform on runtime
	\trident's usefulness would be drastically limited  %wouldn't be particularly useful 
	if it forced users to run additional computationally-intensive ionization simulations every time they attempted to create a synthetic spectrum, so \trident comes pre-packaged with all the information it needs to determine the ionization fraction for the ions of all elements up to Zn (the 30th element on the periodic table) \cite{trident}. In more technical terms, \trident is equipped with a set of data tables that it uses to look up pre-computed ionization fraction values given various parameters such as temperature and redshift\footnote{What \trident is really doing is using a linear interoplation of data in the lookup table to produce estimates for ``intermediate'' values (i.e. values not in the table).}. The data in these tables are the result of simulations performed using the CLOUDY \cite{cloudy}, a code used for simulating photoionization. 
	
	To recap, when \trident encounters a dataset that contains the field for a required atom but is missing fields for one or more of its ionization states, it uses Eq.~\ref{eq:trident1} to estimate the presence of the missing ions throughout the dataset volume, then adds a new field containing that information, effectively ``populating'' the volume (or a specific section of the volume) with the missing ions. However, it is feasible (and, in fact, not unlikely) that the dataset in question %does not contain any fields relating to a particular desired element. 
	comes from a simulation that didn't track the presence of a particular atom. %\emph{any} ions of a particular element. 
	In such cases, 
	%did not track the number density for 
	$n_X$ is missing from the dataset, so \trident makes the estimate \cite{trident}
	\begin{equation}\label{eq:trident2}
	    n_X=n_HZ\left(\frac{n_X}{n_H}\right)_\odot, 
	\end{equation}
	where $n_H$ is the number density of hydrogen, $Z$ is metallicity, and $(n_X/n_H)_\odot$ is the solar abundance of the atom in question. 
	
	Using Equations \ref{eq:trident1} and \ref{eq:trident2}, \trident %is able to produce synthetic spectra with data from any simulation so long as it tracks hydrogen. 
	is able to estimate the spatial number density of all elements up to atomic number 30 so long as the dataset being used has a field corresponding to hydrogen. \trident also has methods for producing spectra in the case that a dataset from a simulation that did not track the presence of \emph{any} element (i.e. not even hydrogen) \cite{trident}, but %using a simulation with no elemental data is ill advised for anyone hoping to use \trident for meaningful analysis, and the vast majority of studies within the field involve simulations that track one or more elements.
	simulations containing no elemental data are not discussed in this thesis (and, in fact, are not particularly common in the professional research scene), so the %steps
	process that \trident %takes 
	undergoes to estimate the presence of hydrogen when no such data is available is beyond the scope of this document (though the interested reader is encouraged to have a look at the method paper \cite{trident}). 
	
    \subsection{Adding missing ions}
    
    %Once \trident has produced 
    
    I mentioned in the previous section that, once \trident has produced its estimates for the density of each relevant ion, it ``populates'' the dataset with that ion by adding a corresponding \yt field. What I have glossed over in this description is the rather nuanced process by which \trident goes about accomplishing this task. In general, there are two types of datasets that \trident interacts with: those originating from grid-based simulation codes (AMR) and those originating from particle-based codes (SPH) (see Sec.~\ref{sec:sph}). %\footnote{This is a lie. There are other types of codes, but at the time of writing computational astrophysics research is dominated by AMR and SPH, and so --for the purposes of this document-- it is acceptable to ignore the others.}. \note{HERE BE DRAGONS: now you have to talk about AMR vs SPH in a prior chapter.}  %Since these two types of simulation code represent matter differently, their corresponding datasets also store information differently. 
    As discussed in Ch.~\ref{ch:changa}, these two types of code represent matter differently (as particles or as a density grid, as their names imply). The way that their output data files store information reflects this dichotomy. Since \trident works with both types of data files, it must somehow resolve this discrepancy, which, as it turns out, is done by piggybacking off of \yt \cite{trident}. \yt was originally designed as a visualization tool specifically for the AMR code ENZO (\cite{enzo}) \cite{yt}, and though its current version is incredibly versatile in terms of the kinds of codes it can do visualization for, it still favors a grid-based paradigm in handling data \cite{trident}. As such, when loading a dataset from an SPH simulation, \yt actually converts the particle data into a variable-resolution grid structure\footnote{Variable-resolution meaning there are more grid boxes in regions with higher particle density \cite{trident}. %The data structure in question is referred to as an ``Octree,'' for the interested reader.}
    } using an Octree structure similar to that discussed in Sec.~\ref{sec:barnes-hut}. 
    
    Unfortunately, \trident's \texttt{ion\_balance} module (the element of the code responsible for generating and adding missing ion fields) has to undergo a more computationally intensive process for SPH datasets than for AMR datasets \cite{trident}. In the latter case, since the data were always in grid-format, \trident can only consider grid cells along the path of the quasar sightline, meaning that \texttt{ion\_balance} doesn't have to consider the \emph{entire} simulation volume, which saves a significant amount of computation time \cite{trident}. In the case of SPH, however, only considering the presence of ions along the line of sight can lead to significant errors \cite{trident}, meaning that \trident must first create missing ion fields for the entire SPH volume before depositing them into their appropriate grid cells in the \yt Octree conversion \cite{trident}. This process adds to computation time, but not unreasonably so, and the avoidance of additional errors is certainly worth the cost. 
    
    \subsection{Synthetic Spectra}\label{sec:synthetic-spectra}
    
\begin{figure}
        \centering
        \includegraphics[scale=0.5]{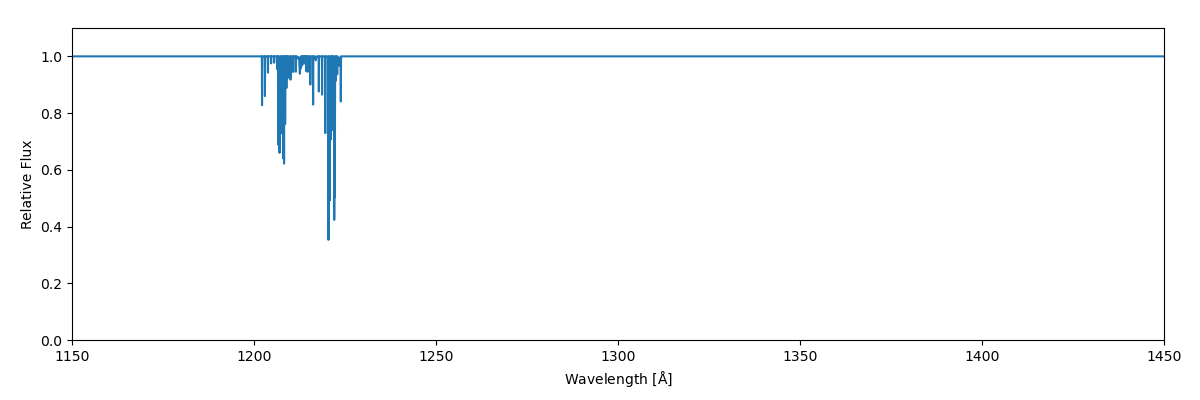}
        \caption[A ``raw'' spectrum produced using an Enzo dataset according to the annotated example on \trident's documentation website: \linebreak\href{https://trident.readthedocs.io/en/latest/}{https://trident.readthedocs.io/en/latest/}. The ion fields considered are H, C, N, O, and Mg. Absorption features can be seen in the 1150 to 1250 Angstrom range.]{A ``raw'' spectrum produced using an Enzo dataset according to the annotated example on \trident's documentation website: \href{https://trident.readthedocs.io/en/latest/}{https://trident.readthedocs.io/en/latest/}. The ion fields considered are H, C, N, O, and Mg. Absorption features can be seen in the 1150 to 1250 Angstrom range.}
        \label{fig:example spec raw}
    \end{figure}

    Once the dataset has been populated with the appropriate ion fields by \texttt{ion\_balance}, \trident traverses the user specified sightline from the start point to end point, depositing absorption profiles onto a ``raw'' spectrum (Fig.~\ref{fig:example spec raw}) according to the elements encountered along the way and their physical properties (temperature, etc.) \cite{trident}. The precise method by which this is accomplished is beyond the scope of this document, but involves a process in spectroscopy known as a Voigt profile calculation. Readers can learn more about the absorption depositing process in Sections 2.4 and 2.5 of the \trident method paper \cite{trident}.
    
\begin{figure}
        \centering
        \includegraphics[scale=0.5]{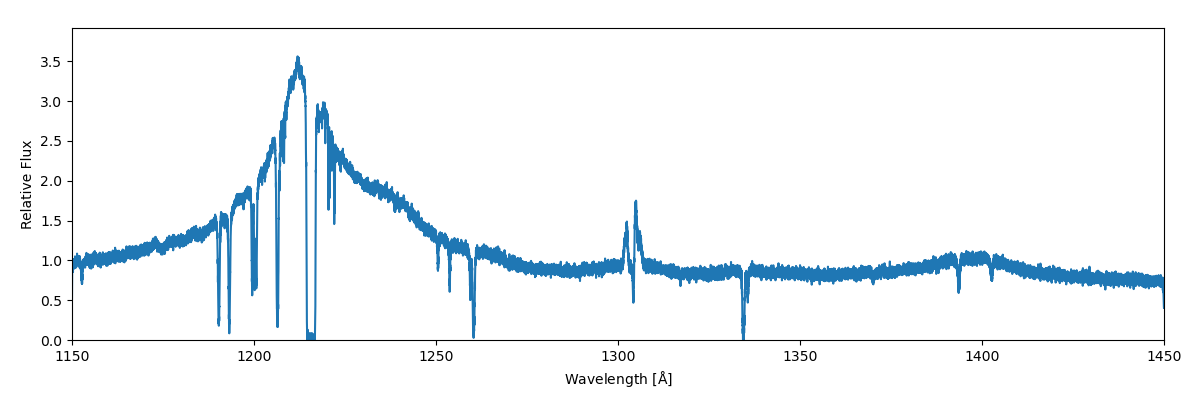}
        \caption[The same spectrum depicted in Fig.~\ref{fig:example spec raw}, with the addition of a quasar background and Milky Way foreground. Source: \linebreak\href{https://trident.readthedocs.io/en/latest/}{https://trident.readthedocs.io/en/latest/}.]{The same spectrum depicted in Fig.~\ref{fig:example spec raw}, with the addition of a quasar background and Milky Way foreground. Source: \href{https://trident.readthedocs.io/en/latest/}{https://trident.readthedocs.io/en/latest/}.}
        \label{fig:example spec}
    \end{figure}

    Once the absorption features have been deposited onto the raw spectrum, users are able to add additional features that would be present in observationally obtained spectra, such as a quasar source spectrum and Milky Way foreground \cite{trident-doc}. The result of this process in the context of an Enzo dataset is depicted in Fig.~\ref{fig:example spec}.
    
\begin{figure}
        \centering
        \includegraphics[scale=0.6]{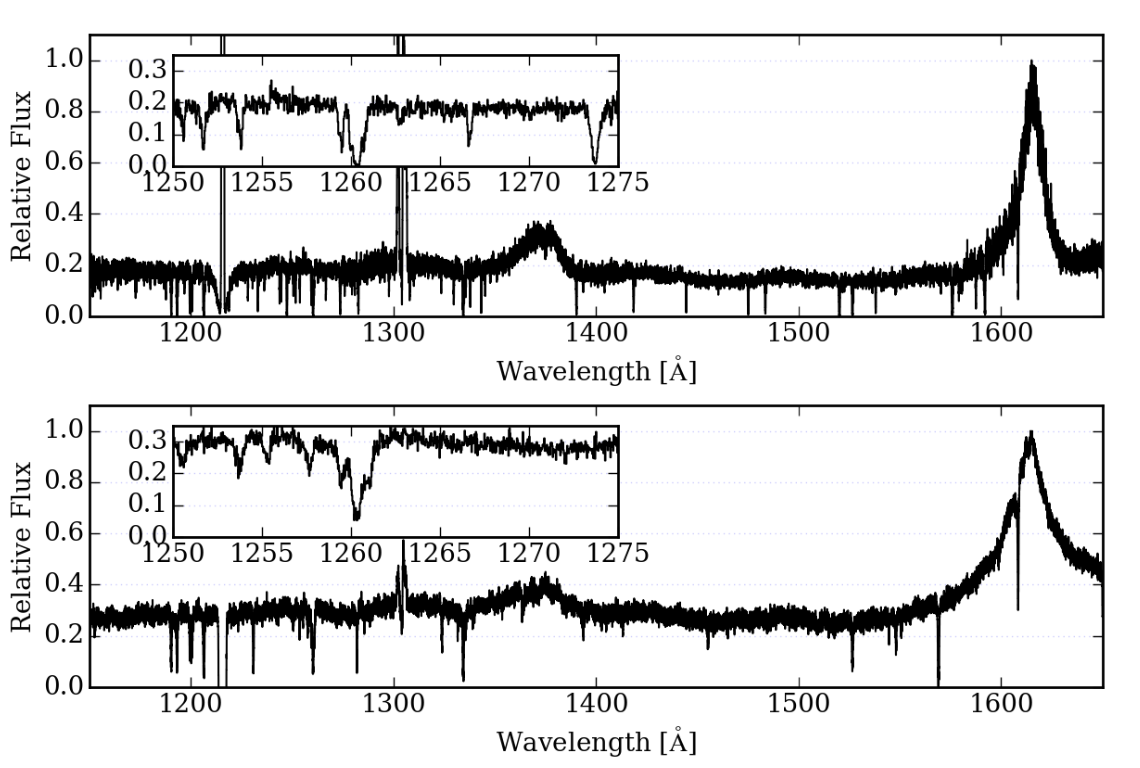}
        \caption{A comparison between an observed quasar absorption (top) spectrum and one generated using trident (bottom) at similar redshifts ($z\sim0.3295$). The observed spectrum was obtained using the Cosmic Origins Spectrogaph on board the Hubble space telescope, and was made publically available by (C. W. Danforth, et al. \textbf{817}, 111 (2016), \href{https://arxiv.org/abs/1402.2655}{arXiv:1402.2655 [astro-ph.CO] }). The synthetic spectrum was produced in \trident using data from an Enzo simulation (B. D. Smith, E. J. Hallman, J. M. Shull, and B. W. O’Shea, \textbf{731}, 6 (2011), \href{https://arxiv.org/abs/1009.0261}{arXiv:1009.0261 [astro-ph.CO] }). This figure was borrowed from the \trident method paper (C. B. Hummels, B. D. Smith, and D. W. Silvia, \textbf{847}, 59 (2017), \href{https://arxiv.org/abs/1612.03935}{arXiv:1612.03935 [astro-ph.IM]}).}
        \label{fig:spectra comparison}
    \end{figure}

    Fig.~\ref{fig:spectra comparison} depicts a comparison between an observationally obtained spectrum \cite{danforth} and a synthetic spectrum produced in \trident \cite{trident} using an Enzo dataset \cite{whim}. While it is interesting to note the qualitative similarities between these two graphs, it is arguably more important to note that both exhibit %incredibly similar absorption profiles, with absorption features occurring 
    crucial similarities with respect to major absorption features \cite{trident}, such as those occurring %for both 
    just before the 1255 Angstrom point, as well as around 1260 Angstroms. These major features can be seen distinctly in the upper left sub-graphs of the two plots, which show the spectra zoomed-in to the wavelength range 1250-1275 Angstroms. This figure provides a crucial demonstration of \trident's viability as a tool for facilitating comparison between results from computational and observational studies \cite{trident}.
    
	\chapter{Results}\label{ch:results}
	
	%Will probably need to start this chapter (or write a separate chapter/section) going into more detail about the data that I am using (AGORA Cal IV). AGORA paper probably good as a source for some details about this. 
	
	The results of this thesis were primarily demonstrative of \trident's viability as a tool for future work at Reed. Using an output file from a ChaNGa simulation performed as part of AGORA's third code comparison paper \cite{agora3} at redshift $z\sim4$, I was able to produce a series of synthetic quasar absorption spectra with \trident. There are a number of problems with these spectra that will need to be addressed in future work before meaningful comparison against observational data can be performed. These issues are discussed in this section, and are also summarized in Ch.~\ref{ch:conclusion}.
	
	\begin{figure}
	    \centering
	    \includegraphics[scale=0.5]{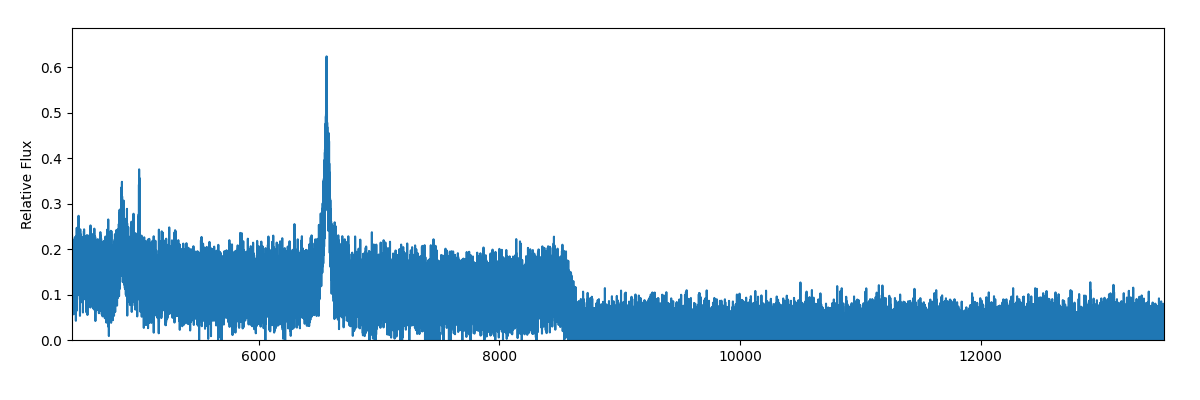}
	    \caption{An initial synthetic quasar spectrum generated using a ChaNGa data snapshot from (S. Roca-F\`{a}brega, et al. \textbf{917}, \href{https://arxiv.org/abs/2106.09738}{64 (2021),
arXiv:2106.09738 [astro-ph.GA]}) with a sightline spanning the body diagonal of the cubic simulation volume. The horizontal axis represents wavelength in Angstroms.}
	    \label{fig:spec_prelim}
	\end{figure}
	
	Fig.~\ref{fig:spec_prelim} depicts the first quasar absorption spectrum that I was able to obtain with \trident using the AGORA data \cite{agora3}. This being an incredibly early test of \trident's compatibility with the dataset, I rather arbitrarily decided to define my start and end points such that the sightline would span the length of the simulation volume's body diagonal, cutting directly through the galaxy\footnote{I say ``arbitrarily,'' but my real motivation was that in my initial attempts at generating spectra, \trident gave me error messages informing me that the sightlines I had chosen didn't have any intersecting ion fields. I chose the easiest calculable line passing through the galaxy in an attempt to ensure that there \emph{would} be intersecting ion fields, which was ultimately successful, as demonstrated by Fig.~\ref{fig:spec_prelim}.} (located at the origin). The spectrum was specifically produced with H, C, N, O, and Mg as the specified contributing ion species. I selected the wavelength range of ``detection'' so that the spectrum would include all possible absorption features associated with the ion species present (see Appendix \ref{appendix:specrange} for the code I used to determine this wavelength range). This spectrum represents my first successful attempt at using \trident on a ChaNGa dataset, and essentially serves as a ``proof-of-concept'' that %simulations originating from ChaNGa can be used in tandem with \trident 
	students at Reed can use simulation data from ChaNGa in tandem with \trident for the purposes of analysis. That being said, there are a number of obvious problems with this spectrum. First, the wavelength range I chose does not correspond to %the observational instruments that are being used in contemporary studies of the CGM
	either the far ultraviolet mode (diffraction gratings in the range of 1150-1230 \si{\angstrom}) or the near ultraviolet mode (diffraction gratings in the range of 1700-3200 \si{\angstrom}) of COS \cite{cos} \cite{cos-nasa}, which is one of the primary pieces of observational equipment used in contemporary surveys of the CGM \cite{tumlinson} and so far the only piece of instrumentation whose measurements \trident is equipped to mimic \cite{trident-doc}. This problem effectively arises from the fact that the lowest redshift data I had available to me at the time of writing was at $z\sim4$, whereas CGM surveys via COS (such as COS-Halos \cite{halos1} \cite{halos2}) generally probe low-redshift galaxies at $z<1$ \cite{halos1} \cite{tumlinson}. In essence, in order to get results that are even comparable to those that might be obtained using COS, absorption features need to be redshifted into the UV range which would require lower redshift simulation data. 
	
	Another issue with the result depicted in Fig.~\ref{fig:spec_prelim} is the %relatively 
	low flux to noise ratio it exhibits. This issue is so severe that, at the graph's current scale%\footnote{Though I argue that it is not merely an issue of scale, I will cover this subject further in my discussion of of my later results.}
	, one cannot make out any distinct absorption features\footnote{At least not visually, though from the appearance of the graph I would be surprised if any distinct absorption features could be found algorithmically.}. %Redshift is one of the primary differences between 
	
\begin{figure}
	    \centering
	    \includegraphics[scale=0.5]{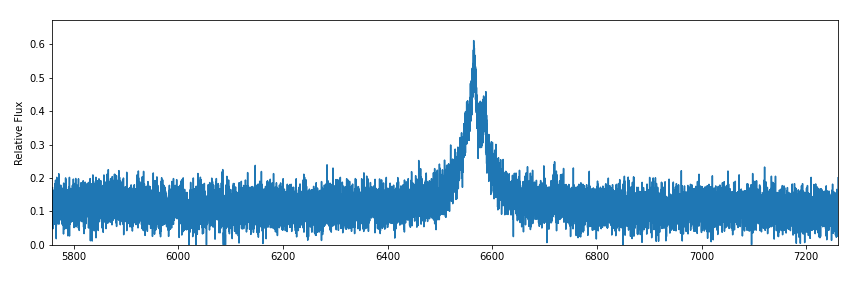}
	    \caption{A spectrum resulting from a quasar sightline with a length of 600 kpc and an impact parameter of 30 kpc from the center of the galaxy. The horizontal axis represents wavelength in Angstroms. Only ion species of H, C, N, O, and Mg are considered during spectrum generation. Data obtained from the ChaNGa simulation published in (S. Roca-F\`{a}brega, et al. \textbf{917}, \href{https://arxiv.org/abs/2106.09738}{64 (2021),
arXiv:2106.09738 [astro-ph.GA]}).}
	    \label{fig:spec30}
	\end{figure}

\begin{figure}
	    \centering
	    \includegraphics[scale=0.5]{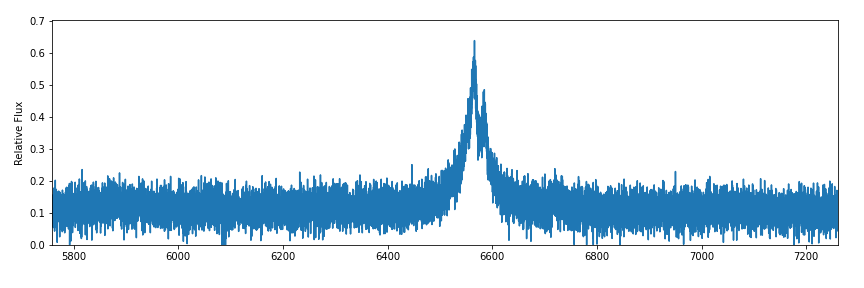}
	    \caption{A spectrum resulting from a quasar sightline with a length of 600 kpc and an impact parameter of 50 kpc from the center of the galaxy. The horizontal axis represents wavelength in Angstroms. Only ion species of H, C, N, O, and Mg are considered during spectrum generation. Data obtained from the ChaNGa simulation published in (S. Roca-F\`{a}brega, et al. \textbf{917}, \href{https://arxiv.org/abs/2106.09738}{64 (2021),
arXiv:2106.09738 [astro-ph.GA]}).}
	    \label{fig:spec50}
	\end{figure}

\begin{figure}
        \centering
        \includegraphics[scale=0.5]{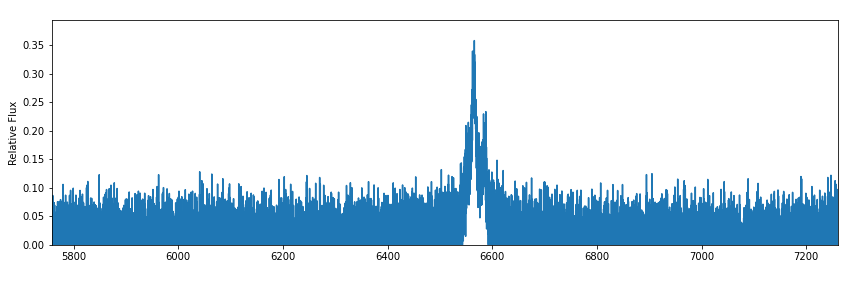}
        \caption{A spectrum resulting from a quasar sightline with a length of 600 kpc and an impact parameter of 100 kpc from the center of the galaxy. The horizontal axis represents wavelength in Angstroms. Only ion species of H, C, N, O, and Mg are considered during spectrum generation. Data obtained from the ChaNGa simulation published in (S. Roca-F\`{a}brega, et al. \textbf{917}, \href{https://arxiv.org/abs/2106.09738}{64 (2021),
arXiv:2106.09738 [astro-ph.GA]}).}
        \label{fig:spec100}
    \end{figure}

\begin{figure}
        \centering
        \includegraphics[scale=0.5]{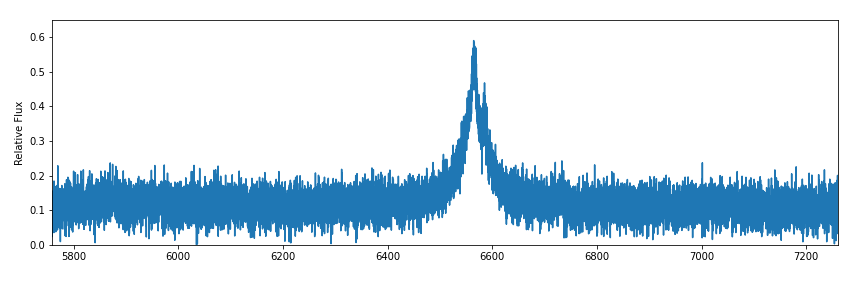}
        \caption{A spectrum resulting from a quasar sightline with a length of 600 kpc and an impact parameter of 500 kpc from the center of the galaxy. The horizontal axis represents wavelength in Angstroms. Only ion species of H, C, N, O, and Mg are considered during spectrum generation. Data obtained from the ChaNGa simulation published in (S. Roca-F\`{a}brega, et al. \textbf{917}, \href{https://arxiv.org/abs/2106.09738}{64 (2021),
arXiv:2106.09738 [astro-ph.GA]}).}
        \label{fig:spec500}
    \end{figure}

	%The flux to noise ratio is an issue that appears 
	In an attempt to address the issues with my initial graph, I was more judicious in choosing both the sightlines and the wavelength ranges for my later results. Figures \ref{fig:spec30}, \ref{fig:spec50}, \ref{fig:spec100}, and \ref{fig:spec500} show synthetic spectra obtained from \trident using the same ChaNGa dataset. The sightlines that produced these spectra had impact parameters of 30 kpc, 50 kpc, 100 kpc, and 500 kpc respectively, and all had length 600 kpc. The parameters of the first two were selected to be relatively consistent with Iryna Butsky's work using \trident with ChaNGa datasets \cite{butsky}, while the latter two were chosen simply to see whether greater impact parameters would have a meaningful affect on the noisiness of the resulting spectrum. In all four cases, the wavelength range was chosen by applying Eq.~\ref{eq:wavelength shift} to the maximum and minimum ``measured'' wavelengths in Fig.~\ref{fig:example spec} in the hopes of getting a narrower range that would also include a number of absorption features. As with my preliminary graph (Fig.~\ref{fig:spec_prelim}), all four of these graphs exhibit high levels of noise with relatively low flux, and no absorption features can be discerned from visual inspection.
	
	It is my belief that the issue presented by the flux to noise ratios in my findings emerges as a result of the high redshift of the simulated galaxy, since redshift is one of the primary differences between the dataset used in this project and the data used to produce the graph in Fig.~\ref{fig:example spec}, as well as the primary difference between the former and other studies using \trident such as \cite{strawn}. %and that used for other studies using \trident such as \cite{strawn}
	An alternative explanation could be that the issue somehow arises from differences between codes. The graph depicted in Fig.~\ref{fig:example spec} comes from a simulation with the code Enzo \cite{enzo} and \cite{strawn} uses ART \cite{art}, both of which are grid-based (AMR) codes, whereas ChaNGa is an SPH code. 
	%is derived from data produced using the simulation code, and \cite{strawn} 
	%results reported in \cite{strawn} both originate from grid-based (AMR) codes, whereas the data I am using are from ChaNGa (an SPH code, see Sec.~\ref{sec:sph}).
	%The graph depicted in Fig.~\ref{fig:example spec} and results reported in \cite{strawn} both originate from grid-based (AMR) codes, whereas the data I am using are from ChaNGa (an SPH code, see Sec.~\ref{sec:sph}). 
	Thus, the treatment of fluid dynamics constitutes another major difference %beyond dataset redshift
	between the three datasets. However, there are a number of reasons why I suspect this is not the explanation for the issues with my results. %to not be the case. 
	First, the data I am using was produced by AGORA for a paper comparing results from a number of simulation codes at either side of the SPH-AMR dichotomy %(including the codes used for the Fig.~\ref{sec:sph}) data and for \cite{strawn}) 
	(including the code used for Fig.~\ref{sec:sph}, that used in \cite{strawn}, \emph{and} ChaNGa) \cite{agora3}. This paper \emph{did} report differences among the codes with respect to the CGM properties they produced, %but the paper does not
	but these differences %largely pertain 
	were largely relegated to CGM kinematics, which would not seem to justify the extreme differences between my results and those shown in Fig.~\ref{fig:example spec}. Additionally, analysis of the CGM in a ChaNGa simulation via \trident is something that has been done before: it is the subject of Ch. 4 of \cite{butsky}. In the case of \cite{butsky}, however, the simulation file used was at $z<0.25$ \cite{butsky}. In fact, low redshift appears to be a common characteristic among most CGM analyses using \trident. The Enzo simulation that produced Fig.~\ref{fig:example spec} had a redshift of $z\approx0.006$, and the data used by \cite{strawn} was at $z\sim1$. Because of this relative consistency with respect to redshift, it is my belief that future inquiries into the simulated CGM at Reed should use lower redshift data. If this does not eliminate the aforementioned issues with my results, it will at least rule out high redshift as a culprit. 
	%, but in this case the data used was at $z<0.25$ \cite{butsky}.
	
	%, %as well as its primary difference from 
	%and with data used in other studies using \trident such as \cite{strawn}. Another major difference between the dataset 
	
	%Comparing my result in Fig.~\ref{fig:spec_prelim}
	
	%I determined this wavelength range using a python helper function which I called \texttt{findShiftedLambda()} 
	%The code that I used to determine the desired wavelength range for a given set of elements with a particular dataset is provided in Appendix \ref{appendix:specrange}.

	\chapter{Conclusion and Future Work}\label{ch:conclusion}
	
	This document has provided an overview of the missing baryon problem, its relation to the CGM, and the role that N-body simulations can play in furthering our understanding of the mystery (Ch.~\ref{ch:cgm}). Additionally, I have provided a conceptual outline of ChaNGa \cite{menon}, a modern SPH N-body code (Ch.~\ref{ch:changa}), as well as discussions regarding analysis and visualization of simulation data with \yt \cite{yt} and \trident \cite{trident} (Ch.~\ref{ch:trident}). As presented in Ch.~\ref{ch:results}, my results from generating synthetic absorption spectra with \trident using ChaNGa data are less than satisfactory, particularly in terms of noisiness and low relative flux. Since contemporary research into the CGM appears to focus largely on low-redshift galaxies commonly with $z\lesssim1$ in both observational \cite{halos1} \cite{halos2} \cite{tumlinson} and computational \cite{strawn} \cite{butsky} contexts, it is very possible that future work related to this project could benefit from using lower-redshift data. This could also lead to ease of comparison with observational data, since COS, a major tool in modern spectroscopic studies of the CGM \cite{tumlinson}, is best suited for observing the CGM of low-redshift galaxies \cite{cos}. Ultimately, despite their issues, my results do provide good justification for further inquiries into the CGM at Reed, since they demonstrate that \trident \emph{can} be used in tandem with ChaNGa to produce synthetic quasar absorption spectra.
	
	Possible extensions to the project largely center around applying the principles and tools I have presented in order to perform CGM analysis. As previously noted, future work would likely benefit from lower-redshift data. Fortunately, the AGORA simulation from which I acquired my data is planned to run down to a redshift of $z\sim2$ for work related to future a publication \cite{agora3}, and it is possible that a Reed student could request access to this data for use in a thesis project. While AGORA's target redshift for the upcoming paper is still larger than the $z\lesssim1$ that is commonly encountered in the literature, it does come much \emph{closer} that the $z\sim4$ data that I used. It would be interesting to see whether %better results can be obtained 
	results obtained from $z\sim2$ simulation data would %yield more satisfactory
	prove more satisfactory than those presented here. 
	
	Alternatively, future students could combine my work with that of Mitchell Burdorf \cite{mitch}, who wrote a thesis in 2021 in which he presents results from a ChaNGa cosmological simulation. Burdorf suggests performing a Zoom-in simulation as a possible extension of his work, and his thesis provides an excellent starting point for research into accomplishing just that. If a student was able to build off of Burdorf's thesis by performing a Zoom-in simulation at Reed to a low enough redshift, they could then feasibly use \yt and \trident to perform an analysis of the CGM. 
	
	Assuming a student is able to get less noisy data with distinct absorption features by using \trident on simulation data, they could then %perform quantitative analysis on said data using a process known as Voigt profile analysis to extract physical information in a manner that imitates observational  as suggested in \cite{trident}. 
	attempt to to extract physical information in a manner that imitates observational techniques, as suggested in \cite{trident}, through a procedure known as Voigt profile analysis. The \trident method paper \cite{trident} points to \cite{romeel-voigt} and \cite{automated-voigt} as appropriate resources for understanding algorithms that can used to extract Voigt profile information from synthetic data produced using \trident.
	
	Finally, successful application of Voigt profile analysis to a \trident spectrum could be extended into a project involving analysis of randomly generated sightlines through a simulated galaxy's CGM. This procedure is used to account for the fact that scientists on Earth do not have control %over which quasar sightlines can be used 
	the sightlines that can be used for CGM analysis in the real universe, and more information can be found in \cite{strawn} as well as in Ch. 4 of \cite{butsky}, with the latter providing a good description of the algorithm used to produce random sightlines. Those who are not interested in writing their own code for such a project can take advantage of \texttt{quasarscan} \cite{quasarscan}, the documentation for which can be found at \href{https://quasarscan.readthedocs.io/en/latest/index.html}{https://quasarscan.readthedocs.io/en/latest/index.html}.
	
	%less noisy 
	
	%that his work can be extended 
	
	%Mitchell Burdorf \cite{mitch} wrote a thesis at Reed in 2021 for which he performed a cosmological N-body simulation. Burdorf suggests that others could extend his work to perform Zoom-in simulations by t
	
	%This is further supported by the fact that COS, a major tool in modern observational studies of the CGM \cite{tumlinson}, is best suited for observing the CGM at lower redshifts 
	
	%results could be improved by working with lower-redshift datasets. Since COS 
	
	%Since contemporary research into the CGM
	
	%, using both observational and synthetic data,
	
	%supernova and stellar winds 
	
	%and therefore must have at one point been part of the galaxy. %These heavier elements, therefore, are most likely the remnants of matter that has been ejected from the galaxy
	%As outflows of baryonic matter pass out of the galaxy and through the CGM, some particles get left behind as ``tracers'' which can serve as clues for the origin 
	
	%More precisely, the CGM is defined to be the medium outside of a galaxy's Interstellar Medium (ISM) but still within its virial radius \textbf{[TUMLINSON \& RYDEN]}
	
	%, though there is some ambiguity with regards to the precise boundaries of the CGM, it is generally defined as being

    \appendix
      \chapter{Determining Wavelength Range for Custom Spectra}\label{appendix:specrange}
      
      This appendix contains the code from a file that I wrote to help find custom wavelength ranges that would capture all potential absorption features for a specified set of ion species within a provided dataset. The code uses Eq.~\ref{eq:wavelength shift}, and takes advantage of a file in \trident's directory tree called \texttt{lines.txt}. This file provides a table of absorption features for an extensive set of ions in their rest frame (i.e. without redshift), and can be found from the \trident directory at \texttt{trident/data/line\_lists/lines.txt}. This code functions by reading \texttt{lines.txt} in order to create a dictionary called \texttt{absorption\_dict} that is indexed by element name (e.g. `Mg'), and whose entries are two-tuples representing the minimum and maximum wavelengths of absorption features associated with the ion species of that element. This dictionary is then used by a function called \texttt{findShiftedLambda()} that takes two arguments: \texttt{data}, which is a data file loaded into \yt via the \texttt{yt.load()} function, and \texttt{element}, which is a character array representing the names of elements whose ion species are to be considered (e.g. `H' or `Mg').
      %that takes as argument a data file loaded into \yt and an character array representing the names of the ion species to consider (e.g. 'H' or 'Mg')
      
\begin{singlespace}
\begin{lstlisting}[language=Python, style=mystyle]
# File: specrange.py
#

import yt
import trident

# Ranges of rest wavelengths for absorption features:
#
# <Element> : [<min_wavelength>, <max_wavelength>]
#
# Obtained from "lines.txt" on the trident GitHub page.
#

line_list = trident.__file__[:-11] + "data/line_lists/lines.txt"

f_lines = open(line_list).readlines()

absorption_dict = {}

for line in f_lines[1:]:
    line = line.split(None)
    try:
        absorption_dict[line[0]].append(float(line[2]))
    except KeyError:
        absorption_dict[line[0]] = [float(line[2])]
    
for elem in absorption_dict:
    absorption_dict[elem] = (min(absorption_dict[elem]),
                             max(absorption_dict[elem]))


# Function for determining the wavelength range of
# absorption features, taking into account current redshift.
# Takes inputs "data" (YT data set) and "element" (string
# representing an element in the lambda0 dictionary).
#
lambda0 = absorption_dict

def findShiftedLambda(data, element):
    z = data.current_redshift
    lambda_min = (1 + z) * lambda0[element][0]
    lambda_max = (1 + z) * lambda0[element][1]
    return([lambda_min, lambda_max])


\end{lstlisting}
\end{singlespace}
      
    %\chapter{Additional Spectra}\label{appendix:final}

%This is where endnotes are supposed to go, if you have them.
%I have no idea how endnotes work with LaTeX.

  \backmatter % backmatter makes the index and bibliography appear properly in the t.o.c...

% if you're using bibtex, the next line forces every entry in the bibtex file to be included
% in your bibliography, regardless of whether or not you've cited it in the thesis.
    \nocite{*}

% Rename my bibliography to be called "Works Cited" and not "References" or ``Bibliography''
% \renewcommand{\bibname}{Works Cited}

%    \bibliographystyle{bsts/mla-good} % there are a variety of styles available; 
%  \bibliographystyle{plainnat}
% replace ``plainnat'' with the style of choice. You can refer to files in the bsts or APA 
% subfolder, e.g. 
 %\bibliographystyle{APA/apa-good}  % or
 %\nocite{apsrev41control}
 \bibliographystyle{apsrev}
 \bibliography{thesis}
 % Comment the above two lines and uncomment the next line to use biblatex-chicago.
 %\printbibliography[heading=bibintoc]

% Finally, an index would go here... but it is also optional.
\end{document}